\documentclass[draft]{agujournal2019}
\usepackage{url} %this package should fix any errors with URLs in refs.
\usepackage{lineno}
\usepackage[inline]{trackchanges} %for better track changes. finalnew option will compile document with changes incorporated.
\usepackage{soul}
\usepackage{setspace}
\usepackage{gensymb}
\usepackage{comment}
\usepackage{multirow}
\usepackage{amsmath}
\usepackage{float}
\usepackage{adjustbox}
\usepackage{graphicx}
\usepackage{bm} % Make symbols and text bold
\usepackage{mfirstuc}
\usepackage{upgreek} % Package to write non-italic greek symbols
\usepackage{lscape} % Change certain page to landscape mode
\usepackage{xcolor} % Colored text
\usepackage{ragged2e} % Package for text alignment

\usepackage{natbib}
\usepackage{makecell}
\justifying
\newcommand{\citeg}[1]{\citep[e.g.,][]{#1}}
\usepackage{float}

\draftfalse
 
\journalname{JAMES}

\begin{document}

%% ------------------------------------------------------------------------ %%
%  Title
%
% (A title should be specific, informative, and brief. Use
% abbreviations only if they are defined in the abstract. Titles that
% start with general keywords then specific terms are optimized in
% searches)
%
%% ------------------------------------------------------------------------ %%

% Example: \title{This is a test title}

\title{Data Imbalance, Uncertainty  Quantification, and Generalization via Transfer  Learning in Data-driven Parameterizations: Lessons from the Emulation of Gravity Wave Momentum Transport in WACCM}

%% ------------------------------------------------------------------------ %%
%
%  AUTHORS AND AFFILIATIONS
%
%% ------------------------------------------------------------------------ %%

% Authors are individuals who have significantly contributed to the
% research and preparation of the article. Group authors are allowed, if
% each author in the group is separately identified in an appendix.)

% List authors by first name or initial followed by last name and
% separated by commas. Use \affil{} to number affiliations, and
% \thanks{} for author notes.
% Additional author notes should be indicated with \thanks{} (for
% example, for current addresses).

% Example: \authors{A. B. Author\affil{1}\thanks{Current address, Antartica}, B. C. Author\affil{2,3}, and D. E.
% Author\affil{3,4}\thanks{Also funded by Monsanto.}}

\authors{Y. Qiang Sun\affil{1}, Hamid A. Pahlavan\affil{1}, Ashesh Chattopadhyay\affil{1}, Pedram Hassanzadeh\affil{1}, Sandro W. Lubis\affil{1,3}, M. Joan Alexander\affil{2}, Edwin Gerber\affil{4}, Aditi Sheshadri\affil{5}, Yifei Guan\affil{1}}

\affiliation{1}{Rice University, Houston, TX, 77005}
\affiliation{2}{NorthWest Research Associates, Boulder, CO, 80301, USA}
\affiliation{3}{Pacific Northwest National Laboratory, Richland, WA, 99354, USA}
\affiliation{4}{New York University, New York, NY 10012, USA}
\affiliation{5}{Stanford University, Stanford, CA, 94305, USA}

%\affiliation{=number=}{=Affiliation Address=}
%(repeat as many times as is necessary)

%% Corresponding Author:
% Corresponding author mailing address and e-mail address:

% (include name and email addresses of the corresponding author.  More
% than one corresponding author is allowed in this LaTeX file and for
% publication; but only one corresponding author is allowed in our
% editorial system.)

% Example: \correspondingauthor{First and Last Name}{email@address.edu}

\correspondingauthor{Y. Qiang Sun and Pedram Hassanzadeh}{ys91@rice.edu and pedramh@uchicago.edu}
%\correspondingauthor{}{pedramh@uchicago.edu}

%% Keypoints, final entry on title page.

%  List up to three key points (at least one is required)
%  Key Points summarize the main points and conclusions of the article
%  Each must be 140 characters or fewer with no special characters or punctuation and must be complete sentences

% Example:
% \begin{keypoints}
% \item	List up to three key points (at least one is required)
% \item	Key Points summarize the main points and conclusions of the article
% \item	Each must be 140 characters or fewer with no special characters or punctuation and must be complete sentences
% \end{keypoints}

\begin{keypoints}

\item WACCM’s orographic, convective, and frontal gravity wave parameterizations are emulated using neural nets to inform future modeling efforts

\item Out-of-distribution generalization (extrapolation) of the neural nets under $4\times$CO$_2$ forcing is enabled via transfer learning with $1\%$ new data

\item Data imbalance is addressed via resampling and weighted loss; uncertainty quantification via Bayesian, dropout, and variational methods
\end{keypoints}

%% ------------------------------------------------------------------------ %%
%
%  ABSTRACT and PLAIN LANGUAGE SUMMARY
%
% A good Abstract will begin with a short description of the problem
% being addressed, briefly describe the new data or analyses, then
% briefly states the main conclusion(s) and how they are supported and
% uncertainties.

% The Plain Language Summary should be written for a broad audience,
% including journalists and the science-interested public, that will not have 
% a background in your field.
%
% A Plain Language Summary is required in GRL, JGR: Planets, JGR: Biogeosciences,
% JGR: Oceans, G-Cubed, Reviews of Geophysics, and JAMES.
% see http://sharingscience.agu.org/creating-plain-language-summary/)
%
%% ------------------------------------------------------------------------ %%

%% \begin{abstract} starts the second page

\begin{abstract}
Neural networks (NNs) are increasingly used for data-driven subgrid-scale parameterization in weather and climate models. While NNs are powerful tools for learning complex nonlinear relationships from data, there are several challenges in using them for parameterizations. Three of these challenges are 1) data imbalance related to learning rare (often large-amplitude) samples; 2) uncertainty quantification (UQ) of the predictions to provide an accuracy indicator; and 3) generalization to other climates, e.g., those with higher radiative forcing. Here, we examine performance of methods for addressing these challenges using NN-based emulators of the Whole Atmosphere Community Climate Model (WACCM) physics-based gravity wave (GW) parameterizations as the test case. WACCM has complex, state-of-the-art parameterizations for orography-, convection- and frontal-driven GWs. Convection- and orography-driven GWs have significant data imbalance due to the absence of convection or orography in many grid points. We address data imbalance using resampling and/or weighted loss functions, enabling the successful emulation of parameterizations for all three sources. We demonstrate that three UQ methods (Bayesian NNs, variational auto-encoders, and dropouts) provide ensemble spreads that correspond to accuracy during testing, offering criteria on when a NN gives inaccurate predictions. Finally, we show that accuracy of these NNs decreases for a warmer climate ($4\times$CO$_2$). However, the generalization accuracy is significantly improved by applying transfer learning, e.g., re-training only one layer using $\sim 1\%$ new data from the warmer climate. The findings of this study offer insights for developing reliable and generalizable data-driven parameterizations for various processes, including (but not limited) to GWs.

\end{abstract}

\section*{Plain Language Summary}
Scientists are increasingly using machine learning methods, especially neural networks (NNs), to improve weather and climate models. However, it can be challenging for a NN to learn rare, large-amplitude events, because they are infrequent in training data. Also, NNs need to express their confidence (certainty) about a prediction and work effectively across different climates, e.g., warmer climates due to increased CO$_2$. Traditional NNs often struggle with these challenges. Here, we share insights gained from emulating the complex physics-based parameterization schemes for gravity waves in a state-of-the-art climate model. We propose specific strategies for addressing imbalanced data, uncertainty quantification (UQ), and making accurate predictions across various climates. For instance, to manage data balance, one such strategy involves amplifying the impact of infrequent events in the training data. We also demonstrate that several UQ methods could be useful in determining the accuracy of predictions. Furthermore, we show that NNs trained on simulations of the historical period do not perform as well in warmer climates. However, we improve the NNs’ performance by employing transfer learning using limited data from warmer climates. This study provides lessons for developing robust and generalizable approaches for using NNs to improve models in the future.

%% ------------------------------------------------------------------------ %%
%
%  TEXT
%
%% ------------------------------------------------------------------------ %%

%%% Suggested section heads:
% \section{Introduction}
%
% The main text should start with an introduction. Except for short
% manuscripts (such as comments and replies), the text should be divided
% into sections, each with its own heading.

% Headings should be sentence fragments and do not begin with a
% lowercase letter or number. Examples of good headings are:

% \section{Materials and Methods}
% Here is text on Materials and Methods.
%
% \subsection{A descriptive heading about methods}
% More about Methods.
%
% \section{Data} (Or section title might be a descriptive heading about data)
%
% \section{Results} (Or section title might be a descriptive heading about the
% results)
%
% \section{Conclusions}

\section{Introduction}

Small-scale processes such as moist convection, gravity waves, and turbulence are key players in the variability of the climate system and its response to increased greenhouse gases. However, as these processes cannot be resolved, entirely or partially, by the coarse-resolution general circulation models (GCMs), they need to be represented as functions of the resolved dynamics via subgrid-scale (SGS) parameterization schemes  \citeg{Kimetal2003, Stensrud2007, prein2015review}. Many of these parameterization schemes are based on heuristic approximations and simplifications, introducing large parametric and epistemic uncertainties in GCMs~\citep{Schneider2017, hourdin2017art, palmer2019stochastic}. 

Recently, there has been a growing interest in developing data-driven SGS parameterizations for different complex processes in the Earth system using machine learning (ML) techniques, particularly deep neural networks (NNs). Promising results have been demonstrated in a wide range of idealized applications, including prototype systems \citep{maulik2019subgrid, gagne2020machine, Rasp2020online, chattopadhyay2020data, frezat2022posteriori, guan2022stable, pahlavan2023explainable}, ocean turbulent processes \citep{bolton2019applications,zhang2023implementation}, moist convection in the atmosphere \citep{GormanDwyer2018,brenowitzBretherton2019,yuval2020stable, BeuclerPritchard2021cloud, iglesiassuarez2023causallyinformed}, radiation \citep{Krasnopolsky2005,belochitskiKrasnopolsky2021,songetal2021radiation}, and microphysics \citep{SeifertRasp2020, Gettelman2021}. The ultimate promise of data-driven parameterizations, learned from observation-derived data and/or high-fidelity high-resolution simulations, is that they might have smaller parametric/structural errors, thus reducing the biases of GCMs and producing more reliable climate change projections~\citeg{Schneider2017, Reichstein2019, schneider2021accelerating}. 

However, there are major challenges in developing trustworthy, interpretable, stable, and generalizable data-driven parameterizations that can be used for such climate change projection efforts. Discussing and even listing all of these challenges is well beyond the scope of this paper. Well-known challenges such as interpretability and stability have been extensively discussed in a number of recent studies~\citeg{McGovern2019, beck2019deep, Brenowitz2020instability,   balaji2021climbing, clare2022explainable, mamalakis2022investigating, guan2022stable, subel2023explaining, pahlavan2023explainable}. Here, we focus on three other key issues:

\begin{enumerate}
    \item Data imbalance, and related to that, learning rare/extreme events, 
    \item Uncertainty quantification (UQ) of the NN-based SGS parameterization outputs, 
    \item Out-of-distribution (OOD) generalization (e.g., extrapolation to climates with higher radiative forcings).
\end{enumerate}

Below we briefly discuss the importance of 1-3 and the current state-of-the-art methods in addressing them in the climate and ML literature. Data imbalance is a well-known problem in the ML literature, especially in the context of classification tasks \citeg{Japkowicz2002, WuChang2003, Chawla_etal_2004,  Sunetal2009DB, huang2016learning, ando2017deep, BUDA2018249,  Johnson2019}. The problem becomes particularly significant when one aims to learn rare/extreme events~\citep{maalouf2011robust, maalouf2014weighted, baldi2014searching, LiuDNNextreme2016, GormanDwyer2018, QiMajda2020, chattopadhyay2020analog, PhysRevFluids.8.040501, Finkeletal2023, ShamekhGentine2023PNAS}. For example, suppose we aim to learn the binary classification of the 99 percentile of temperature anomalies using a NN. In this case, label 0 (no extreme) will constitute $99 \%$ of the training (or testing) set while label 1 (extreme) will be just $1 \%$. With many common loss functions such as mean squared error (MSE) or root-mean squared-error (RMSE), training a NN will result in one that predicts 0 for any sample (extreme or no extreme) while having a seemingly high accuracy of $99 \%$ (of course, other metrics such as precision/recall will show the shortcoming, see \citet{chattopadhyay2020analog}). The most common remedy to this problem for classification tasks is resampling. An example is down-sampling non-extreme cases by a factor of 100, which effectively balances the dataset. 

In addition to \textit{classification} tasks, Data imbalance also presents a significant challenge in \textit{regression} tasks required for parameterization schemes in climate models. As highlighted by \citet{chantry2021}, such imbalances contributed to the unsuccessful emulation of their orographic gravity wave parameterization (GWP) scheme, largely because orography affects the gravity wave (GW) drag in only a fraction of the grid columns. This challenge also persists in emulating GWP for non-orographic GWs, especially when GWs are intricately linked to their sources. For instance, the presence of zero convective GW drag at numerous grid points due to the absence of convection creates a notably imbalanced dataset. This issue will be explored further in the results section. In regression tasks, data imbalance may also manifest in the form of difficulty in learning large-amplitude (extreme) outputs, which are rare and constitute only a small fraction of the training set. In the case of GWs, Observations have shown that gravity wave amplitudes are highly intermittent such that the largest 10\% events alone can contribute more than 50\% of the total momentum flux \citep{hertzog2012intermittency}, so the extreme events will contribute an outsized fraction of the total drag. Nonetheless, poorly learning these large-amplitude outputs, like drag forces, can result in instabilities ~\citeg{guan2022stable}. Addressing data imbalance in climate applications has received relatively limited attention. In this study, we propose several remedies based on resampling techniques and weighted loss functions, demonstrating their advantages in enabling successful emulations of all GWP schemes and improving the learning of rare/extreme events.

Quantifying the uncertainties in outputs from NN-based parameterization schemes is essential when employing these schemes, particularly for high-stakes decision-making tasks such as climate change projections. Crucially, during testing when we are unable to directly determine a prediction's accuracy, we need a UQ method that can provide a credible {\it confidence level} for each prediction, serving as a reliable indicator of its accuracy. During inference, the output of an NN can be inaccurate for various reasons, including poor approximation (e.g., due to poor NN architecture), poor within-distribution generalization (e.g., for inputs that are rare events), or poor optimization (collectively referred to as {\it epistemic uncertainty}), as well as because of OOD generalization errors due to input samples from a distribution different from that of the training set \citep{ABDAR2021243,  lu2021learning, krueger2021out, miller2021accuracy, shen2021towards, wu2021quantifying, ye2021towards, zhang2021can,  subel2023explaining}. Quantifying the level of uncertainty would then allow us to avoid using a data-driven parameterization scheme when it is inaccurate due to one of the aforementioned reasons~\citep{maddox2019simple, zhu2019physics, li2022uncertainty, psaros2023uncertainty}. In the context of data-driven parameterization in climate modeling, the two most challenging sources of uncertainty are rare/extreme events and OOD generalization errors. The latter is a concern, particularly when the GCM is used for climate change studies (see below for more discussions). 

Developing UQ methods for NNs is an active area of research in the ML community, and there is not a generally applicable rigorous method yet. For instance, techniques like Markov-Chain Monte Carlo can be prohibitively expensive, especially when dealing with high-dimensional systems ~\citep{oh2005data, ballnus2017comprehensive, chen2019new}. For a comprehensive review in the context of scientific ML, refer to \citet{psaros2023uncertainty}. The topic has also started to increasingly gain attention in the climate literature ~\citep{guillaumin2021stochastic, Gordon2022, haynes2023creating, barnes2023sinh}. In this study, we will assess the performance of three common UQ methods (Bayesian, dropout, and variational NNs) by analyzing the relationships between uncertainty and accuracy during inference testing. We will also consider scenarios involving OOD generalization errors resulting from global warming.

As already mentioned above, OOD generalization (extrapolation to a test data distribution different from that of the training set) is a major challenge for applications involving non-stationarity, like a changing climate. Studies have already shown that the lack of OOD generalization in data-driven parameterizations leads to inaccurate and unstable simulation~\citep{rasp2018deep, GormanDwyer2018,chattopadhyay2020data,   guan2022stable,nagarajan2020understanding}. A general and powerful method for improving the OOD generalization capability of NNs is transfer learning (TL), which involves re-training a few or all of the layers of a NN using a small amount of data from the new system~\citep{yosinski2014transferable}. This approach has already shown remarkable success in enabling data-driven parameterization schemes to extrapolate across the parameter space (e.g., to $100 \times $ higher Reynolds number) in canonical test cases~\citep{chattopadhyay2020data,subel2021data,guan2022learning,subel2023explaining,zhang2023implementation}. In particular, \citet{subel2023explaining} introduced SpArK (Spectral Analysis of Regression Kernels and Activations) showing that re-training even one layer can lead to successful OOD generalization, although this optimal layer, unlike the rule of thumb in the ML literature, may not be the deepest but the shallowest hidden layer. Here, we further leverage these studies and show how TL can enable OOD generalization of data-driven parameterization schemes in state-of-the-art GCMs.

The methods used in this study and the learned lessons apply to a broad range of processes and applications in climate modeling. However, the results are presented for a single test case, that is based on the emulation of complex physics-based GWP schemes in version 6 of the Whole Atmosphere Community Climate Model (WACCM), a state-of-the-art GCM~\citep{Gettelman2019}. Here, we use the emulations of current physics-based parameterization schemes as a stepping stone towards learning data-driven parameterizations from observations and high-fidelity simulations by testing ideas for addressing items 1-3 listed earlier. Furthermore, developing better representations of un- and under-resolved GWs in GCMs is an important problem on its own \citep{Kimetal2003, Alexander2010QJRMS, Achatz2022}. A number of recent studies have taken the first steps in learning data-driven GWP from observations and high-resolution simulations \citep{Matsuoka2020, Amiramjadi2022, Sunetal2023WRF, Dongetal2023}, though careful and time-consuming steps are needed in producing, analyzing, and using such data. Furthermore, two recent studies focused on emulators of simpler GWP schemes in a forecast model and idealized GCM have readily shown the usefulness of lessons learned from emulators \citep{chantry2021, Espinosa2022, hardiman2023machine}. This further motivates the focus on using emulators for testing ideas for addressing data imbalance, UQ, and OOD generalization.

This paper is structured as follows. Section \ref{sec: Method} introduces the WACCM simulations and the NN architectures used in this study. The findings, detailed in Section \ref{sec: Result}, emphasize the insights gained in addressing data imbalance and UQ, alongside OOD generalization of the emulators under warmer climate conditions. Consistent with \citet{chantry2021}, we find that using an NN to emulate the parameterization of orographic GWs is significantly more challenging than non-orographic GWs. This necessitated additional steps to achieve reasonable offline performance, as detailed in Section \ref{ORO}. To the best of our knowledge, this stands as the first NN-based emulation of orographic GWs to address the challenges in \citet{chantry2021}. Finally, we provide a concluding summary in Section \ref{Summary}.

\section{Data and Methods} \label{sec: Method}

\subsection{The Whole Atmosphere Community Climate Model (WACCM)}

The NCAR's WACCM version 6 introduced in \cite{Gettelman2019} is used in this study. WACCM has state-of-the-art GWP schemes for GWs from three different sources: orography (OGWs), convection (CGWs), and fronts (FGWs). These complex sources make the emulation of the GWP schemes in WACCM a challenging task. This is, therefore, a suitable test case to investigate ideas for learning rare events, UQ, and OOD generalization to benefit the future efforts for the much more complex task, that is learning data-driven GWP schemes from observations and/or high-resolution GW-resolving simulations \citep{Amiramjadi2022, Sunetal2023WRF}. 

The configuration of the WACCM used in this study is identical to the public version in \cite{Gettelman2019}, with a horizontal resolution of $0.95 {^\circ} \times 1.25 {^\circ}$ and 70 vertical levels. The two non-orographic GWP schemes in WACCM both follow \cite{Richter2010}, yet allow separate specifications of FGW and CGW sources. 
For OGWs, WACCM uses an updated planetary boundary layer form drag scheme from \cite{Beljaars2004}, near-surface nonlinear drag processes following \cite{Scinocca2000}, and a ridge-finding algorithm to define orographic sources based on \cite{Bacmeister1994}. A full documentation of WACCM OGWs can also be found in \cite{Kruse2022}.

 We conduct two sets of simulations: A 10-year pre-industrial ``control'' run, and a 10-year pseudo-global-warming ``future'' run with 4$\times$CO$_2$ and uniform +4 K sea-surface temperature increases. In each run, we save, on the native grid, all the inputs and outputs for each of the three GWPs every 3 hours to capture the diurnal cycle. A complete list of these inputs/outputs, which are used in the training of the NN-based emulators, is presented in \ref{sec:appA}. 
 
 We train separate NNs for emulating the three GWP schemes that have different sources. We use the first 6 years of the control run for training and the last 4 years for validation (years 7 and 8) and testing (years 9 and 10). With a grid resolution of $\sim$1${^\circ}$, there are 55,296 columns for each time snapshot, resulting in approximately 960 million input/output columns during the 6-year training period. Given the strong temporal correlation between the 3-hourly samples, we perform sub-sampling on both the training and validation data to reduce the dataset size. To accomplish this, we begin by shuffling all the input/output column pairs in time at each latitude/longitude grid point. Then, we randomly select 2,000 input/output pairs at each location for training and 500 pairs for validation.

\begin{figure}[t]
    \centering
    \includegraphics[width=33pc]{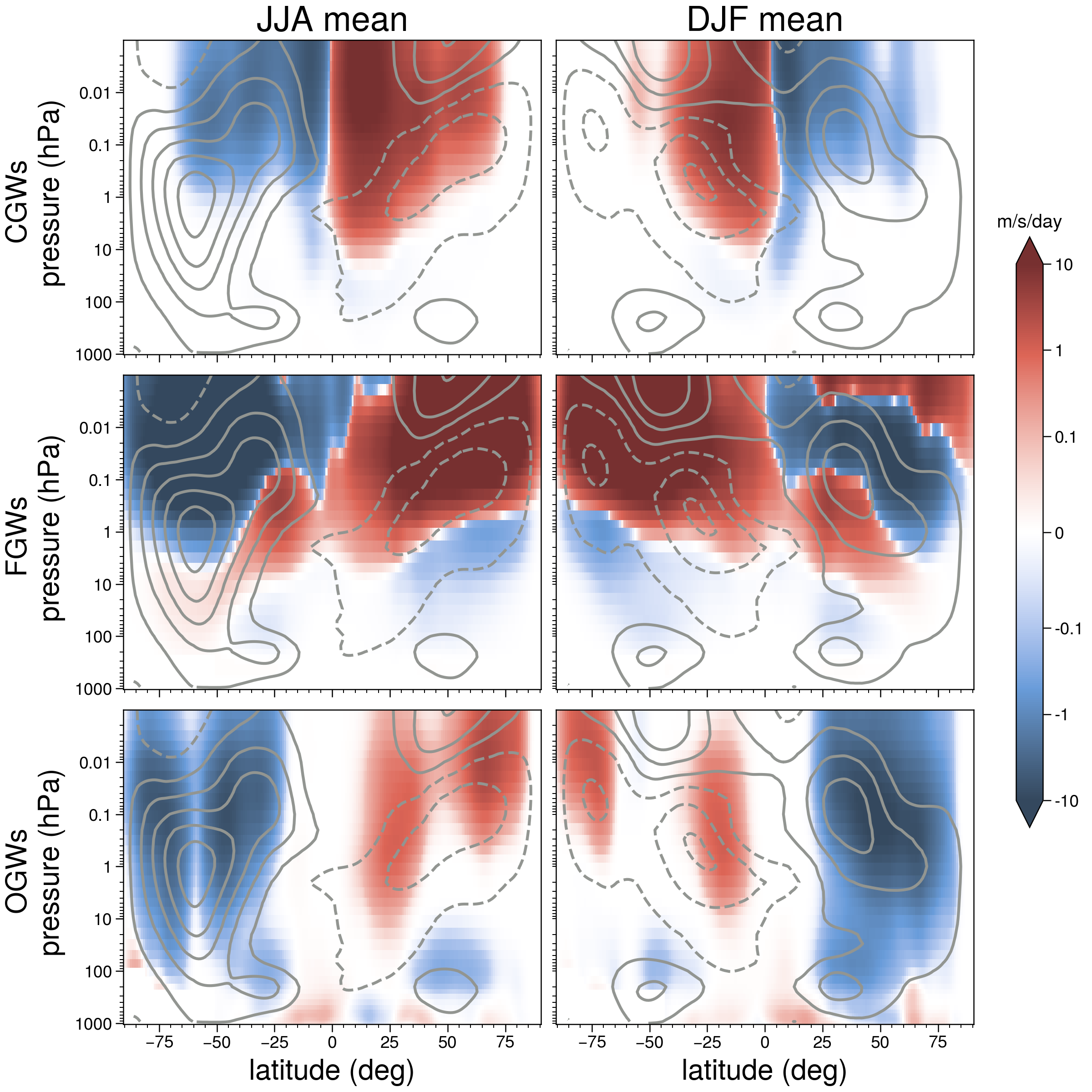}
    \caption{Climatology of zonal-mean GWD during summers (JJA) and winters (DJF) from all 3 sources in the control (pre-industrial) WACCM simulations. Top: CGWs; middle: FGWs; bottom: OGWs. The climatology of the zonal-mean zonal wind is also shown (grey lines), with an interval of 20 m/s. Dashed lines indicate negative values. Zero lines are omitted.}
    \label{fig: climate}
\end{figure}

To give the readers a general idea of the parameterized GWs and large-scale circulation in WACCM, Figure \ref{fig: climate} shows the zonal-mean climatology for zonal GW drag/forcing, hereinafter referred to as GWD, arises from the divergence of gravity wave momentum transport (fluxes), from all three sources, computed from the 6-year training period in the control run. The zonal-mean zonal wind climatology is also shown. Seasonal dependency for both the GWD and the circulation is observed in the simulations. At levels below 100 hPa, the tendencies of non-orographic GW  are relatively small compared to those from OGWs; however, their amplitudes increase significantly at higher altitudes. While the parameterized effect of GWs is generally to decelerate the zonal flow, there are exceptions, notably in regions like the equatorward flanks of the stratospheric polar night jets, where FGWs can accelerate the flow. For more information on the GWP schemes and circulations in WACCM, see \citet{GarciaOGW-WACCM2017} and \citet{Gettelman2019}.

\subsection{The NNs and UQ}
\label{sec:ANN}
\subsubsection{The Deterministic Fully Connected NN}
Here we briefly describe the general structure of the NN-based regression models trained as emulators for GWP schemes. For the deterministic artificial NN, denoted as ANN in this study, we use multilayer perceptrons (MLP). MLPs, which are feedforward fully connected NNs, take inputs through successive layers of linear transformation and non-linear activation functions to produce an output, so as to learn a functional relationship between the input and output (Figure~\ref{fig: ANN}a). Deep MLPs have multiple layers of weights, which are optimized over many samples of input-output data pairs. Such MLPs are thus very powerful in terms of learning complicated functional relationships. Generally, we can write the governing equations of an MLP as
\begin{eqnarray}
z^\ell=\sigma \left(W^\ell z^{\ell-1}+b^\ell\right),
\label{eqn. neural network function}
\end{eqnarray}
where $z^\ell$ is the activation (output) of layer $\ell$, $W^\ell$ is the weight matrix connecting layers $\ell$ and $\ell-1$, and $b^\ell$ is the bias at layer $\ell$, which allows the network to fit the data even when all input features are equal to 0. $\sigma$ is the non-linear activation function.

In this study, we employ the same NN structure while training three distinct NNs, each for GWP originating from one of the three unique GW sources. The input layer contains the same input variables (see \ref{sec:appA}) used by the WACCM GWPs across all vertical levels. There are 10 hidden layers in total (Figure~\ref{fig: ANN}a), and there are $500$ neurons in each hidden layer. In the output layer, both zonal and meridional GWD are predicted. The activation function in each layer, $\sigma$, is chosen to be swish \citep{ramachandran2017searching}, except for the output layer, where it is linear. During training, $W^\ell$ and $b^\ell$ are randomly initialized and learned by minimizing a loss function using an ADAM optimizer, with a fixed learning rate of $\alpha=0.0001$. One of the loss functions used here is the common MSE, i.e.,
\begin{equation}~\label{eq: Loss_mse}
%\mathcal{L}(y, \hat{y}; {\Theta}) = || y\left({\Theta}\right)  - \hat{y} ||_{MSE} 
% \mathcal{L}({\Theta}) = \frac{1}{N \times M} \sum_{i=1}^{N} \sum_{z=1}^{M} \left({\rm\bf{NN}}\left({x}_{i},{\Theta}\right) - {y}_{iz} \right)^2.
\mathcal{L}({\Theta}) = \frac{1}{n} \sum_{i=1}^{n} \,\,  \Bigl\| \mathbf{NN}\left({x}_{i},{\Theta}\right) -  {y}_{i} \Bigl\|_2^2
\end{equation}
Here, $n$ is the number of training samples and $\|.\|_2$ is the $L_2$ norm. For training sample $i$, vector $x_i$ contains all the inputs to the NN (\ref{sec:appA}), vector $y_{i}$ contains the true zonal and meridional GWD at each vertical level, and ${\Theta} = \{\theta_j\}_{j = 1 \cdots p}$ denotes the trainable parameters, i.e., the weights ($p\approx 3\times 10^6$). 

\subsubsection{The UQ Methods and Metrics} \label{sec: UQ methods}
Although deterministic NNs are powerfully expressive and can exhibit high out-of-sample predictive skills, they do not provide estimates of the uncertainty associated with their predictions. As mentioned earlier, currently there is no rigorous method to estimate the uncertainty of an NN prediction. That said, a variety of techniques have been developed for UQ in NNs, though the validity and usefulness of the estimated uncertainty for scientific applications remain subjects of ongoing investigations ~\citeg{psaros2023uncertainty,haynes2023creating}. In this paper, we use three different and widely used approaches to perform UQ from the ML literature: Bayesian neural network (BNN), dropout neural network (DNN), and variational auto-encoder (VAE). A brief overview of these approaches is provided below.

\begin{figure}[ht]
    \centering
    \includegraphics[width=35pc]{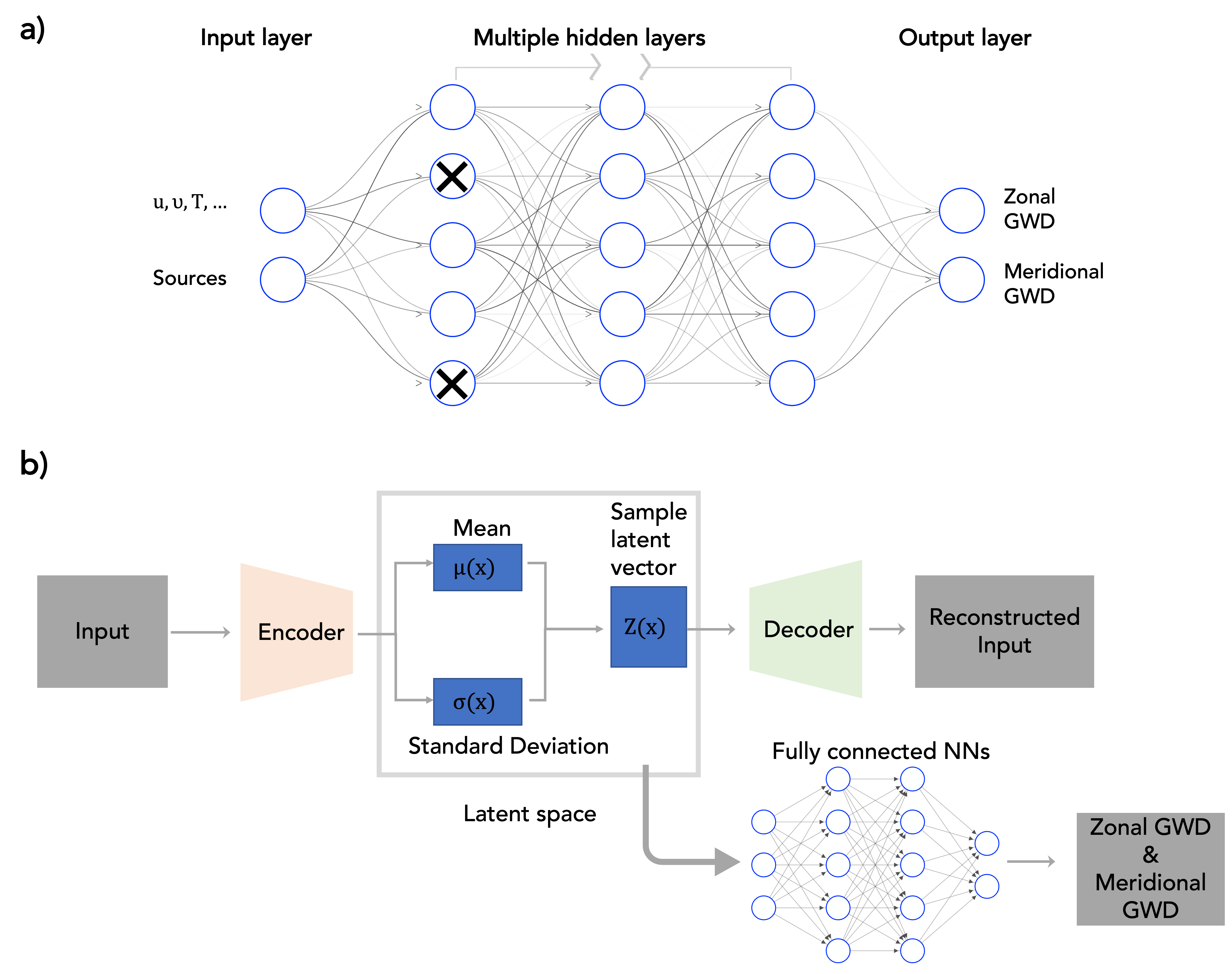}
    \caption{Schematics of the NN-based emulators and different training/re-training strategies used in this study. (a) Schematic for the MLP and DNN. The inputs of the NN are connected through successive layers of neurons (blue circles) to the output (GWDs). A fully connected MLP NN is trained from randomly initialized weights and biases in all layers.  A DNN is the same but some connections are randomly eliminated during training and inference (black crosses). In TL, only some of the layers of a previously trained MLP are re-trained using new data. (b) Schematic for the VAE. A low-dimensional latent space is constructed and then used as the input for the additional fully connected NNs, which is similar to the one in (a).}
    \label{fig: ANN}
\end{figure}

\textit{Bayesian neural network (BNN)}: A BNN combines the deterministic NN described earlier and in Figure~\ref{fig: ANN}a with Bayesian inference \citep{blundell2015weight}. Simply speaking, a BNN estimates distributions of the weights, rather than point values (as in a deterministic NN).  The posterior distributions in the BNN (i.e., the distributions of the weights and biases) are calculated using the Bayes rule. In this study, we follow the standard practice and assume that all variational forms of the posterior are normal distributions. Furthermore, to accelerate the training process, we use the normal distribution $\mathcal{N} (\mu, 1)$ for all the priors in the BNN (where $\mu$ is obtained from parameters of the trained deterministic NN). Note that while we are assuming normal distributions for the trainable parameters, the predictions generated by BNN can fit different distributions due to the use of nonlinear activation functions. The resulting distribution of the predictions during inference gives an estimate of their uncertainty.

\textit{Dropout neural network (DNN)}: A DNN is developed by randomly eliminating all outgoing connections from some of the nodes (Figure~\ref{fig: ANN}a) in each hidden layer of a deterministic NN during the training and the inference \citep{srivastava2014dropout}. The fraction of nodes ``dropped'' on average in each layer is called the dropout ratio. Mathematically, Equation~(\ref{eqn. neural network function}) can be reformulated for a DNN as:
\begin{eqnarray}~\label{eqn. DROPOUT}
z^\ell=\sigma \left(D^\ell W^\ell z^{\ell-1}+b^\ell \right),
\end{eqnarray}
where the dropout matrix $D^\ell$ is a square diagonal binary matrix of integers 0 or 1. The diagonal elements of $D^\ell$ follow a Bernoulli distribution where the probability of zero is the dropout ratio.
    
Dropout was initially developed as a regularization technique to prevent over-fitting in NNs. However, \citet{gal2016dropout} showed that training a NN with the dropout technique approximates a Bayesian NN. In this study, we use a dropout rate of 0.1, which is incorporated in all hidden layers, but we also investigate the sensitivity of the DNN to different dropout rates, as later shown in \ref{sec:appB}. Note that the random dropping out is also used during inference, leading to a distribution for each prediction. 

\textit{Variational auto-encoder (VAE)}: A typical VAE \citep{Kingma2014} consists of two NNs (Figure~\ref{fig: ANN}b): an encoder that transforms the input into a lower-dimensional latent space, parameterized by a normal probability distribution, and a decoder that inverts this transformation and produces the original input. The difference between the decoder's output and the original input drives the learning process of the encoder and decoder, while the parameterized lower-dimensional latent space provides the uncertainty of this transformation. The VAE was developed for generative reconstructions of data by simply drawing samples from the latent space. The VAE is basically a dimension-reduction method. Many variants, however, have been proposed for more specific purposes.  In this study, following \cite{Foster2021}, we add a third NN, as illustrated in Figure~\ref{fig: ANN}b, that randomly draws samples from the parameterized latent space as inputs, and predicts the zonal and meridional GWDs as outputs. The difference between the predicted GWDs and the true GWDs drives the learning of the third NN. Consequently, the loss for the entire network consists of three components: the loss between the reconstructed input and the original input, the Kullback–Leibler (KL) divergence between the distribution of the latent space and a normal distribution, and the loss between the predicted GWDs by the third NN and the true GWDs.

For a specific input, each of these three UQ methods discussed above can be run multiple times, generating an ensemble of predictions with different realizations of the weights by drawing from the trained distribution. This is in contrast to the deterministic NN that provides just a single-valued prediction for a given input. These ensembles can then be used to quantify the uncertainty associated with that prediction. We expect that the RMSE of the ensemble mean should exhibit approximately a 1-1 relationship with the ensemble spread (i.e., the standard deviation of the ensemble members). To investigate this relationship, we use the spread-skill plot \citep{delle2013probabilistic}. Detailed calculations behind the spread-skill plot can be found in \ref{sec:appUQ}, where we also introduce two metrics: spread-skill reliability (SSREL) and overall spread-skill ratio (SSRAT), both of which summarize the information presented in the spread-skill plot.

\subsection{Transfer Learning}
Transfer learning refers to leveraging/reusing information (weights) from an already well-trained base NN to effectively build a new NN for a different system from which only a small amount of training data is available~\citep{yosinski2014transferable, Tanetal2018, chattopadhyay2020data}. For our purpose, which is improving OOD generalization to the warmer climate, the TL procedure is as follows. For any of the NNs described earlier (e.g., the one in Figure~\ref{fig: ANN}a), we train them from randomly initialized weights and biases with data from the control simulations. The NN will work well during inference for test samples from the control but not from future (warmer climate) simulations (as shown in the Results section). To address this, TL is applied wherein most of the NN's weights are kept constant, and only one or two hidden layers are re-trained using a limited dataset from the future simulation. Although this small dataset is insufficient for training an entire NN from random initialization, careful and correct selection of hidden layers for re-training, as discussed in \cite{subel2023explaining}, allows the development of an NN that accurately adapts to the new system, i.e., the future climate conditions.

Here, we re-train the NN-based emulator that was initially trained on the control data with new data from only 1 month (30 consecutive days) of integration (1.4\% of 6 years simulation for the initial training) of WACCM model under future forcing (4$\times$CO$_2$). We have explored different choices of layers to re-train with the same amount of new data and found that re-training the first hidden layer yields the best results, consistent with \citet{subel2023explaining}. Therefore, the results with only re-training the first hidden layer are shown in Section \ref{sec: Result} unless stated otherwise.

\section{Results} \label{sec: Result}

\subsection{Data Imbalance}

\begin{figure}[ht]
    \centering
    \includegraphics[width=20pc]{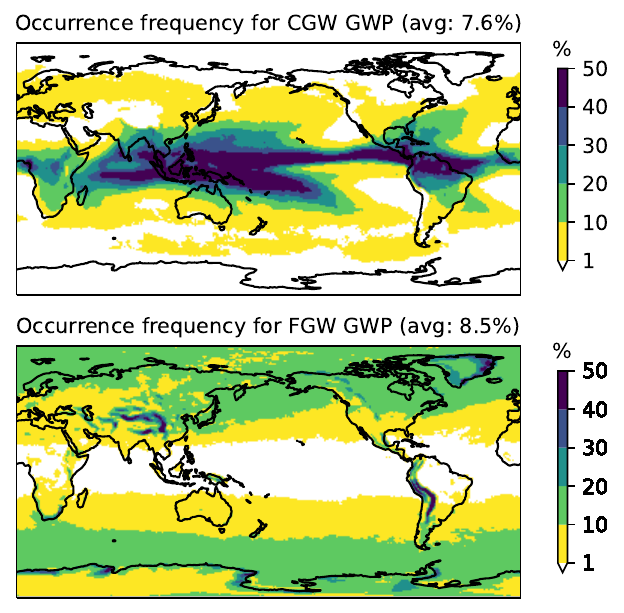}
    \caption{Distribution of occurrence frequency for CGWs (top) and FGWs (bottom) in the WACCM pre-industrial control simulations, based on the average of the 6-year training dataset.}
    \label{fig: non-zero ratio}
\end{figure}

As discussed earlier, the physics-based GWP schemes in WACCM are directly linked to their sources. This means they only produce non-zero values when their respective sources are active. For example, in a specific grid box, CGWs only register non-zero values when there is active convection within that box. The heterogeneous and sometimes intermittent nature of these sources leads to a dataset that is significantly imbalanced. Figure~\ref{fig: non-zero ratio} shows global maps of the occurrence frequency of non-zero GWD for CGWs and FGWs. On average, only $7.6\% $ of all GCM columns yield non-zero CGWs, primarily located in the tropics. Similarly, for FGWs, only $8.5\% $ of all columns have non-zero outputs, but unlike CGWs, the majority of these are located in mid-to-high latitudes, particularly along the storm track region. For the OGWs in WACCM, data imbalance presents a greater challenge, to be discussed in a later section. While it is possible to simply separate zero and non-zero columns for emulation work where we know the truth, this approach falls short with real-world data, which is the main purpose of this study.

In addition to their sources, several other factors specific to GWD data exacerbate the data imbalance problem. In the case of each GCM column with non-zero GW activity, momentum fluxes are generally concentrated at a few critical height levels rather than being smoothly distributed throughout the entire column. This further restricts the effective occurrence frequency of non-zero values. Moreover, GWs exhibit significant intermittency, where a small portion of large-amplitude GWs often dominates the morphology of the observed global GW momentum flux distribution \citep{hertzog2012intermittency, geller2013comparison}. Therefore, it is crucial for NNs to not only accurately identify the columns that produce GWDs but also to effectively learn and recognize rare and extreme GWDs.

Given the complexity of the GWD dataset, different normalization methods are considered in this study. The first method, dubbed “NORM1”, is the typical normalization used in ML practices, which calculates elemental means and standard deviations for each feature (i.e., input variable at a given model level) and normalizes both inputs and outputs by these values (e.g., \cite{Espinosa2022}). With this approach, the same relative changes in wind at each level are treated equally in the input. The loss function in Equation~(\ref{eq: Loss_mse}) also penalizes the same relative error in GWD at each level equally. The second method, referred to as ``NORM2``, is designed with the physics of GWD in mind. For the velocity inputs ($u, v$) and the tendency outputs (GWD), each column is normalized by one single value, which is the largest standard deviation from all model levels. Additionally, the mean values for these variables, are retained (e.g., $u_{norm2}(x, y, z, t) = u(x, y, z, t) / max(std(u))$ ~). Unlike NORM1, the original wind profile's structure is preserved in NORM2, and large GWD values at certain heights maintain a relatively larger value after this normalization. For all other input variables, NORM2 is identical to NORM1. Compared to NORM1, NORM2 places more emphasis on large GWD values and penalizes the NN more for missing these significant tendencies. These two normalization methods are also employed in \cite{chantry2021}, who found similar performance from these methods with the non-orographic GWPs.

\begin{figure}[ht]
    \centering
    \makebox[\textwidth][c]{\includegraphics[width=35pc]{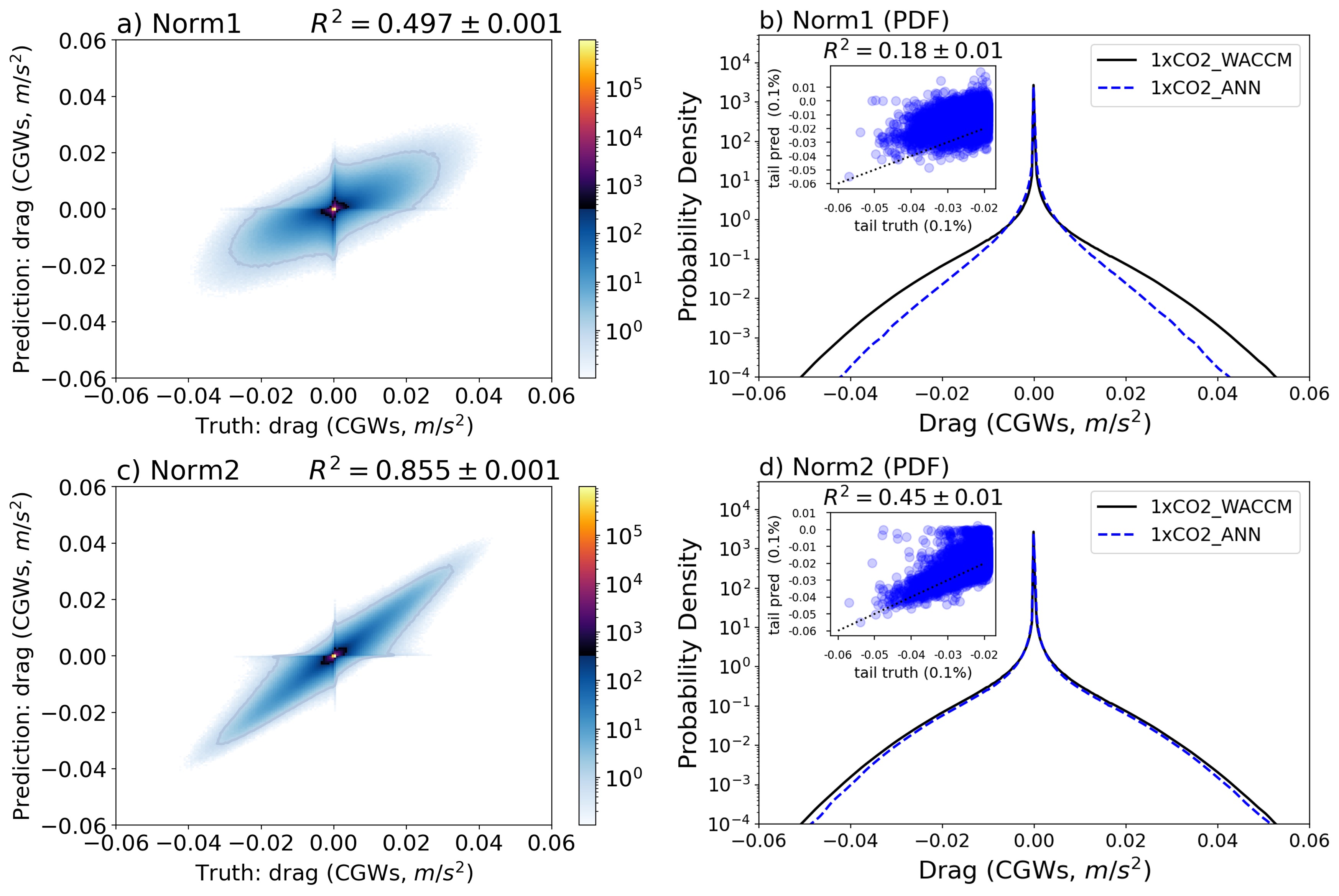}}
    \caption{Data imbalance for GWD due to CGWs and the emulation results with two different normalization methods. a) A 2D histogram displaying the emulated GWD due to CGWs and the truth, with the training dataset normalized using NORM1; b) Distribution of the original convective GWD (black line) and the predicted values (blue line) with NORM2. The scatter plot in the corner represents the tail part only, including points with the top 0.1\% amplitudes; c) Similar to a), but for NORM2; d) Similar to b), but for NORM2.
    The $R^2$ uncertainty range is estimated by dividing the test data into 10 segments, calculating the metric for each segment, and then computing the standard deviation (STD).}
    \label{fig: NORM1+2}
\end{figure}

Figure~\ref{fig: NORM1+2} shows the performance of the emulations for CGWs with the two normalization methods. When employing NORM1, the conventional approach seen in prior ML practices, and also our initial attempts, the emulator's performance is poor. Although the NN demonstrates some skill, its predictions tend to cluster around zero. However, when the second normalization method (NORM2) is employed, the emulation results show significant improvement, in contrast to the findings of \cite{chantry2021}. We attribute this improvement to the more pronounced data imbalance in our dataset, and it is likely a consequence of NORM2's emphasis on modeling the large GWD values. Nonetheless, emulating the tail of the probability density function (PDF) (rare events) remains poor, as evidenced by the tails in Figure~\ref{fig: NORM1+2}c, primarily due to the predominance of zero GWD columns in the training dataset. To more effectively address the data imbalance issue in these regression tasks, we further propose two approaches here:

\begin{enumerate}

    \item Resampling the data  (ReSAM): \,In this approach, we limit the number of training sample pairs with zero GWD to be equal to the number of samples with non-zero GWD. This significantly reduces the number of columns with zero GWD, thus mitigating the data imbalance issue. Additionally, this sub-sampling reduces the total size of the training dataset, which, in turn, enhances the training speed (approximately sevenfold). While resampling methods have been well-established in the ML literature, they have mainly been used for classification problems. Their application to regression problems in climate research has not been extensively explored.
    
    \item Weighted loss function (WeLoss): \,Instead of assigning the same weight to all sample pairs in the loss function, we modify the weight for each column based on the PDF of its maximum GWD amplitude. This adjustment allows us to re-formulate the loss function defined in Equation~(\ref{eq: Loss_mse}) as
    
\begin{equation}~\label{eq: weight_WeLoss}
%\mathcal{L}(y, \hat{y}; {\Theta}) = \frac{ \sum_{i=1}^{N} \left[ W_i * \left( y_i\left({\Theta}\right) - \hat{y_i} \right)^2 \right]} {\sum_{i=1}^{N} W_i}
%
\mathcal{L}({\Theta}) = \frac{1}{n} \sum_{i=1}^{n} \,\,  \Bigl\| \mathbf{W_i}\bigl\{\mathbf{NN}\left({x}_{i},{\Theta}\right) -  {y}_{i} \bigr\} \Bigl\|_2^2
\end{equation}
where
\begin{equation}
 \mathbf{{W_i}} =  \frac{1}{ PDF(max(\left |{y_i}(z) \right |))}
%Loss^{col} \propto \frac{1}{ PDF(max(\left |\hat{y}(z) \right |))} || y\left({\Theta}\right)  - \hat{y} ||_{MSE}. 
\end{equation}   

    Note that, in practice, we lack knowledge of the precise PDF for the maximum GWD within each column. Therefore, we employ a histogram with 20 bins as an alternative. Given the fact that large-amplitude GW events are rare, the WeLoss approach incentivizes the NN to prioritize these significant events.
    
\end{enumerate}

\begin{figure}[htbp]
    \centering
    \makebox[\textwidth][c]{\includegraphics[width=36pc]{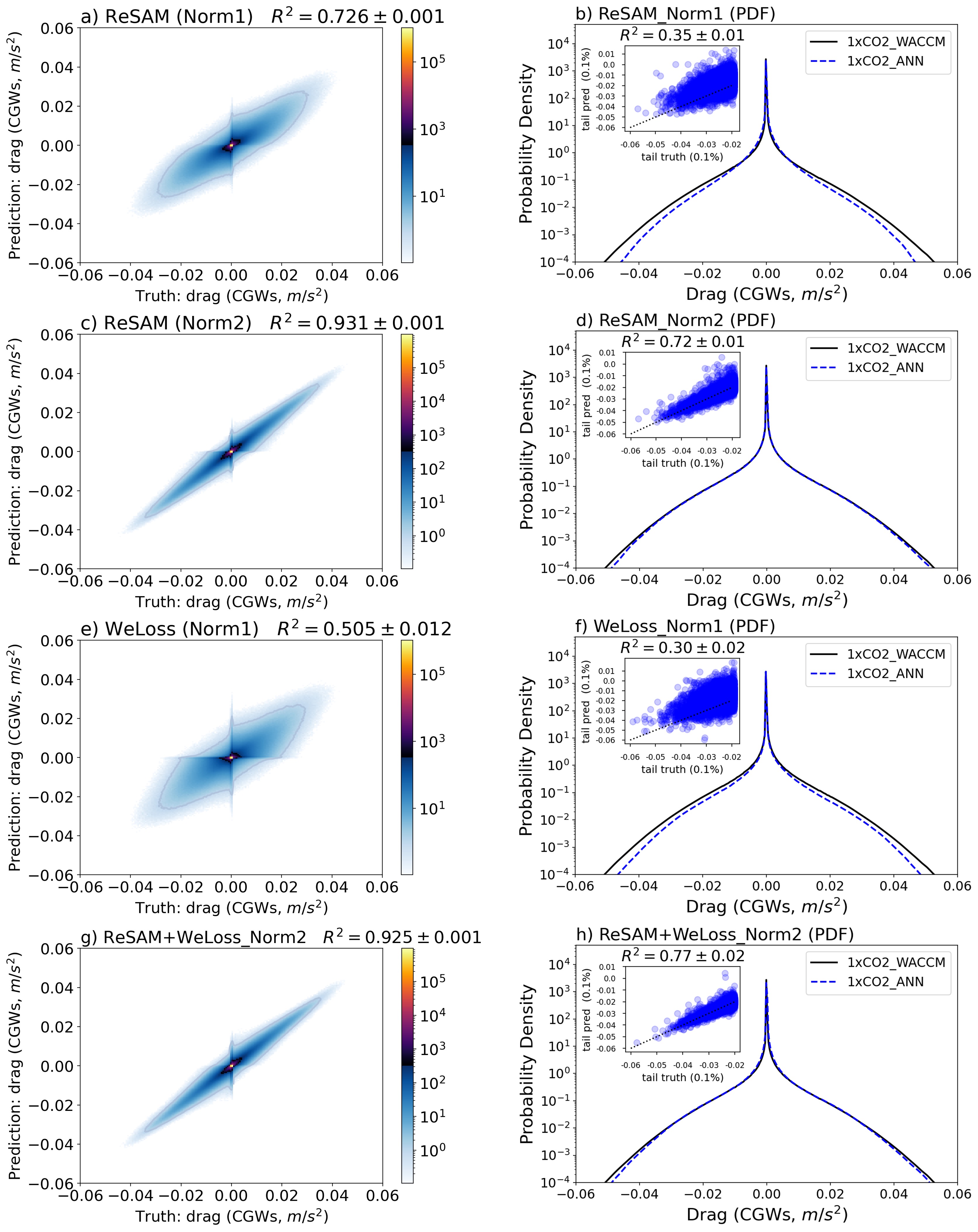}}
    \caption{Similar to Figure~\ref{fig: NORM1+2}, but for CGWs with the proposed ReSAM and WeLoss methods. a) A 2D histogram for the emulation with resampled data (ReSAM) after using Norm1; b) Distribution of the emulated GWD due to CGWs similar to Figure~\ref{fig: NORM1+2}b, but with ReSAM applied; c) Similar to a), with training data normalized using Norm2; d) Similar to b), with training data normalized using Norm2; e) Similar to a), but with the WeLoss approach; f) Similar to b), but with the WeLoss approach; g) Similar to c), after applying both ReSAM and WeLoss methods together; h) Similar to d), after applying both ReSAM and WeLoss methods.}
    \label{fig: ImBal conv}
\end{figure}

When we apply the ReSAM approach to balance the training dataset (after normalization with NORM1 or NORM2), the emulation results significantly improve, as shown in Figure~\ref{fig: ImBal conv}. In fact, when considering the R-squared value between the NN prediction and the ground truth, the ReSAM approach with NORM2 yields the best results. However, as the training dataset is still predominantly composed of zeros and small GWD values due to the intermittence of the GWs, examining the emulation results for only large amplitude GW events (e.g., the top 0.1\% in Figure~\ref{fig: ImBal conv}d) reveals less satisfactory performance ($R^2 = 0.72$). Regarding the WeLoss approach, it has a more limited impact on improving the R-squared value of the emulation (as shown in Figure~\ref{fig: ImBal conv}e). However, it proves valuable in capturing the tails of the PDF and, thus, rare events (as depicted in Figure~\ref{fig: ImBal conv}f). Moreover, as ReSAM and WeLoss represent distinct operations, they can be effectively combined when constructing a NN. The result of this combined approach for emulating the CGWs can be found in Figures~\ref{fig: ImBal conv}g and~\ref{fig: ImBal conv}h. While the R-squared value for the entire distribution only marginally changes (0.925 vs. 0.931 with ReSAM only), the performance of the emulation for the tail part has been improved ($R^2$ increased to 0.77).

\begin{figure}[htbp]
    \centering
    \makebox[\textwidth][c]{\includegraphics[width=36pc]{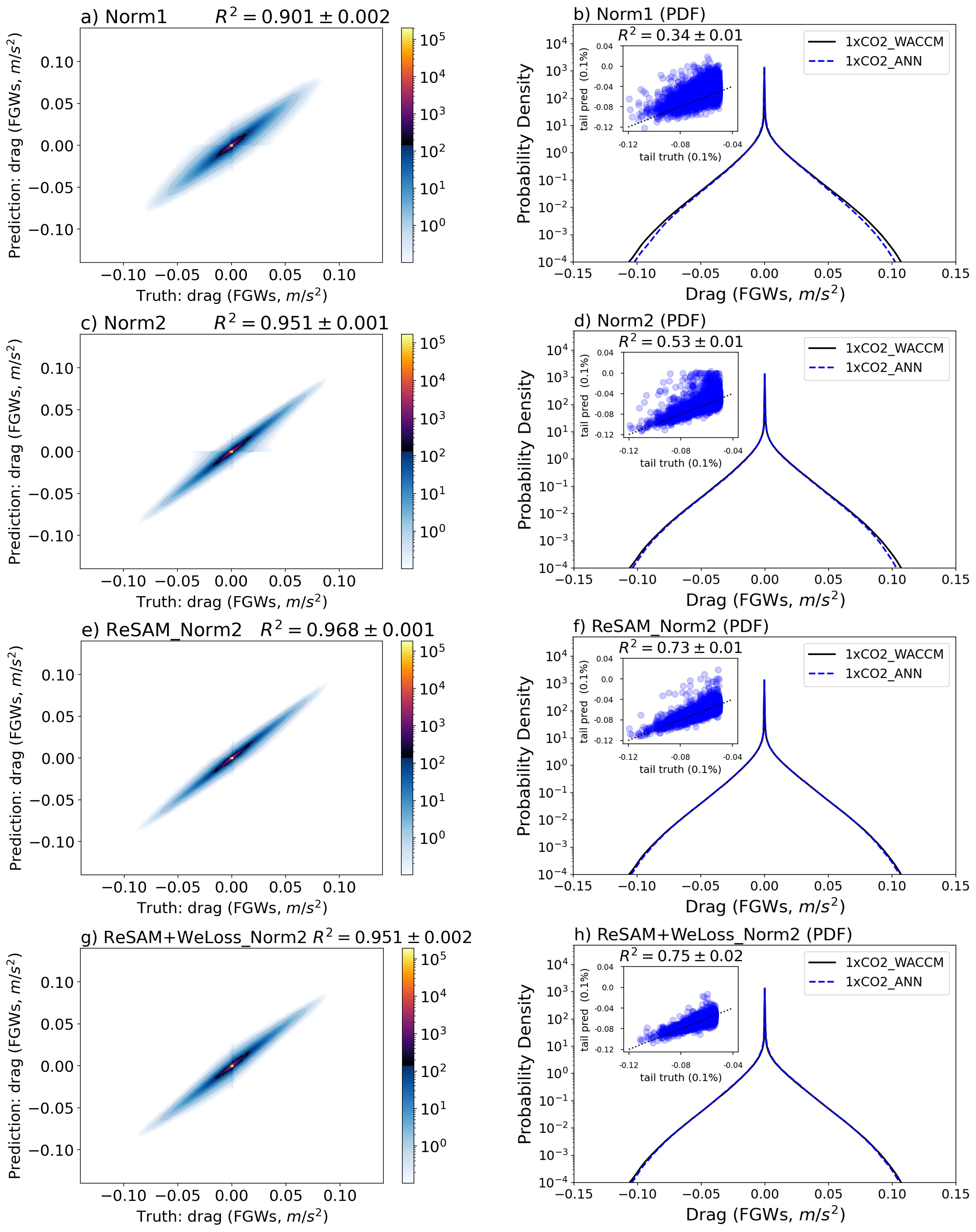}}
    \caption{Similar to Figure~\ref{fig: ImBal conv}, except for FGWs. a) A 2D histogram for the emulation using Norm1, without ReSAM or WeLoss; b) Distribution of the GWD due to FGWs with NORM1, similar to Figure~\ref{fig: NORM1+2}b; c) Similar to a), with training data normalized using Norm2; d) Similar to b), with training data normalized using Norm2; e) Similar to c), but with the ReSAM approach; f) Similar to d), but with the ReSAM approach; g) Similar to Figure~\ref{fig: ImBal conv}g, applying both ReSAM and WeLoss methods to the FGWs; h) Similar to Figure~\ref{fig: ImBal conv}h, applying both ReSAM and WeLoss methods to the FGWs.}
    \label{fig: ImBal front}
\end{figure}

Similarly, Figure~\ref{fig: ImBal front} presents the offline emulation results for the FGWs. The conclusions drawn for CGWs generally hold true. However, data imbalance in FGWs is less pronounced compared to CGWs, which simplifies the task of emulating FGWs. Even without any resampling or changes to the normalization or (see Figure~\ref{fig: ImBal front}a), we achieve reasonable emulation results ($R^2 = 0.9$). One contributing factor is the wider spatial distribution of FGWs compared to CGWs (refer to Figure~\ref{fig: non-zero ratio}). Additionally, the source of FGWs (frontogenesis function) in WACCM exhibits a much more continuous nature compared to precipitation and diabatic heating. As the data imbalance issue is less severe for FGWs, the performance with different normalization methods becomes more similar, echoing findings from \cite{chantry2021} who emulated non-orographic GWs (including convective and frontal GWs) together. 

In summary, data imbalance can pose challenges when learning from data that closely resembles real-world data (further discussed in the subsequent section on emulating OGWs). Proper resampling techniques can significantly enhance the NNs' performance by improving dataset balance. Furthermore, modifying the loss function to penalize the NNs more for missing extreme values can further improve performance at the tails of the PDF. For the remainder of the paper, unless otherwise specified, we continue to employ the ReSAM approach and the standard loss function with NORM2 unless stated otherwise.

\subsection{Uncertainty Quantification} \label{sec: UQ}

\begin{figure}[htbp]
    \centering
    \makebox[\textwidth][c]{\includegraphics[width=37pc]{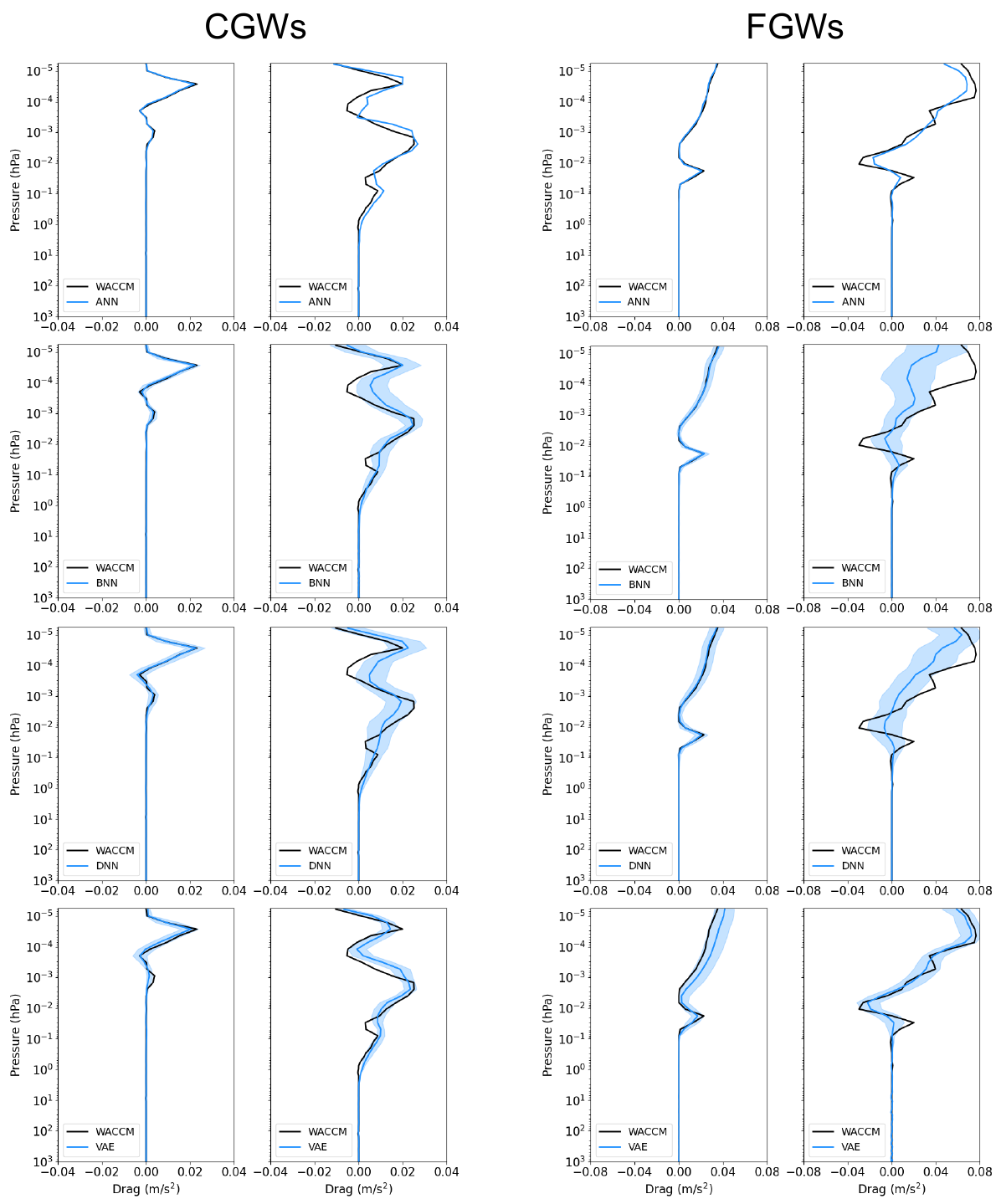}}
    \caption{Sample profiles of zonal GWD as predicted by various NNs, as indicated. The true profile is shown by the black line, while the blue solid line represents the mean of 1000 ensemble members. The shaded region indicates the 95\% confidence interval. In each pair of CGWs and FGWs profiles, the left column provides examples with low estimated uncertainty, corresponding to instances of low error. Conversely, the right column illustrates cases with high uncertainty when the error is high.}
    \label{fig: UQ profiles}
\end{figure}

As outlined in subsection \ref{sec: UQ methods}, we employ three different methods (i.e., BNN, DNN, and VAE) to quantify the uncertainty of predictions during inference (testing). For this purpose, an ensemble of 1000 members is generated by running each UQ-equipped NN 1000 times for each input from the testing set. Figure~\ref{fig: UQ profiles} presents sample profiles of zonal GWD derived from the deterministic NN (ANN) and the three UQ-equipped NNs, alongside the true GWD profiles from WACCM. Note that these examples have not been used in the training or validation process. It is evident from the figure that all three UQ-equipped NNs show reasonable skill in predicting the complex profiles of GWD due to CGWs and FGWs (also reflected in R-squared in Table~\ref{table: spread-skill}), albeit with a slight decrease in accuracy compared to ANN. As discussed earlier, a valuable uncertainty estimate should correspond closely with the NN's test accuracy, providing insights into when to trust the NN's prediction during inference. Such a relationship can be seen in a few randomly chosen GWD profiles that's shown in Figure~\ref{fig: UQ profiles}. In each pair of CGW and FGW profiles, the left column shows the estimated uncertainty is also low when the prediction error is low, indicating the NN's confidence in its accurate predictions. In contrast, the right column, which generally represents more complex profiles, exhibits the NN's less accurate predictions, and increased uncertainty, highlighted by the wider confidence intervals.

\begin{figure}[htbp]
    \centering
    \makebox[\textwidth][c]{\includegraphics[width=36pc]{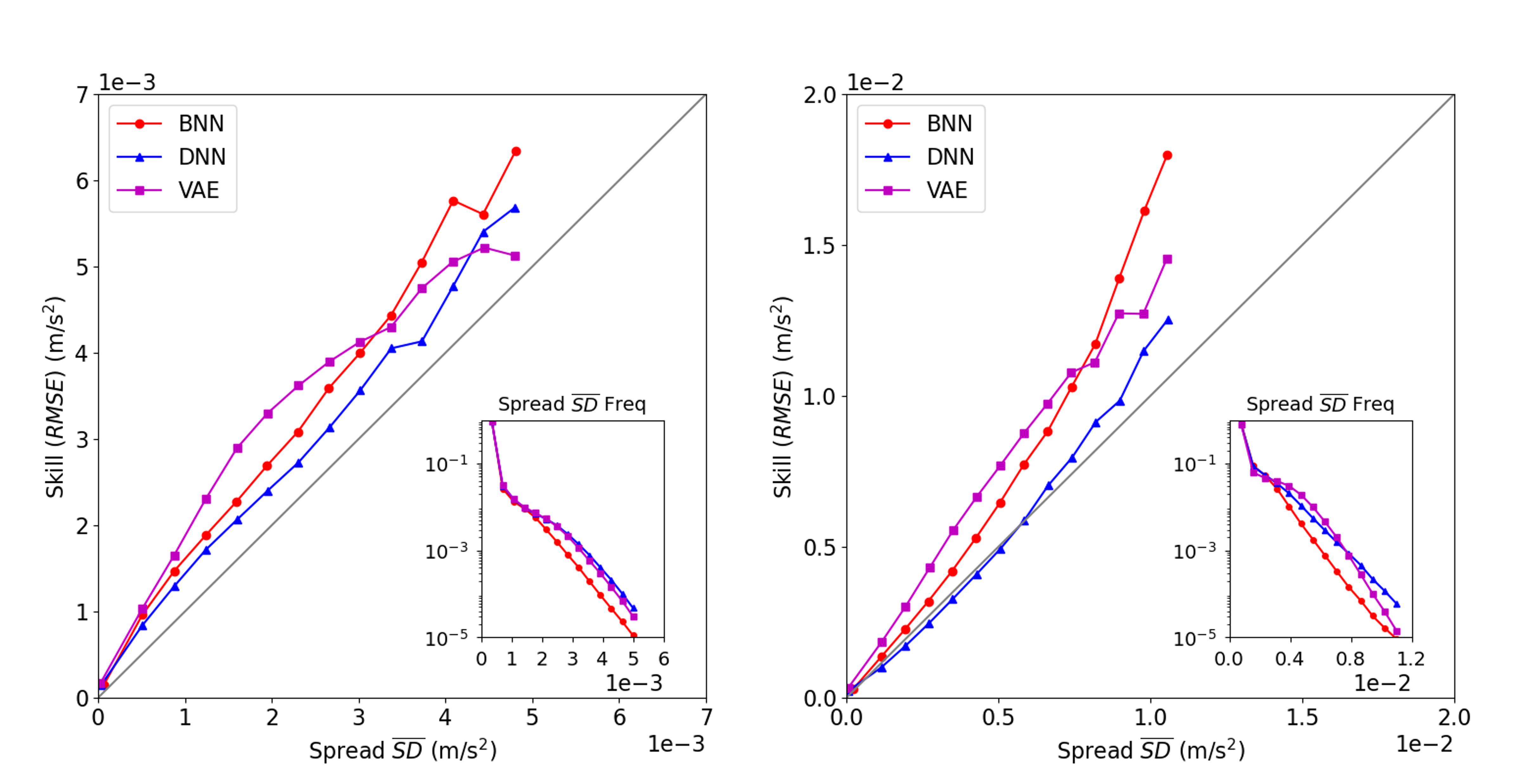}}
    \caption{Spread-skill plot for GWD due to (left) CGWs, and (right) FGWs. The diagonal 1:1 line represents the perfect spread-skill line. Points above (below) this line correspond to spread values where the model is overconfident (underconfident). The inset histogram shows how often each spread value occurs. See \ref{sec:appUQ} for a detailed discussion on the calculations of the spread-skill plot.}
    \label{fig: error vs. Uncertainty}
\end{figure}

While Figure ~\ref{fig: UQ profiles} demonstrates the performance of the UQ methods for just a few GWD profiles, the spread-skill plots shown in Figure~\ref{fig: error vs. Uncertainty} offer a broader perspective based on 60,000 profiles, following the calculations detailed in \ref{sec:appUQ}. It is evident from the plots that all three UQ methods produce reasonably informative uncertainty estimates, as their curves closely align with the 1-to-1 line. In the case of CGWs, all data points are above the 1:1 line, indicating a slight overconfidence (underdispersiveness) across all three UQ methods, with the DNN being slightly closer to the 1-to-1 line. For the FGWs, the DNN demonstrates slightly better performance, although it marginally drops below the 1-to-1 line in the first few bins, indicating a slight underconfidence. Notably, it can be seen from the spread frequency inset that the vast majority of the data points are within the first few bins, for which both spread and skill values are small, and they are generally closer to the 1-to-1 line.

It should also be noted that for the large values of model spread ($\overline{\mathrm{SD}}$), there is only a very limited number of data points, as is evident from the inset histograms. Consequently, the standard deviation (STD) can become a misleading measure of spread because of the non-normal distributions.

To summarize the quality of the spread-skill plots for the three UQ methods, we explore the metrics introduced in subsection \ref{sec: UQ methods} and \ref{sec:appUQ} (see Table~\ref{table: spread-skill}). The R-squared value for the ensemble mean prediction is also given to show the accuracy of each UQ method. Based on SSREL, whose ideal value is zero, BNN shows the best performance for both CGWs and FGWs. However, if we check SSRAT, where 1 is the optimal number, DNN is the best among these three methods. This discrepancy can be explained by a closer look at the Equations~\eqref{eqn. SSREL} and~\eqref{eqn. SSRAT}. SSREL, which is a bin-weighted average difference, is most sensitive to the performance of the NN in the first bin, where the vast majority of the data points are located (see the inset histograms in Figure~\ref{fig: error vs. Uncertainty}), while SSRAT is more influenced by larger values of spread and skill. Accordingly, the VAE shows the highest values of SSREL, which is indicative of its sub-optimal performance in the first bin, where there are small values of spread and skill.

In the results presented in Figure~\ref{fig: error vs. Uncertainty} and Table~\ref{table: spread-skill}, each height level of a GWD profile is considered as an individual sample. A zonal GWD profile, with its 70 vertical levels, thus constitutes 70 distinct samples. While analyzing these samples offers insights into the NN's overall performance by averaging statistics across numerous profiles, our primary interest is often in the uncertainty associated with an individual GWD profile. This uncertainty can then aid in determining whether to trust/use the NN's prediction for that particular GWD profile. Therefore, we use Equation~\eqref{eqn. spread-skill-sample} to assess the relationship between uncertainty and test accuracy for each GWD profile. Furthermore, to estimate uncertainty, here we use the interquartile range (IQR) to reduce the influence of outliers.

\begin{landscape}
\begin{figure}[t]
    \centering
    \includegraphics[width=42pc]{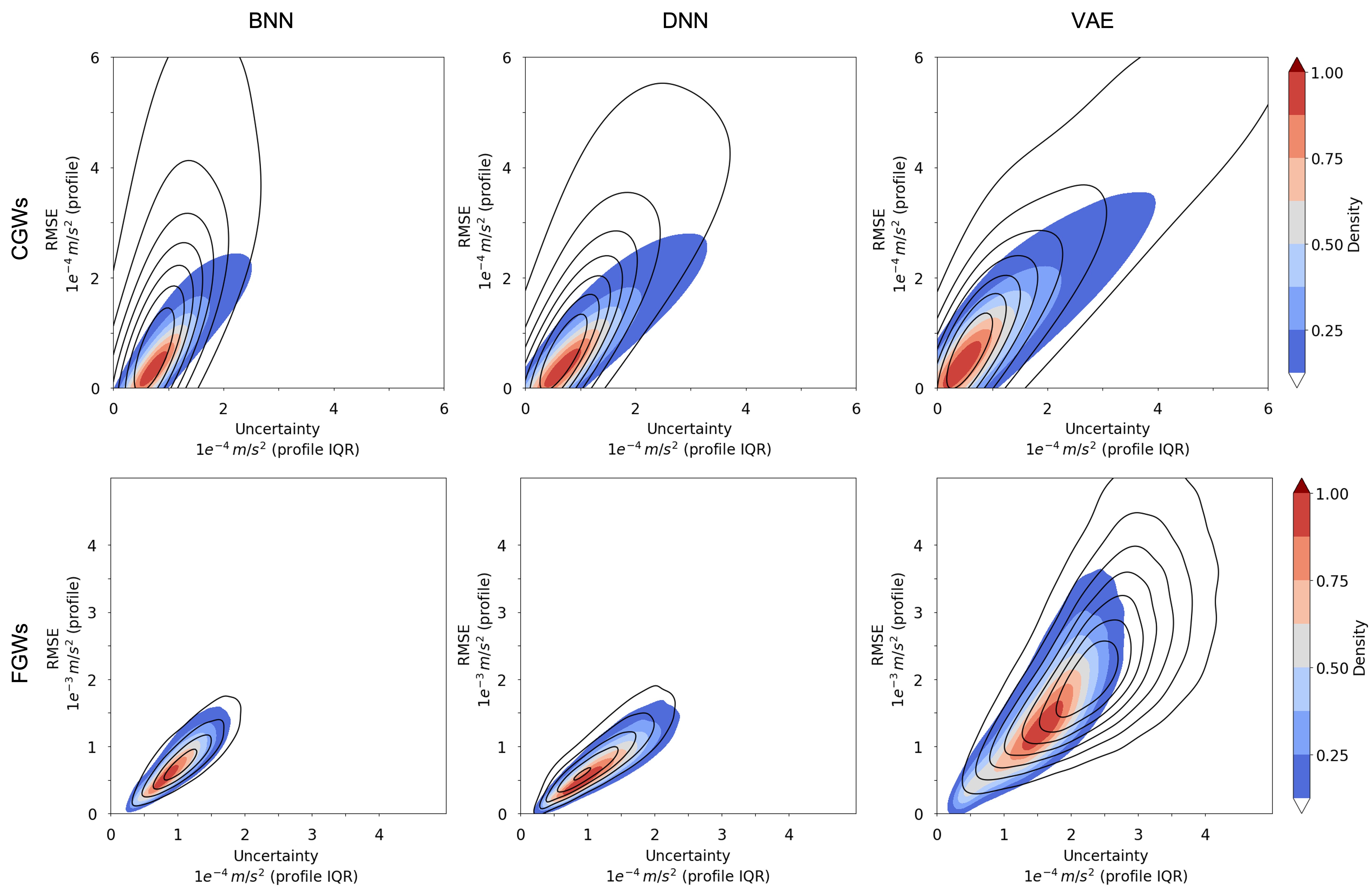}
    \caption{Gaussian kernel density for spread, defined as interquartile range (IQR) versus RMSE for (top) CGWs, and (bottom) FGWs, when validated against out-of-sample data from the current climate (color shading) and out-of-distribution data from the future, warmer climate (black lines). Results are shown for BNN (left), DNN (middle), and VAE (right). All NNs are exclusively trained on data from the current (control) climate.} 
    \label{fig: spread-skill-sample}
\end{figure}
\end{landscape}

Figure \ref{fig: spread-skill-sample} shows the Gaussian kernel density of spread against RMSE for all 60,000 profiles, as indicated by the color shading. The $x$-axis represents the IQR of each GWD profile, while the $y$-axis denotes its corresponding RMSE. A strong correlation between the two is observed across all three UQ methods. Consequently, GWD profiles with larger uncertainties often coincide with larger errors. Figure~\ref{fig: spread-skill-sample} also shows a close similarity between BNN and DNN. In contrast, VAE tends to provide marginally larger uncertainties, especially for FGWs. This is consistent with VAE's slightly reduced accuracy as indicated in Table \ref{table: spread-skill}.  Overall, given the monotonic relationship between the uncertainty and test error, these results show that all three UQ methods provide useful and informative uncertainty for with-distribution test samples. A user can set a threshold on uncertainty based on their tolerance for error (RMSE) and decide whether they trust the NN for a given input sample.

%One potential application of reliable UQ for parameterizations, as discussed in the Introduction, is that the high uncertainty can indicate low accuracy, which for example, could be a result of input samples from a different climate.

% Black lines in Figure \ref{fig: spread-skill-sample} show the performance of the NNs when trained on the control simulation and tested with input data from the future climate. 

The results presented so far show the performance of the UQ methods based on the testing data, i.e., data from the current climate. However, the effective performance of UQ methods can also be tested (perhaps more meaningfully) on OOD data, e.g., data from a warmer climate. This is particularly relevant for climate change studies. Accordingly, we evaluate the performance of these trained NNs with input data from the future climate, as depicted by the black lines in Figure \ref{fig: spread-skill-sample}. For FGWs, the spread-skill relationship remains largely similar, especially for BNN and DNN. This suggests that, based on their uncertainties, we can still reliably estimate the error in the NN predictions for FGWs for the warming climate. A similar pattern is observed for the VAE, though it exhibits increased uncertainties and higher errors with OOD data. As shown in a later section, for FGWs, the NNs generalize to the warmer climate without any further effort.

In contrast, for CGWs, given the same level of uncertainty, the error in NN predictions increases significantly for the OOD data compared to that from the current climate, which means the spread-skill relationship, especially for the BNN and DNN, fails to generalize to the OOD data. From this perspective, VAE performs better, showing that for the same level of uncertainty, the increase in error is not as substantial as in BNN and DNN. The VAE also yields considerably higher uncertainty estimates for future climate, which may aid in the detection of OOD data. The observed discrepancies in the performance of the NNs for CGWs and FGWs hint at different levels of their generalizability, a topic we will delve into more deeply in the following subsection.

 \begin{table}[t]
 \caption{ Evaluation scores for the three UQ methods. See Section 2 for more details.}
 \centering
 \begin{tabular}{| l | c | c | c | c | c | c |}
 \hline
   & \multicolumn{3}{|c|}{CGWs} & \multicolumn{3}{|c|}{FGWs} \\
 \hline
  & BNN & DNN & VAE & BNN & DNN & VAE \\
  \hline
   SSREL (1e-4)  & 1.29 & 1.48 & 2.14 & 1.20 & 1.69 & 5.21 \\
   SSRAT  & 0.73 & 0.82 & 0.72 & 0.69 & 0.93 & 0.69 \\
   R-squared & 0.90 & 0.86 & 0.87 & 0.94 & 0.92 & 0.89  \\
 \hline
 \end{tabular}
 \label{table: spread-skill}
 \end{table}

In summary, while the three UQ methods provide credible and valuable uncertainty estimates for the current climate, the BNN and DNN are confidently wrong in estimating CGWs in a warmer climate although VAE shows some promising results. This problem is common among various UQ techniques as pointed out in the ML literature: they frequently show overconfidence when assessed with OOD data \citeg{ovadia2019can}. The optimal UQ method selection depends on the specific metric of interest and the intended application. While BNN is more broadly used in the literature and gives the best accuracy, DNN could achieve similar performance and is often more practical given its simplicity. On the other hand, VAE seems to perform better when applied to OOD data, at least in the one test case here. These observations warrant further research in the future using multiple test cases and climate-relevant applications. We also note here that each method has multiple tuning hyperparameters to optimize its uncertainty quantification. Consequently, the discrepancies among the three methods could potentially be mitigated with proper hyperparameter tuning (as discussed in \ref{sec:appB}).

\subsection{Out-of-distribution (OOD) Generalization via Transfer Learning}

As previously discussed, the GWP schemes in WACCM are coupled to their sources, which might change in a warmer climate. Specifically, under $4\times$CO$_2$ forcing, we expect changes in both the amplitude and the phase speed distribution of GWs, in particular for the CGWs, due to their built-in sensitivities to changes in the convection. Consequently, the physics scheme in WACCM produces slightly stronger GWD for CGWs, especially in the tail of the distribution. This intensified GWD results in a shorter quasi-biennial oscillation (QBO) period in WACCM. However, it is important to recognize that the response of the QBO to climate change differs across various general circulation models \citep{Richter2022}.

\begin{figure}[tp]

    \makebox[\textwidth][c]{\includegraphics[width=36pc]{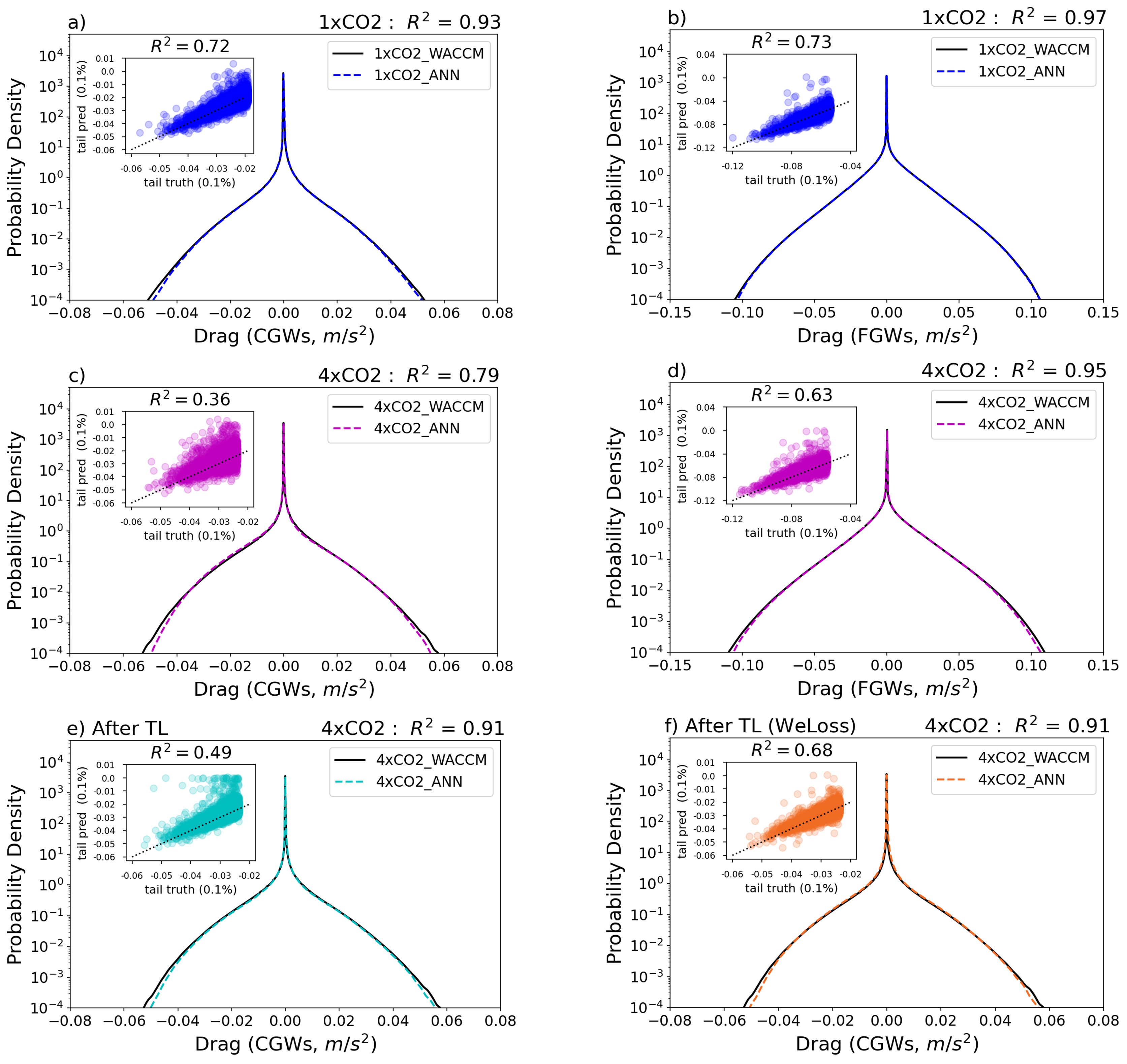}}
    \caption{NN performance for pre-industrial and warming scenarios for different sources (a,c,e,f: CGWs ;\,  b,d: FGWs). \textbf{a)} PDF of GWD due to CGWs in WACCM simulation and the predicted CGWs using NN emulator, scatter plot shows points for the tail part only. \textbf{b)} same as (a), but for FGWs. \textbf{c)} same as (a), but for the warming scenario, \textbf{d)} same as (b) but for the warming scenario. \textbf{e)} same as (c) but after applying transfer learning to the first hidden layer of the NN with 1-month WACCM simulation data under warming scenario ($\sim$ 1\% of the size of the training data)  \textbf{f)} same as (e) but with the weighted loss function used when we conduct transfer learning (WeLoss).}
    \label{fig: ANN-TL}
\end{figure}

The intensification of the CGWs in future climate simulations presents an opportunity to study how NNs handle the OOD data. Our findings in the UQ section already suggest increased prediction errors when testing NNs with OOD data, which raises concerns about their applicability in climate change studies. To more thoroughly investigate this issue, we conduct additional evaluations on our ANNs, by applying them to data samples from future climate simulations, as illustrated in Figure~\ref{fig: ANN-TL}. It is clear that the ANN for the CGWs does not generalize well, evidenced by a decrease in $R^2$ from 0.93 to 0.79. The ANN particularly struggles to capture the increase in GWD in the tail, with $R^2$ for the tails decreasing from 0.72 to 0.36. As a result, it seems unlikely that this emulator will accurately reproduce changes in the circulation under different climate conditions, such as the shorter QBO period resulting from future warming in WACCM.

In contrast to CGWs, the amplitude of FGWs shows a less marked increase in the future climate, and their PDF distribution closely resembles that of the control simulations. As a result, the ANN demonstrates better generalizability for FGWs when it is tested against future climate data, as seen in Figure~\ref{fig: ANN-TL}d. There is only a slight decrease in the ANN's performance, with $R^2$ dropping from 0.97 to 0.95.

Two factors can contribute to the considerable OOD generalization errors in an NN when applied across two distinct systems. First, the input-output relationship might vary between the two systems. Second, the input variables in the new system could originate from a distribution different from that of the original system (regardless of whether the input-output relationship remains the same or changes). The former is hard to quantify in a high-dimensional dataset. The latter can be quantified using similarity distances. To help us better understand these differences between the OOD generalizability of CGWs and FGWs, we assess the similarity between their input and output data distributions from control and future climate simulations using the Mahalanobis distance ($D$). The Mahalanobis distance is a measure of the distance between a data point and a distribution \citep{Lingetal2015}. Specifically, it is a multi-dimensional generalization of the idea of measuring how many standard deviations away a point is from the mean of the distribution. The application of Mahalanobis distance in understanding the source of OOD generalization errors in data-driven parameterization was previously demonstrated in~\cite{guan2022stable} for a simple turbulent system.

% this is bassed on the average distance of the points that are more than 3 sigma away from the mean.
 \begin{table}[ht]

 \caption{Change of Mahalanobis distance based on the ratio of the average distance of the points that are more than 3 standard deviations away from the mean. The choice of the variables here is based on \ref{sec:appA}, showing $u, v, T,$ and source function contain most of the information needed for the NN. } 
 \centering
  \resizebox{34pc}{!}{\begin{tabular}{l | c | c | c | c | c | c}
 \hline
  Variables  & $u$ & $v$ & $T$ &  \makecell{Source \\ (diabatic heating for CGWs, \\ frontogenesis for FGWs)} & Zonal drag & Meridional drag \\

 \hline
   Distance (Convection)  & 1.03 & 1.00 & 1.19 & 3.62  & 1.42 & 1.44\\
   
   Distance (Front)   & 1.03 & 0.96 & 1.50 & 1.10 & 1.00 & 1.00\\
   
 \hline
 \end{tabular}}
 \label{table: 2}
 \end{table}

To use the Mahalanobis distance, we first calculate the mean and covariance matrix of the training data from the control run. We then analyze the distribution of Mahalanobis distances in this training data, setting a baseline value, referred to as \(D_{ctrl}\). This baseline is the average distance for data points that deviate by more than 3 standard deviations from the mean. This choice aims to focus on outliers for which extrapolation is more challenging. Following this, we apply the same process to the data points in the future climate dataset, denoted as \(D_{warm}\). Table~\ref{table: 2} presents the ratio of \(D_{warm}\) for the warming scenario to \(D_{ctrl}\) for the control scenario for selected variables. When this ratio is close to 1.0, it suggests minimal changes in this variable's distribution under a warming scenario. Note that the NNs trained based only on these variables demonstrate performance comparable to NNs trained on all variables (not shown), which is why we only focus on these few key variables.

The results reveal that among the various variables significantly contributing to the emulation of CGWs, diabatic heating (source of CGWs) is the sole variable that exhibits substantial changes from the control to the warming scenario. Conversely, changes in variables used to emulate FGWs are considerably smaller. This outcome suggests that the likely reason for the better generalizability of FGWs is that the input distribution remains almost unchanged (and the input-output relationship, which is the physics scheme, remains the same too).

To improve the generalizability of the emulator for CGWs, we explore TL, a technique introduced earlier and proven to be a powerful tool for improving the OOD generalizability of data-driven parameterization in canonical turbulent flows \citeg{guan2022stable,subel2023explaining}. Rather than re-training the entire NN for future climate scenarios, we only re-train, follwoing~\cite{subel2023explaining}, just a portion of the NN, thereby requiring only a small fraction of the data. Figure~\ref{fig: ANN-TL}e showcases the emulation results after only re-training the first hidden layer of ANN using data from the first month of the WACCM simulation in the $4\times$CO$_2$ scenario, which amounts to approximately 1\% of the original training dataset. After applying TL, the performance of the emulator in the warming scenario significantly improves, with $R^2$ rising from 0.79 to 0.91, nearly matching its performance in the control simulations ($R^2$ = 0.93). However, the improvement in the PDF tails is less pronounced, showing only a modest increase in $R^2$ from 0.36 to 0.51. This is likely due to the limited number of large-amplitude GW events within the one-month period. Instead of using more data from the future climate (which is challenging to obtain in a realistic situation), we leverage the WeLoss approach, described earlier, during re-training. This modification results in a significant improvement in the tail, with $R^2$ increasing from 0.51 to 0.68.  Note that this improvement in the tail is critical, as inadequate learning of these rare but large-amplitude GWDs can result in significant errors and instabilities.

We would like to point out that during the TL experiments, we have examined the effects of re-training each individual hidden layer of the NN. Our findings indicate that re-training the first layer yields the best results, which aligns with the findings in \cite{subel2023explaining}. Re-training the last layer only brings marginal improvements to the NN (not shown). Notably, our experiments involving re-training the first two layers did not result in further performance enhancements, suggesting that the number of neurons is not the primary factor contributing to the varied performance observed when re-training different layers.

\begin{figure}[tp]
    \centering
    \includegraphics[width=36pc]{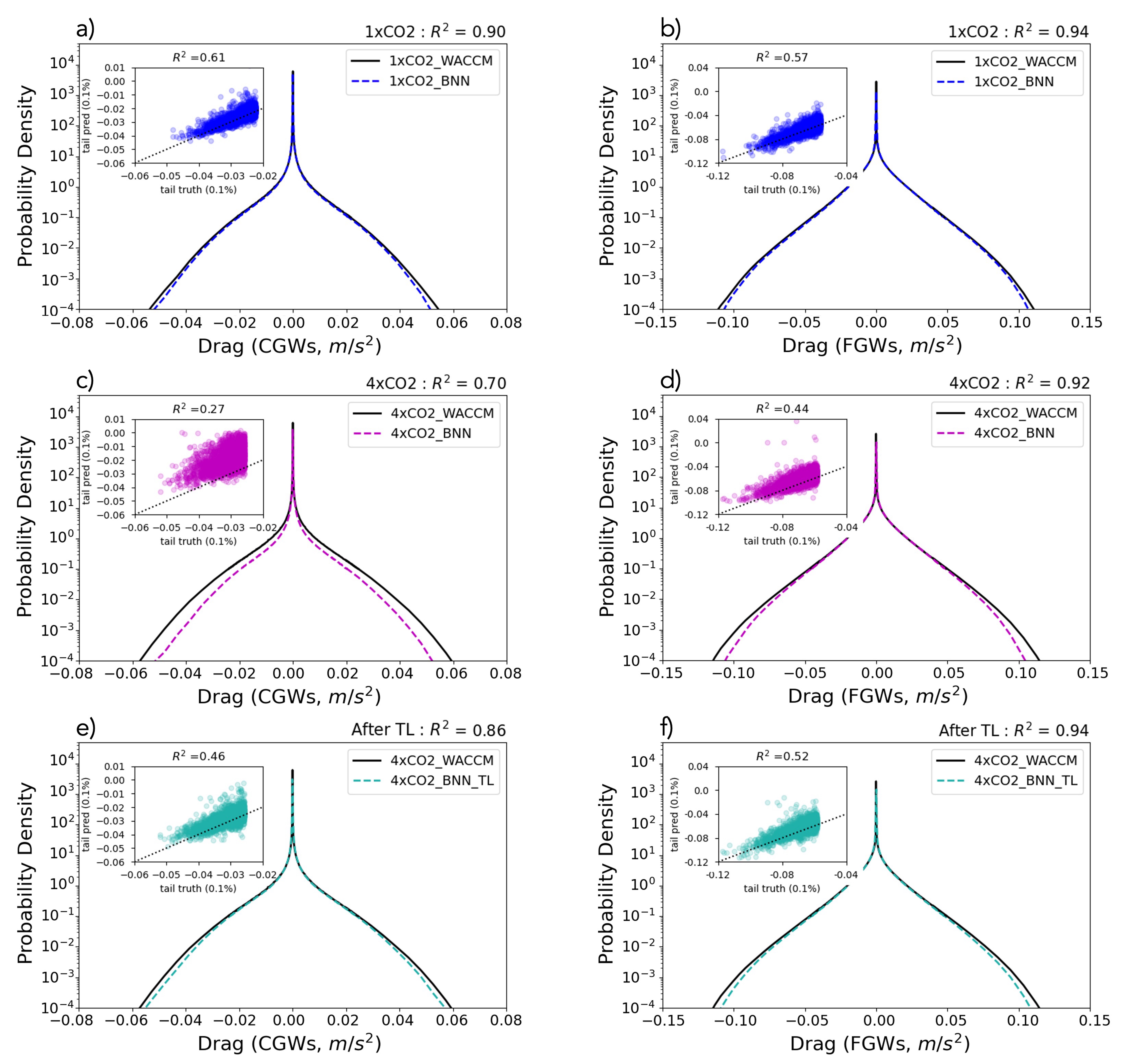}
    \caption{ Panels (a) to (e) are the same as those in Figure~\ref{fig: ANN-TL} but for BNN. Panel (f) shows emulation for FGWs under warming scenario after applying transfer learning to the first hidden layer of the NN.}
    \label{fig:BNN pdf}
\end{figure}

Similar results regarding TL are also observed with other NNs used in this study. For instance, Figure~\ref{fig:BNN pdf} presents the same plot as Figure~\ref{fig: ANN-TL}, but for the BNN. It is evident that BNN also struggles with generalization to OOD data, as could also be interpreted based on the results presented in section~\ref{sec: UQ}. It is also the case for DNN and VAE (not shown). Overall, when these NNs are tested against the $4\times$CO$_2$ future climate data, their accuracy is not better than the deterministic ANN. However, methods with UQ, especially the VAE (see Figure~\ref{fig: spread-skill-sample}), could potentially indicate the increased uncertainty when testing with input data from the $4\times$CO$_2$ integration. These results underscore the necessity of re-training the NNs using TL.

\section{Emulation of Orographic GWs (OGWs)} \label{ORO}

\begin{figure}[ht]
    \centering
    \includegraphics[width=33pc]{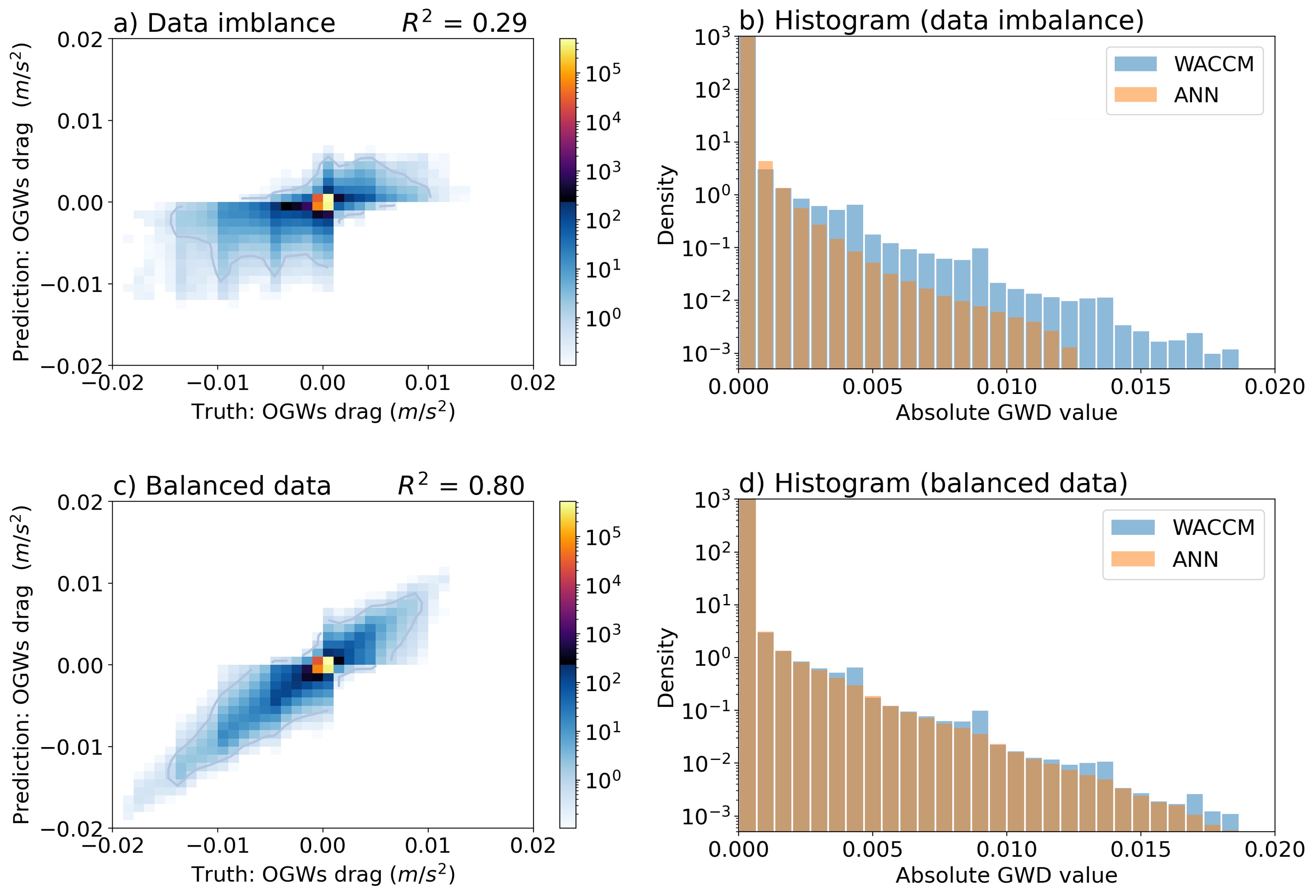}
    \caption{Performance of the emulator for OGWs when trained with all columns over land (panel a) \& panel b)) and balanced training data with a balanced number of large-drag columns (column maximum $>$ 1 STD of all GWD from OGWs) and small-drag columns (panel c) \& panel d). }
    \label{fig:oro2}
\end{figure}

Similar to \cite{chantry2021}, our initial attempts to emulate OGWs did not succeed, primarily due to the presence of a pronounced data imbalance. Notably, the physics-based scheme responsible for OGW generation operates exclusively over terrestrial regions. However, it is surprising that the issue of data imbalance continues to persist, even when we limit our NN training and testing exclusively to columns located over land (Figure~\ref{fig:oro2}a). Still, the emulated OGW drag often remains close to zero and completely fails to predict the rare events (Figure~\ref{fig:oro2}b), which poses a considerable hurdle for the emulator's performance. Further investigations reveal that the key to this problem lies in the highly localized nature of orographic GWD, where significant drag is observed only at a handful of specific locations. Furthermore, even within these limited regions, GWD exhibits a significant intermittent behavior. To help our understanding, we also conducted a $K$-means clustering analysis, categorizing GWD data for all land-based columns (Table ~\ref{table: ORO cluster}). Among the 6 clusters, cluster 4 accounts for a staggering $97.51\%$ of the dataset. Remarkably, all samples within this cluster exhibit exceptionally weak orographic GWD, as evidenced by the cluster center's maximum GWD amplitude, which is two orders of magnitude smaller than that of other clusters.

\begin{table}[ht]
 \caption{ Clustering analysis for OGWs. Analysis is done for all columns over land in the training data.}
 \centering
 \begin{tabular}{| c | c | c |}
 \hline
  Cluster  & \makecell{Frequency (\%) in the \\ training data} & \makecell{Maximum GWD amplitude \\ of cluster center} \\
 \hline
   c1   & 0.18 & 8.7 e-3 m/s$^2$  \\
   c2   & 0.13 & 4.4 e-3 m/s$^2$  \\
   c3   & 0.93 & 3.6 e-3 m/s$^2$  \\
   \textbf{c4}   & \textbf{97.51} & 2.8 \textbf{e-5} m/s$^2$  \\
   c5   & 0.15 & 2.1 e-3 m/s$^2$  \\
   c6   & 1.10 & 4.3 e-3 m/s$^2$  \\
 \hline
 \end{tabular}
 \label{table: ORO cluster}
 \end{table}

To overcome this persistent data imbalance in the OGWs, we first separate all columns over land into large-drag columns (with column maximum greater than one STD of all GWD from OGWs) and small-drag columns. We then perform subsampling on the latter group only to create a more balanced dataset. To improve NN training, we also include all columns from the 6-year simulation to augment the sample size of the large-drag columns. Figures~\ref{fig:oro2}c and~\ref{fig:oro2}d illustrate the performance after re-balancing the dataset. Notably, the result represents a substantial improvement, evidenced by an $R^2$ increase from 0.29 to 0.80, and also a significant improvement in the accuracy for rare events. While we acknowledge that this skill remains lower than what is achieved for CGWs and FGWs, it already signifies a reasonable NN. Furthermore, we posit that by incorporating additional training data (either by extending the WACCM model integration or simply augmenting the data with OGWs scheme only), we can further improve our emulation results. The possibility of achieving superior emulation outcomes through the adoption of an alternative NN architecture is also possible, although such exploration is beyond the scope of this paper.

\section{Summary and Discussion} \label{Summary}

Through the emulation of complex GWPs in a state-of-the-art atmospheric model (WACCM), we have elucidated and explored solutions for three critical challenges in the development of ML-based data-driven SGS schemes for climate applications: data imbalance, UQ, and OOD generalizability under different climates. A brief summary is provided below:

\begin{enumerate}
    \item In the presence of non-stationary, and highly imbalanced datasets, such as those encountered in WACCM, specialized approaches (e.g., resampling and weighted loss function) are essential to enhance the performance of data-driven models. Through resampling, we have successfully trained a robust NN emulator for OGWs, a challenging task as demonstrated in \cite{chantry2021}. The effectiveness of the trained emulator is also significantly influenced by the choice of the loss function used during training. In our case, while a weighted loss function (WeLoss) does not improve the overall $R^2$ score, it yields significant improvements in the emulation results for the PDF tails of the GWD. This finding aligns with those in \cite{Gomez2022HeatWave}, where their custom loss function, tailored to emphasize extreme events, led to substantial improvements in predicting heatwaves.
    
    \item All three UQ methods employed in this study provide reasonable uncertainty estimates for GWD prediction for the current climate. The spread-skill plots (refer to Figures~\ref{fig: error vs. Uncertainty} and~\ref{fig: spread-skill-sample}) indicate that greater uncertainty corresponds to a larger prediction error. Yet, the reliability of UQ methods diminishes when they are challenged with OOD data. Both BNN and DNN used in this study tend to be overconfident in estimating CGWs in a warmer climate, thereby struggling to identify OOD samples. The VAE, on the other hand, yields rather promising results in providing useful UQ for OOD data. Given the variations in different methods, the metrics selected to assess the SGS model will play a significant role in determining the choice for the UQ methods. We also note that further optimization of tunable parameters within each UQ method could affect their performance (refer to~\ref{sec:appUQ}).
    
    \item Our findings illustrate the challenges SGS schemes face in generalizing to OOD data and extrapolating to new climates. Nonetheless, the TL approach has proven highly effective in aiding an NN to extrapolate to different climates. For CGWs in WACCM, the physics-based scheme exhibits larger GWD under $4\times$CO$_2$ forcing, primarily due to an increase in diabatic heating from convection. With only one month of simulation data from this future warming scenario (representing approximately 1\% of the original training dataset), we successfully reduce its OOD generalization error through re-training the first layer of the NN, following the findings of \cite{subel2023explaining}. Additionally, we have illustrated the value of metrics like the Mahalanobis distance in assessing the potential OOD generalizability of NNs.
    
\end{enumerate}

We would like to emphasize that these challenges are often intertwined. For instance, addressing data imbalance in CGWs is a prerequisite for obtaining an accurate NN model, which, in turn, impacts UQ and OOD generalizability assessments. Moreover, there exists a strong link between UQ and OOD generalizability evaluations: if the NN struggles with OOD generalization, performing poorly with OOD data, the reliability of UQ for such data (e.g., data from a warmer climate) also becomes questionable. This presents a substantial challenge for UQ methods, especially for climate change research where reliable UQ methods are crucial.

This study has primarily focused on offline skill assessment. We acknowledge that good offline performance (at least in terms of common metrics such as $R^2$) is not necessarily an indicator of stable and accurate online (coupled to climate model) performance~\citep{ross2022benchmarking,guan2022stable}, though more strict metrics such as $R^2$ of the PDF tails might better connect the offline and online performance~\citep{pahlavan2023explainable}. However, for the purpose of this study, which is to provide a testbed to test ideas for data imbalance, UQ, and OOD generalization with transfer learning, the offline tests, particularly using the several metrics we have used, suffice. That said, the main reason that we have not provided online results is that coupling various complex NNs, with the same framework, to complex climate models (e.g., WACCM) without slowing down the model is a challenging and time-consuming task~\citep{Espinosa2022}, and this is work in progress.

Emulating complex GWPs within the WACCM provided a unique opportunity to address three critical challenges in developing ML-based, data-driven SGS schemes for climate science applications. However, it is crucial to acknowledge that such emulated schemes essentially adopt the limitations inherent in the physics-based schemes. Addressing these limitations, the next step is to harness high-resolution data from GW-resolving simulations, which are carefully validated against observational data. A library of such high-resolution simulations, notably of convectively generated GWs using the Weather Research and Forecasting (WRF) model, is now established \citep{Sunetal2023WRF}, alongside additional global high-resolution simulations~\citep{wedi2020baseline, polichtchouk2023resolved, kohler2023comparing}. The next phase involves integrating the approaches outlined in this study with the data from these GW-resolving simulations to develop a stable, trustworthy, and generalizable data-driven GWP scheme. This scheme is then expected to overcome the limitations of physics-based GWPs and potentially incorporate features like the transient effect \citep{Boloni2021, Kim2021JAS} and lateral propagation of GWs \citeg{Sato2009}—marking a significant advancement towards next-generation GWP schemes.

%%%%%%%%%%%%%%%%%%%%%%%%%%%%%%%%
%% Optional Appendix goes here
%
% The \appendix command resets counters and redefines section heads
%
% After typing \appendix
%We also plan to apply online learning for the true data-driven SGS %neural network with high-resolution gravity wave data generated in Sun et al. (2023).

\appendix

\section{Input/output variables  for the physics-based GWP schemes and their emulators} \label{sec:appA}
We use the exact same inputs as those of each GWP scheme in the WACCM for the training of the NN-based emulator of that scheme. These inputs are listed in Table~\ref{tab:A1}. As for the outputs, we only consider the zonal and meridional drag forcings.  The GWPs in WACCM also estimate additional effects of the GWs that result in changes of temperature profile and vertical diffusion. These outputs are not considered in our emulations.  

\begin{table}[th]
\caption {List of the input and output variables for the NNs trained as emulators of the GWP schemes in WACCM. The numbers in parentheses in front of each variable are the number of vertical levels for that variable. Note that each input and output is a 1D column at a given latitude/longitude grid point. Diabatic heating in WACCM is provided by the cumulus scheme. The topography variables listed in the table are mxdis (height estimates for ridges), hwdth (width of ridges), clngt (length of ridges), angll (orientation of ridges), and anixy (anisotropy of ridges).}
\centering
\resizebox{\textwidth}{!}{
\begin{tabular}{|c|c|c|c|c|}
\hline
         \multirow{2}{*}{GWP} & \multicolumn{3}{c|}{Input} & \multirow{2}{*}{Output} \\
\cline{2-4}
              & pressure levels  & surface level  & forcing  &         \\
\hline
         CGWs & \multirow{3}{*}{ \makecell{$u (70), $\\ $v (70),$ \\$T (70),$ \\ $z (70),$ \\$ \rho (71),$\\ Brunt–Väisälä frequency $N$ (70), \\  dry static energy $DSE$ (70)} }& \multirow{3}{*}{ \makecell{ lat (1),\\ lon (1), \\ $P_{surface}$ (1),} }   & diabatic heating (70)  &  \multirow{3}{*}{ \makecell{ zonal drag \\ GWD$_x$ (70), \\ meridional drag \\ GWD$_y$ (70),} }    \\
\cline{1-1}
\cline{4-4}
         FGWs &  &    & \makecell{frontogenesis \\ function (70) }  & \\
\cline{1-1}
\cline{4-4}
         OGWs &  &     & \makecell{mxdis (16), \\ hwdth (16),\\ clngt (16),\\ angll (16),\\ anixy (16),}  & \\
\hline
\end{tabular}
}
\label{tab:A1}
\end{table}

%\begin{figure}[tp]
%    \centering
%    \includegraphics[width=33pc]{Figs/PictureA1.jpg}
%    \caption{Offline performance for the NNs when trained using all the input variables versus using only $u, v, T,lat, lon$ and forcing functions for a) CGWs and b) FGWs. } \label{fig: A1}
%\end{figure}
%"""

From Table~\ref{tab:A1}, one can guess that some input variables are correlated with each other. Consequently, it is plausible that the trained NNs may have spurious connections. Preliminary tests further support this notion, indicating that employing only $u, v, T,$ and the forcing function as inputs yields comparable offline skill (results not presented here).

%Figure~\ref{fig: A1} shows that we could get a comparable skill, for both CGWs and FGWs, if we only use $u,  v, T, lat, lon $ and the forcing function (diabatic heating for CGWs, and the frontogenesis function for FGWs) as the inputs of the NNs.

\section{Tuning UQ-equipped NNs} \label{sec:appB}
\begin{figure}[tp]
    \centering
    \includegraphics[width=35pc]{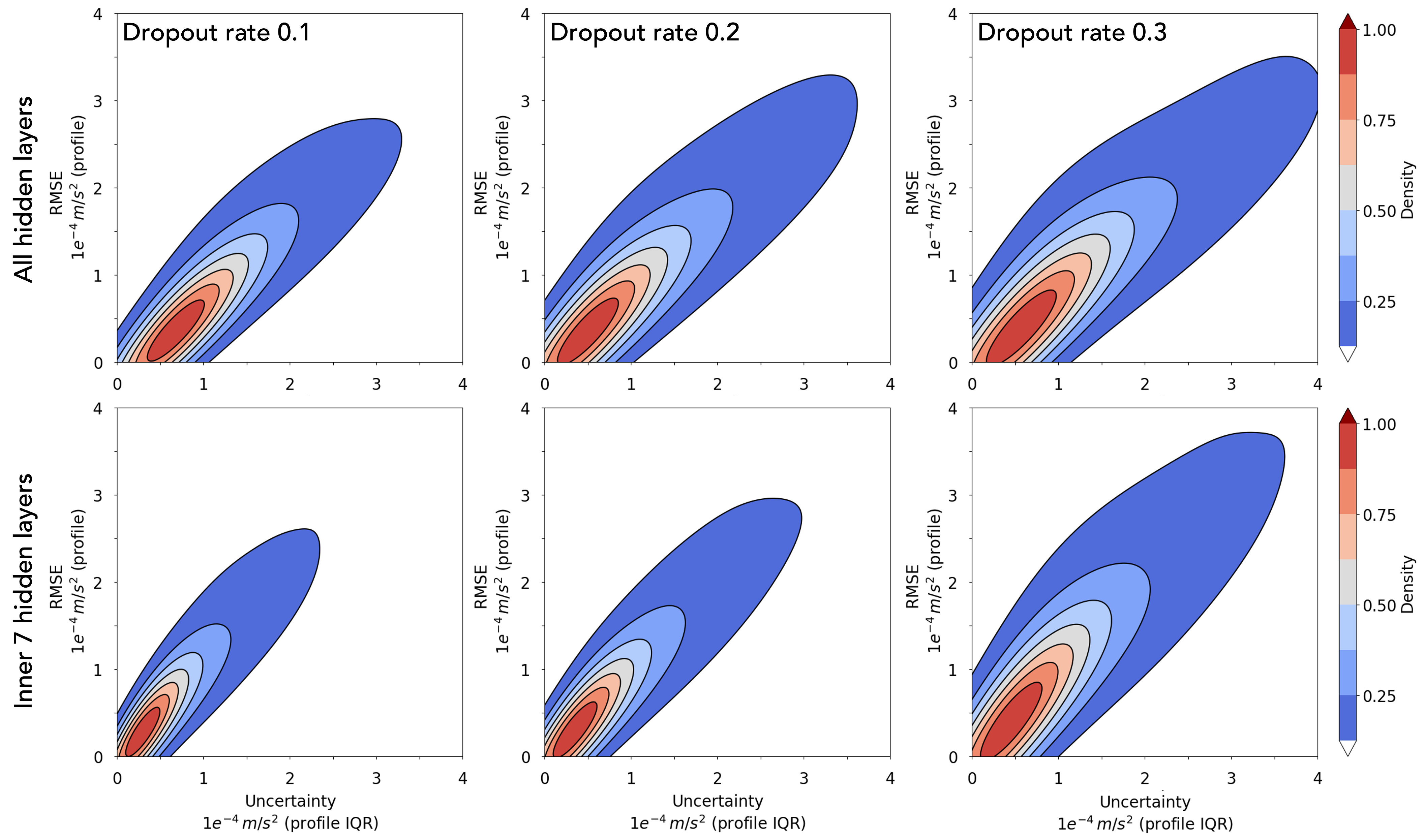}
    \caption{Similar to Figure~\ref{fig: spread-skill-sample} but for DNN only with different dropout rates, which are applied to different numbers of hidden layers. }
    \label{fig: DNN-tuning}
\end{figure}

In addition to the hyperparameters of the deterministic NNs, designing an architecture for UQ often demands additional hyperparameter optimization. For instance, for the DNN, decisions need to be made regarding the number of neurons to drop out (dropout rate). While less common, one can also choose whether to apply dropout to all hidden layers or only selected ones. Variations in the dropout rate and the layers to which dropout is applied can influence the final configuration and performance of the DNN. Figure~\ref{fig: DNN-tuning} illustrates these effects. As we increase the number of dropped neurons (whether through a higher dropout rate or by subjecting more layers to dropout), the uncertainty in the DNN predictions tends to rise. Yet, there is a persistent pattern in the relationship between spread (IQR) and RMSE across the various plots in Figure \ref{fig: DNN-tuning}. Specifically, as spread increases, RMSE concurrently grows, consistent with the insights highlighted in Figure~\ref{fig: spread-skill-sample}.

In the case of BNN or VAE, even though there is no dropout rate, there are distinct tuning opportunities available. For instance, with the VAE, one might consider applying dropout to the NN emulator. Moreover, given that the loss function in VAE comprises three components, decisions can be made regarding which component to penalize more heavily, allowing for nuanced adjustments to its performance.

\section{The UQ metrics} \label{sec:appUQ}
Each point in the spread-skill plot corresponds to one specific bin of ensemble spread ($\overline{\mathrm{SD}_k}$), which is defined as the average standard deviation of the ensemble members. We first separate the spread using a pre-selected number of bins (a subjective choice of 15 is used here). Then for the $k^{th}$ bin:
  
 \begin{eqnarray}~\label{eqn. spread-skill}
 \begin{cases}
 
    \mathrm{RMSE}_k = \left[ \frac{1}{N_k} \sum_{i=1}^{N_k} \left( \hat{y_i} - \overline{y_i} \right)^2  
 \right]^\frac{1}{2}\\
 
  \overline{\mathrm{SD}_k} =  \frac{1}{N_k} \sum_{i=1}^{N_k} \left[\frac{1}{M-1} \sum_{j=1}^{M}  \left(  \overline{y_i} - y_{ij} \right)^2  
  \right]^\frac{1}{2}\\
  
  \overline{y_i} = \frac{1}{M} \sum_{j=1}^{M} y_{ij}
  \end{cases}
  \end{eqnarray}
$\hat{y_i}$ is the observed value for the $i^{th}$ example, $\overline{y_i}$ is the mean prediction for the $i^{th}$ example, $y_{ij}$ is the $j^{th}$ prediction for the $i^{th}$ example, $N_k$ is the total number of examples in the $k^{th}$ bin, and $M$ is the ensemble size. Following \cite{haynes2023creating}, we summarize the quality of the spread-skill plot by two measures: spread-skill reliability (SSREL) and overall spread-skill ratio (SSRAT). SSREL is the bin-weighted mean distance from the 1-to-1 line: 

 \begin{eqnarray}~\label{eqn. SSREL}
    \mathrm{SSREL} = \sum_{k=1}^{K} \frac{N_k}{N}|\mathrm{RMSE}_k - \overline{\mathrm{SD}_k}|
    \end{eqnarray}
where $N$ is the total number of examples, $K$ is the total number of bins, and other variables are as in Equation~\ref{eqn. spread-skill}. SSREL varies from $[0, \infty)$, and the ideal value is 0. On the other hand, SSRAT is averaged over the whole dataset:

 \begin{eqnarray}~\label{eqn. SSRAT}
    \mathrm{SSRAT} = \frac{\overline{\mathrm{SD}}}{\mathrm{RMSE}}
\end{eqnarray}
SSRAT also varies from $[0, \infty)$, and the ideal value is 1. SSRAT $>$ 1 indicates the model is under-confident on average, while SSRAT $<$ 1 indicates that the model is overconfident on average.

In Equation \eqref{eqn. spread-skill}, each level of a GWD profile is considered as an individual sample. As discussed earlier, while these samples help assess the model's overall performance, our main interest is often the uncertainty of individual GWD profiles. Such uncertainty informs the trustworthiness of the model's prediction for that specific profile. Accordingly, for each profile, we can compute:

 \begin{eqnarray}~\label{eqn. spread-skill-sample}
 \begin{cases}
 
    \mathrm{RMSE}_{\mathrm{profile}} = \left[ \frac{1}{N_z} \sum_{z=1}^{N_z} \left( \hat{y_z} - \overline{y_z} \right)^2  
 \right]_{\mathrm{profile}}^\frac{1}{2}\\
 
  \mathrm{IQR}_{\mathrm{profile}} =   \left[\frac{1}{N_z} \sum_{z=1}^{N_z}  \left(  y_{z,75th} - y_{z,25th} \right)^2  
  \right]_{\mathrm{profile}}^\frac{1}{2}\\
  
  \overline{y_{z}} = \left[\frac{1}{M} \sum_{j=1}^{M} y_{zj}\right]_{\mathrm{profile}}
  \end{cases}
\end{eqnarray}
where ${N_z}$ is the number of vertical levels for each profile, and $\mathrm{IQR}_{\mathrm{profile}}$ is its interquartile range: $y_{z,25th}$ corresponds with the 25th percentile, and $y_{z,75th}$ corresponds with the 75th percentile. 

% Instead of uncertainty estimate for each bin as in Equation~\eqref{eqn. spread-skill} that provides statistics averaged over many profiles, Equation~\eqref{eqn. spread-skill-sample} could be calculated for each profile separately.

%\textcolor{red}{Working on the Table/Figure for all Inputs and Outputs variables in WACCM and the NN}

%\section{SI info}

%\textcolor{blue}{Additional figures in SI. e.g., different layers we tried for transfer learning
%(first layer gives the best result).}

% will show
% A: Here Is Appendix Title
%
%\appendix
%\section{Here is a sample appendix}

%%%%%%%%%%%%%%%%%%%%%%%%%%%%%%%%%%%%%%%%%%%%%%%%%%%%%%%%%%%%%%%%
%
% Optional Glossary, Notation or Acronym section goes here:
%
%%%%%%%%%%%%%%
% Glossary is only allowed in Reviews of Geophysics
%  \begin{glossary}
%  \term{Term}
%   Term Definition here
%  \term{Term}
%   Term Definition here
%  \term{Term}
%   Term Definition here
%  \end{glossary}

%
%%%%%%%%%%%%%%
% Acronyms
%   \begin{acronyms}
%   \acro{Acronym}
%   Definition here
%   \acro{EMOS}
%   Ensemble model output statistics
%   \acro{ECMWF}
%   Centre for Medium-Range Weather Forecasts
%   \end{acronyms}

%
%%%%%%%%%%%%%%
% Notation
%   \begin{notation}
%   \notation{$a+b$} Notation Definition here
%   \notation{$e=mc^2$}
%   Equation in German-born physicist Albert Einstein's theory of special
%  relativity that showed that the increased relativistic mass ($m$) of a
%  body comes from the energy of motion of the body—that is, its kinetic
%  energy ($E$)—divided by the speed of light squared ($c^2$).
%   \end{notation}

\section*{Open Research}
The data for all the analyses in the main text are available at \url{https://doi.org/10.5281/zenodo.10019987}. The emulator code is available at \url{https://github.com/yqsun91/WACCM-Emulation}. All the raw WACCM output data are available on request from authors. 

\acknowledgments
We thank Andre Souza for insightful discussions. This work was supported by grants from the NSF OAC CSSI program (\#2005123 , \#2004512, \#2004492, \#2004572), and by the generosity of Eric and Wendy Schmidt by recommendation of the Schmidt Futures program to PH, MJA, EG, and AS. PH is also supported by the Office of Naval Research (ONR) Young Investigator Award N00014-20-1-2722. SL is supported by the Office of Science, U.S. Department of Energy Biological and Environmental Research as part of the Regional and Global Climate Model Analysis program area. Computational resources were provided by NSF XSEDE (allocation ATM170020) and NCAR's CISL (allocation URIC0009).

%% ------------------------------------------------------------------------ %%
%% References and Citations

%%%%%%%%%%%%%%%%%%%%%%%%%%%%%%%%%%%%%%%%%%%%%%%
%
% \bibliography{<name of your .bib file>} don't specify the file extension
%
% don't specify bibliographystyle

% In the References section, cite the data/software described in the Availability Statement (this includes primary and processed data used for your research). For details on data/software citation as well as examples, see the Data & Software Citation section of the Data & Software for Authors guidance
% https://www.agu.org/Publish-with-AGU/Publish/Author-Resources/Data-and-Software-for-Authors#citation

%%%%%%%%%%%%%%%%%%%%%%%%%%%%%%%%%%%%%%%%%%%%%%%

\bibliography{main}

\begin{thebibliography}{}

\bibitem [\protect \citeauthoryear {%
Abdar%
\ \protect \BOthers {.}}{%
Abdar%
\ \protect \BOthers {.}}{%
{\protect \APACyear {2021}}%
}]{%
ABDAR2021243}
\APACinsertmetastar {%
ABDAR2021243}%
\begin{APACrefauthors}%
Abdar, M.%
, Pourpanah, F.%
, Hussain, S.%
, Rezazadegan, D.%
, Liu, L.%
, Ghavamzadeh, M.%
\BDBL {}Nahavandi, S.%
\end{APACrefauthors}%
\unskip\
\newblock
\APACrefYearMonthDay{2021}{}{}.
\newblock
{\BBOQ}\APACrefatitle {A review of uncertainty quantification in deep learning: Techniques, applications and challenges} {A review of uncertainty quantification in deep learning: Techniques, applications and challenges}.{\BBCQ}
\newblock
\APACjournalVolNumPages{Information Fusion}{76}{}{243-297}.
\newblock
\begin{APACrefURL} \url{https://www.sciencedirect.com/science/article/pii/S1566253521001081} \end{APACrefURL}
\newblock
\begin{APACrefDOI} \doi{https://doi.org/10.1016/j.inffus.2021.05.008} \end{APACrefDOI}
\PrintBackRefs{\CurrentBib}

\bibitem [\protect \citeauthoryear {%
Achatz%
}{%
Achatz%
}{%
{\protect \APACyear {2022}}%
}]{%
Achatz2022}
\APACinsertmetastar {%
Achatz2022}%
\begin{APACrefauthors}%
Achatz, U.%
\end{APACrefauthors}%
\unskip\
\newblock
\APACrefYearMonthDay{2022}{}{}.
\newblock
{\BBOQ}\APACrefatitle {Gravity Waves and Their Impact on the Atmospheric Flow} {Gravity waves and their impact on the atmospheric flow}.{\BBCQ}
\newblock
\BIn{} \APACrefbtitle {Atmospheric Dynamics} {Atmospheric dynamics}\ (\BPGS\ 407--505).
\newblock
\APACaddressPublisher{Berlin, Heidelberg}{Springer Berlin Heidelberg}.
\newblock
\begin{APACrefURL} \url{https://doi.org/10.1007/978-3-662-63941-2_10} \end{APACrefURL}
\newblock
\begin{APACrefDOI} \doi{10.1007/978-3-662-63941-2_10} \end{APACrefDOI}
\PrintBackRefs{\CurrentBib}

\bibitem [\protect \citeauthoryear {%
Alexander%
\ \protect \BOthers {.}}{%
Alexander%
\ \protect \BOthers {.}}{%
{\protect \APACyear {2010}}%
}]{%
Alexander2010QJRMS}
\APACinsertmetastar {%
Alexander2010QJRMS}%
\begin{APACrefauthors}%
Alexander, M\BPBI J.%
, Geller, M.%
, McLandress, C.%
, Polavarapu, S.%
, Preusse, P.%
, Sassi, F.%
\BDBL {}Watanabe, S.%
\end{APACrefauthors}%
\unskip\
\newblock
\APACrefYearMonthDay{2010}{}{}.
\newblock
{\BBOQ}\APACrefatitle {Recent developments in gravity-wave effects in climatemodels and the global distribution of gravity-wavemomentum flux from observations and models} {Recent developments in gravity-wave effects in climatemodels and the global distribution of gravity-wavemomentum flux from observations and models}.{\BBCQ}
\newblock
\APACjournalVolNumPages{Quarterly Journal of the Royal Meteorological Society}{136}{}{}.
\newblock
\begin{APACrefDOI} \doi{10.1002/qj.637} \end{APACrefDOI}
\PrintBackRefs{\CurrentBib}

\bibitem [\protect \citeauthoryear {%
Amiramjadi%
, Plougonven%
, Mohebalhojeh%
\BCBL {}\ \BBA {} Mirzaei%
}{%
Amiramjadi%
\ \protect \BOthers {.}}{%
{\protect \APACyear {2022}}%
}]{%
Amiramjadi2022}
\APACinsertmetastar {%
Amiramjadi2022}%
\begin{APACrefauthors}%
Amiramjadi, M.%
, Plougonven, R.%
, Mohebalhojeh, A\BPBI R.%
\BCBL {}\ \BBA {} Mirzaei, M.%
\end{APACrefauthors}%
\unskip\
\newblock
\APACrefYearMonthDay{2022}{}{}.
\newblock
{\BBOQ}\APACrefatitle {Using machine learning to estimate non-orographic gravity wave characteristics at source levels} {Using machine learning to estimate non-orographic gravity wave characteristics at source levels}.{\BBCQ}
\newblock
\APACjournalVolNumPages{Journal of the Atmospheric Sciences}{}{}{}.
\newblock
\begin{APACrefDOI} \doi{10.1175/JAS-D-22-0021.1} \end{APACrefDOI}
\PrintBackRefs{\CurrentBib}

\bibitem [\protect \citeauthoryear {%
Ando%
\ \BBA {} Huang%
}{%
Ando%
\ \BBA {} Huang%
}{%
{\protect \APACyear {2017}}%
}]{%
ando2017deep}
\APACinsertmetastar {%
ando2017deep}%
\begin{APACrefauthors}%
Ando, S.%
\BCBT {}\ \BBA {} Huang, C\BPBI Y.%
\end{APACrefauthors}%
\unskip\
\newblock
\APACrefYearMonthDay{2017}{}{}.
\newblock
{\BBOQ}\APACrefatitle {Deep over-sampling framework for classifying imbalanced data} {Deep over-sampling framework for classifying imbalanced data}.{\BBCQ}
\newblock
\BIn{} \APACrefbtitle {Machine Learning and Knowledge Discovery in Databases: European Conference, ECML PKDD 2017, Skopje, Macedonia, September 18--22, 2017, Proceedings, Part I 10} {Machine learning and knowledge discovery in databases: European conference, ecml pkdd 2017, skopje, macedonia, september 18--22, 2017, proceedings, part i 10}\ (\BPGS\ 770--785).
\PrintBackRefs{\CurrentBib}

\bibitem [\protect \citeauthoryear {%
Bacmeister%
, Newman%
, Gary%
\BCBL {}\ \BBA {} Chan%
}{%
Bacmeister%
\ \protect \BOthers {.}}{%
{\protect \APACyear {1994}}%
}]{%
Bacmeister1994}
\APACinsertmetastar {%
Bacmeister1994}%
\begin{APACrefauthors}%
Bacmeister, J\BPBI T.%
, Newman, P\BPBI A.%
, Gary, B\BPBI L.%
\BCBL {}\ \BBA {} Chan, K\BPBI R.%
\end{APACrefauthors}%
\unskip\
\newblock
\APACrefYearMonthDay{1994}{}{}.
\newblock
{\BBOQ}\APACrefatitle {An algorithm for forecasting mountain wave-related turbulence in the stratosphere} {An algorithm for forecasting mountain wave-related turbulence in the stratosphere}.{\BBCQ}
\newblock
\APACjournalVolNumPages{Weather and Forecasting}{9}{}{}.
\newblock
\begin{APACrefDOI} \doi{10.1175/1520-0434(1994)009<0241:AAFFMW>2.0.CO;2} \end{APACrefDOI}
\PrintBackRefs{\CurrentBib}

\bibitem [\protect \citeauthoryear {%
Balaji%
}{%
Balaji%
}{%
{\protect \APACyear {2021}}%
}]{%
balaji2021climbing}
\APACinsertmetastar {%
balaji2021climbing}%
\begin{APACrefauthors}%
Balaji, V.%
\end{APACrefauthors}%
\unskip\
\newblock
\APACrefYearMonthDay{2021}{}{}.
\newblock
{\BBOQ}\APACrefatitle {Climbing down Charney’s ladder: machine learning and the post-Dennard era of computational climate science} {Climbing down charney’s ladder: machine learning and the post-dennard era of computational climate science}.{\BBCQ}
\newblock
\APACjournalVolNumPages{Philosophical Transactions of the Royal Society A}{379}{2194}{20200085}.
\PrintBackRefs{\CurrentBib}

\bibitem [\protect \citeauthoryear {%
Baldi%
, Sadowski%
\BCBL {}\ \BBA {} Whiteson%
}{%
Baldi%
\ \protect \BOthers {.}}{%
{\protect \APACyear {2014}}%
}]{%
baldi2014searching}
\APACinsertmetastar {%
baldi2014searching}%
\begin{APACrefauthors}%
Baldi, P.%
, Sadowski, P.%
\BCBL {}\ \BBA {} Whiteson, D.%
\end{APACrefauthors}%
\unskip\
\newblock
\APACrefYearMonthDay{2014}{}{}.
\newblock
{\BBOQ}\APACrefatitle {Searching for exotic particles in high-energy physics with deep learning} {Searching for exotic particles in high-energy physics with deep learning}.{\BBCQ}
\newblock
\APACjournalVolNumPages{Nature communications}{5}{1}{4308}.
\PrintBackRefs{\CurrentBib}

\bibitem [\protect \citeauthoryear {%
Ballnus%
\ \protect \BOthers {.}}{%
Ballnus%
\ \protect \BOthers {.}}{%
{\protect \APACyear {2017}}%
}]{%
ballnus2017comprehensive}
\APACinsertmetastar {%
ballnus2017comprehensive}%
\begin{APACrefauthors}%
Ballnus, B.%
, Hug, S.%
, Hatz, K.%
, G{\"o}rlitz, L.%
, Hasenauer, J.%
\BCBL {}\ \BBA {} Theis, F\BPBI J.%
\end{APACrefauthors}%
\unskip\
\newblock
\APACrefYearMonthDay{2017}{}{}.
\newblock
{\BBOQ}\APACrefatitle {Comprehensive benchmarking of Markov chain Monte Carlo methods for dynamical systems} {Comprehensive benchmarking of markov chain monte carlo methods for dynamical systems}.{\BBCQ}
\newblock
\APACjournalVolNumPages{BMC Systems Biology}{11}{1}{1--18}.
\PrintBackRefs{\CurrentBib}

\bibitem [\protect \citeauthoryear {%
Barnes%
, Barnes%
\BCBL {}\ \BBA {} DeMaria%
}{%
Barnes%
\ \protect \BOthers {.}}{%
{\protect \APACyear {2023}}%
}]{%
barnes2023sinh}
\APACinsertmetastar {%
barnes2023sinh}%
\begin{APACrefauthors}%
Barnes, E\BPBI A.%
, Barnes, R\BPBI J.%
\BCBL {}\ \BBA {} DeMaria, M.%
\end{APACrefauthors}%
\unskip\
\newblock
\APACrefYearMonthDay{2023}{}{}.
\newblock
{\BBOQ}\APACrefatitle {Sinh-arcsinh-normal distributions to add uncertainty to neural network regression tasks: Applications to tropical cyclone intensity forecasts} {Sinh-arcsinh-normal distributions to add uncertainty to neural network regression tasks: Applications to tropical cyclone intensity forecasts}.{\BBCQ}
\newblock
\APACjournalVolNumPages{Environmental Data Science}{2}{}{e15}.
\PrintBackRefs{\CurrentBib}

\bibitem [\protect \citeauthoryear {%
Beck%
, Flad%
\BCBL {}\ \BBA {} Munz%
}{%
Beck%
\ \protect \BOthers {.}}{%
{\protect \APACyear {2019}}%
}]{%
beck2019deep}
\APACinsertmetastar {%
beck2019deep}%
\begin{APACrefauthors}%
Beck, A.%
, Flad, D.%
\BCBL {}\ \BBA {} Munz, C\BHBI D.%
\end{APACrefauthors}%
\unskip\
\newblock
\APACrefYearMonthDay{2019}{}{}.
\newblock
{\BBOQ}\APACrefatitle {Deep neural networks for data-driven {LES} closure models} {Deep neural networks for data-driven {LES} closure models}.{\BBCQ}
\newblock
\APACjournalVolNumPages{Journal of Computational Physics}{398}{}{108910}.
\PrintBackRefs{\CurrentBib}

\bibitem [\protect \citeauthoryear {%
Beljaars%
, Brown%
\BCBL {}\ \BBA {} Wood%
}{%
Beljaars%
\ \protect \BOthers {.}}{%
{\protect \APACyear {2004}}%
}]{%
Beljaars2004}
\APACinsertmetastar {%
Beljaars2004}%
\begin{APACrefauthors}%
Beljaars, A\BPBI C.%
, Brown, A\BPBI R.%
\BCBL {}\ \BBA {} Wood, N.%
\end{APACrefauthors}%
\unskip\
\newblock
\APACrefYearMonthDay{2004}{}{}.
\newblock
{\BBOQ}\APACrefatitle {A new parametrization of turbulent orographic form drag} {A new parametrization of turbulent orographic form drag}.{\BBCQ}
\newblock
\APACjournalVolNumPages{Quarterly Journal of the Royal Meteorological Society}{130}{}{}.
\newblock
\begin{APACrefDOI} \doi{10.1256/qj.03.73} \end{APACrefDOI}
\PrintBackRefs{\CurrentBib}

\bibitem [\protect \citeauthoryear {%
Belochitski%
\ \BBA {} Krasnopolsky%
}{%
Belochitski%
\ \BBA {} Krasnopolsky%
}{%
{\protect \APACyear {2021}}%
}]{%
belochitskiKrasnopolsky2021}
\APACinsertmetastar {%
belochitskiKrasnopolsky2021}%
\begin{APACrefauthors}%
Belochitski, A.%
\BCBT {}\ \BBA {} Krasnopolsky, V.%
\end{APACrefauthors}%
\unskip\
\newblock
\APACrefYearMonthDay{2021}{}{}.
\newblock
{\BBOQ}\APACrefatitle {Robustness of neural network emulations of radiative transfer parameterizations in a state-of-the-art general circulation model} {Robustness of neural network emulations of radiative transfer parameterizations in a state-of-the-art general circulation model}.{\BBCQ}
\newblock
\APACjournalVolNumPages{Geoscientific Model Development}{14}{12}{7425--7437}.
\newblock
\begin{APACrefURL} \url{https://gmd.copernicus.org/articles/14/7425/2021/} \end{APACrefURL}
\newblock
\begin{APACrefDOI} \doi{10.5194/gmd-14-7425-2021} \end{APACrefDOI}
\PrintBackRefs{\CurrentBib}

\bibitem [\protect \citeauthoryear {%
Beucler%
, Ebert-Uphoff%
, Rasp%
, Pritchard%
\BCBL {}\ \BBA {} Gentine%
}{%
Beucler%
\ \protect \BOthers {.}}{%
{\protect \APACyear {2021}}%
}]{%
BeuclerPritchard2021cloud}
\APACinsertmetastar {%
BeuclerPritchard2021cloud}%
\begin{APACrefauthors}%
Beucler, T.%
, Ebert-Uphoff, I.%
, Rasp, S.%
, Pritchard, M.%
\BCBL {}\ \BBA {} Gentine, P.%
\end{APACrefauthors}%
\unskip\
\newblock
\APACrefYearMonthDay{2021}{05}{}.
\newblock
\APACrefbtitle {Machine Learning for Clouds and Climate (Invited Chapter for the AGU Geophysical Monograph Series "Clouds and Climate").} {Machine learning for clouds and climate (invited chapter for the agu geophysical monograph series "clouds and climate").}
\newblock
\begin{APACrefDOI} \doi{10.1002/essoar.10506925.1} \end{APACrefDOI}
\PrintBackRefs{\CurrentBib}

\bibitem [\protect \citeauthoryear {%
Blundell%
, Cornebise%
, Kavukcuoglu%
\BCBL {}\ \BBA {} Wierstra%
}{%
Blundell%
\ \protect \BOthers {.}}{%
{\protect \APACyear {2015}}%
}]{%
blundell2015weight}
\APACinsertmetastar {%
blundell2015weight}%
\begin{APACrefauthors}%
Blundell, C.%
, Cornebise, J.%
, Kavukcuoglu, K.%
\BCBL {}\ \BBA {} Wierstra, D.%
\end{APACrefauthors}%
\unskip\
\newblock
\APACrefYearMonthDay{2015}{}{}.
\newblock
\APACrefbtitle {Weight Uncertainty in Neural Networks.} {Weight uncertainty in neural networks.}
\PrintBackRefs{\CurrentBib}

\bibitem [\protect \citeauthoryear {%
Bolton%
\ \BBA {} Zanna%
}{%
Bolton%
\ \BBA {} Zanna%
}{%
{\protect \APACyear {2019}}%
}]{%
bolton2019applications}
\APACinsertmetastar {%
bolton2019applications}%
\begin{APACrefauthors}%
Bolton, T.%
\BCBT {}\ \BBA {} Zanna, L.%
\end{APACrefauthors}%
\unskip\
\newblock
\APACrefYearMonthDay{2019}{}{}.
\newblock
{\BBOQ}\APACrefatitle {Applications of deep learning to ocean data inference and subgrid parameterization} {Applications of deep learning to ocean data inference and subgrid parameterization}.{\BBCQ}
\newblock
\APACjournalVolNumPages{Journal of Advances in Modeling Earth Systems}{11}{1}{376--399}.
\PrintBackRefs{\CurrentBib}

\bibitem [\protect \citeauthoryear {%
Brenowitz%
, Beucler%
, Pritchard%
\BCBL {}\ \BBA {} Bretherton%
}{%
Brenowitz%
\ \protect \BOthers {.}}{%
{\protect \APACyear {2020}}%
}]{%
Brenowitz2020instability}
\APACinsertmetastar {%
Brenowitz2020instability}%
\begin{APACrefauthors}%
Brenowitz, N\BPBI D.%
, Beucler, T.%
, Pritchard, M.%
\BCBL {}\ \BBA {} Bretherton, C\BPBI S.%
\end{APACrefauthors}%
\unskip\
\newblock
\APACrefYearMonthDay{2020}{}{}.
\newblock
{\BBOQ}\APACrefatitle {Interpreting and Stabilizing Machine-Learning Parametrizations of Convection} {Interpreting and stabilizing machine-learning parametrizations of convection}.{\BBCQ}
\newblock
\APACjournalVolNumPages{Journal of the Atmospheric Sciences}{77}{12}{4357 - 4375}.
\newblock
\begin{APACrefURL} \url{https://journals.ametsoc.org/view/journals/atsc/77/12/jas-d-20-0082.1.xml} \end{APACrefURL}
\newblock
\begin{APACrefDOI} \doi{https://doi.org/10.1175/JAS-D-20-0082.1} \end{APACrefDOI}
\PrintBackRefs{\CurrentBib}

\bibitem [\protect \citeauthoryear {%
Brenowitz%
\ \BBA {} Bretherton%
}{%
Brenowitz%
\ \BBA {} Bretherton%
}{%
{\protect \APACyear {2019}}%
}]{%
brenowitzBretherton2019}
\APACinsertmetastar {%
brenowitzBretherton2019}%
\begin{APACrefauthors}%
Brenowitz, N\BPBI D.%
\BCBT {}\ \BBA {} Bretherton, C\BPBI S.%
\end{APACrefauthors}%
\unskip\
\newblock
\APACrefYearMonthDay{2019}{}{}.
\newblock
{\BBOQ}\APACrefatitle {Spatially Extended Tests of a Neural Network Parametrization Trained by Coarse-Graining} {Spatially extended tests of a neural network parametrization trained by coarse-graining}.{\BBCQ}
\newblock
\APACjournalVolNumPages{Journal of Advances in Modeling Earth Systems}{11}{8}{2728-2744}.
\newblock
\begin{APACrefURL} \url{https://agupubs.onlinelibrary.wiley.com/doi/abs/10.1029/2019MS001711} \end{APACrefURL}
\newblock
\begin{APACrefDOI} \doi{https://doi.org/10.1029/2019MS001711} \end{APACrefDOI}
\PrintBackRefs{\CurrentBib}

\bibitem [\protect \citeauthoryear {%
Buda%
, Maki%
\BCBL {}\ \BBA {} Mazurowski%
}{%
Buda%
\ \protect \BOthers {.}}{%
{\protect \APACyear {2018}}%
}]{%
BUDA2018249}
\APACinsertmetastar {%
BUDA2018249}%
\begin{APACrefauthors}%
Buda, M.%
, Maki, A.%
\BCBL {}\ \BBA {} Mazurowski, M\BPBI A.%
\end{APACrefauthors}%
\unskip\
\newblock
\APACrefYearMonthDay{2018}{}{}.
\newblock
{\BBOQ}\APACrefatitle {A systematic study of the class imbalance problem in convolutional neural networks} {A systematic study of the class imbalance problem in convolutional neural networks}.{\BBCQ}
\newblock
\APACjournalVolNumPages{Neural Networks}{106}{}{249-259}.
\newblock
\begin{APACrefURL} \url{https://www.sciencedirect.com/science/article/pii/S0893608018302107} \end{APACrefURL}
\newblock
\begin{APACrefDOI} \doi{https://doi.org/10.1016/j.neunet.2018.07.011} \end{APACrefDOI}
\PrintBackRefs{\CurrentBib}

\bibitem [\protect \citeauthoryear {%
Bölöni%
, Kim%
, Borchert%
\BCBL {}\ \BBA {} Achatz%
}{%
Bölöni%
\ \protect \BOthers {.}}{%
{\protect \APACyear {2021}}%
}]{%
Boloni2021}
\APACinsertmetastar {%
Boloni2021}%
\begin{APACrefauthors}%
Bölöni, G.%
, Kim, Y\BPBI H.%
, Borchert, S.%
\BCBL {}\ \BBA {} Achatz, U.%
\end{APACrefauthors}%
\unskip\
\newblock
\APACrefYearMonthDay{2021}{}{}.
\newblock
{\BBOQ}\APACrefatitle {Toward transient subgrid-scale gravity wave representation in atmospheric models. Part I: Propagation model including nondissipative wave mean-flow interactions} {Toward transient subgrid-scale gravity wave representation in atmospheric models. part i: Propagation model including nondissipative wave mean-flow interactions}.{\BBCQ}
\newblock
\APACjournalVolNumPages{Journal of the Atmospheric Sciences}{78}{}{}.
\newblock
\begin{APACrefDOI} \doi{10.1175/JAS-D-20-0065.1} \end{APACrefDOI}
\PrintBackRefs{\CurrentBib}

\bibitem [\protect \citeauthoryear {%
Chantry%
, Hatfield%
, Dueben%
, Polichtchouk%
\BCBL {}\ \BBA {} Palmer%
}{%
Chantry%
\ \protect \BOthers {.}}{%
{\protect \APACyear {2021}}%
}]{%
chantry2021}
\APACinsertmetastar {%
chantry2021}%
\begin{APACrefauthors}%
Chantry, M.%
, Hatfield, S.%
, Dueben, P.%
, Polichtchouk, I.%
\BCBL {}\ \BBA {} Palmer, T.%
\end{APACrefauthors}%
\unskip\
\newblock
\APACrefYearMonthDay{2021}{}{}.
\newblock
{\BBOQ}\APACrefatitle {Machine Learning Emulation of Gravity Wave Drag in Numerical Weather Forecasting} {Machine learning emulation of gravity wave drag in numerical weather forecasting}.{\BBCQ}
\newblock
\APACjournalVolNumPages{Journal of Advances in Modeling Earth Systems}{13}{7}{e2021MS002477}.
\newblock
\begin{APACrefURL} \url{https://agupubs.onlinelibrary.wiley.com/doi/abs/10.1029/2021MS002477} \end{APACrefURL}
\newblock
\APACrefnote{e2021MS002477 2021MS002477}
\newblock
\begin{APACrefDOI} \doi{https://doi.org/10.1029/2021MS002477} \end{APACrefDOI}
\PrintBackRefs{\CurrentBib}

\bibitem [\protect \citeauthoryear {%
Chattopadhyay%
, Nabizadeh%
\BCBL {}\ \BBA {} Hassanzadeh%
}{%
Chattopadhyay%
, Nabizadeh%
\BCBL {}\ \BBA {} Hassanzadeh%
}{%
{\protect \APACyear {2020}}%
}]{%
chattopadhyay2020analog}
\APACinsertmetastar {%
chattopadhyay2020analog}%
\begin{APACrefauthors}%
Chattopadhyay, A.%
, Nabizadeh, E.%
\BCBL {}\ \BBA {} Hassanzadeh, P.%
\end{APACrefauthors}%
\unskip\
\newblock
\APACrefYearMonthDay{2020}{}{}.
\newblock
{\BBOQ}\APACrefatitle {Analog forecasting of extreme-causing weather patterns using deep learning} {Analog forecasting of extreme-causing weather patterns using deep learning}.{\BBCQ}
\newblock
\APACjournalVolNumPages{Journal of Advances in Modeling Earth Systems}{12}{2}{e2019MS001958}.
\PrintBackRefs{\CurrentBib}

\bibitem [\protect \citeauthoryear {%
Chattopadhyay%
, Subel%
\BCBL {}\ \BBA {} Hassanzadeh%
}{%
Chattopadhyay%
, Subel%
\BCBL {}\ \BBA {} Hassanzadeh%
}{%
{\protect \APACyear {2020}}%
}]{%
chattopadhyay2020data}
\APACinsertmetastar {%
chattopadhyay2020data}%
\begin{APACrefauthors}%
Chattopadhyay, A.%
, Subel, A.%
\BCBL {}\ \BBA {} Hassanzadeh, P.%
\end{APACrefauthors}%
\unskip\
\newblock
\APACrefYearMonthDay{2020}{}{}.
\newblock
{\BBOQ}\APACrefatitle {Data-driven super-parameterization using deep learning: Experimentation with multiscale {L}orenz 96 systems and transfer learning} {Data-driven super-parameterization using deep learning: Experimentation with multiscale {L}orenz 96 systems and transfer learning}.{\BBCQ}
\newblock
\APACjournalVolNumPages{Journal of Advances in Modeling Earth Systems}{12}{11}{e2020MS002084}.
\PrintBackRefs{\CurrentBib}

\bibitem [\protect \citeauthoryear {%
Chawla%
, Japkowicz%
\BCBL {}\ \BBA {} Kotcz%
}{%
Chawla%
\ \protect \BOthers {.}}{%
{\protect \APACyear {2004}}%
}]{%
Chawla_etal_2004}
\APACinsertmetastar {%
Chawla_etal_2004}%
\begin{APACrefauthors}%
Chawla, N\BPBI V.%
, Japkowicz, N.%
\BCBL {}\ \BBA {} Kotcz, A.%
\end{APACrefauthors}%
\unskip\
\newblock
\APACrefYearMonthDay{2004}{jun}{}.
\newblock
{\BBOQ}\APACrefatitle {Editorial: Special Issue on Learning from Imbalanced Data Sets} {Editorial: Special issue on learning from imbalanced data sets}.{\BBCQ}
\newblock
\APACjournalVolNumPages{SIGKDD Explor. Newsl.}{6}{1}{1–6}.
\newblock
\begin{APACrefURL} \url{https://doi.org/10.1145/1007730.1007733} \end{APACrefURL}
\newblock
\begin{APACrefDOI} \doi{10.1145/1007730.1007733} \end{APACrefDOI}
\PrintBackRefs{\CurrentBib}

\bibitem [\protect \citeauthoryear {%
Chen%
\ \BBA {} Majda%
}{%
Chen%
\ \BBA {} Majda%
}{%
{\protect \APACyear {2019}}%
}]{%
chen2019new}
\APACinsertmetastar {%
chen2019new}%
\begin{APACrefauthors}%
Chen, N.%
\BCBT {}\ \BBA {} Majda, A\BPBI J.%
\end{APACrefauthors}%
\unskip\
\newblock
\APACrefYearMonthDay{2019}{}{}.
\newblock
{\BBOQ}\APACrefatitle {A new efficient parameter estimation algorithm for high-dimensional complex nonlinear turbulent dynamical systems with partial observations} {A new efficient parameter estimation algorithm for high-dimensional complex nonlinear turbulent dynamical systems with partial observations}.{\BBCQ}
\newblock
\APACjournalVolNumPages{Journal of Computational Physics}{397}{}{108836}.
\PrintBackRefs{\CurrentBib}

\bibitem [\protect \citeauthoryear {%
Clare%
, Sonnewald%
, Lguensat%
, Deshayes%
\BCBL {}\ \BBA {} Balaji%
}{%
Clare%
\ \protect \BOthers {.}}{%
{\protect \APACyear {2022}}%
}]{%
clare2022explainable}
\APACinsertmetastar {%
clare2022explainable}%
\begin{APACrefauthors}%
Clare, M\BPBI C.%
, Sonnewald, M.%
, Lguensat, R.%
, Deshayes, J.%
\BCBL {}\ \BBA {} Balaji, V.%
\end{APACrefauthors}%
\unskip\
\newblock
\APACrefYearMonthDay{2022}{}{}.
\newblock
{\BBOQ}\APACrefatitle {Explainable Artificial Intelligence for Bayesian Neural Networks: Towards trustworthy predictions of ocean dynamics} {Explainable artificial intelligence for bayesian neural networks: Towards trustworthy predictions of ocean dynamics}.{\BBCQ}
\newblock
\APACjournalVolNumPages{arXiv preprint arXiv:2205.00202}{}{}{}.
\PrintBackRefs{\CurrentBib}

\bibitem [\protect \citeauthoryear {%
Delle~Monache%
, Eckel%
, Rife%
, Nagarajan%
\BCBL {}\ \BBA {} Searight%
}{%
Delle~Monache%
\ \protect \BOthers {.}}{%
{\protect \APACyear {2013}}%
}]{%
delle2013probabilistic}
\APACinsertmetastar {%
delle2013probabilistic}%
\begin{APACrefauthors}%
Delle~Monache, L.%
, Eckel, F\BPBI A.%
, Rife, D\BPBI L.%
, Nagarajan, B.%
\BCBL {}\ \BBA {} Searight, K.%
\end{APACrefauthors}%
\unskip\
\newblock
\APACrefYearMonthDay{2013}{}{}.
\newblock
{\BBOQ}\APACrefatitle {Probabilistic weather prediction with an analog ensemble} {Probabilistic weather prediction with an analog ensemble}.{\BBCQ}
\newblock
\APACjournalVolNumPages{Monthly Weather Review}{141}{10}{3498--3516}.
\PrintBackRefs{\CurrentBib}

\bibitem [\protect \citeauthoryear {%
Dong%
\ \protect \BOthers {.}}{%
Dong%
\ \protect \BOthers {.}}{%
{\protect \APACyear {2023}}%
}]{%
Dongetal2023}
\APACinsertmetastar {%
Dongetal2023}%
\begin{APACrefauthors}%
Dong, W.%
, Fritts, D\BPBI C.%
, Liu, A\BPBI Z.%
, Lund, T\BPBI S.%
, Liu, H\BHBI L.%
\BCBL {}\ \BBA {} Snively, J.%
\end{APACrefauthors}%
\unskip\
\newblock
\APACrefYearMonthDay{2023}{}{}.
\newblock
{\BBOQ}\APACrefatitle {Accelerating Atmospheric Gravity Wave Simulations Using Machine Learning: Kelvin-Helmholtz Instability and Mountain Wave Sources Driving Gravity Wave Breaking and Secondary Gravity Wave Generation} {Accelerating atmospheric gravity wave simulations using machine learning: Kelvin-helmholtz instability and mountain wave sources driving gravity wave breaking and secondary gravity wave generation}.{\BBCQ}
\newblock
\APACjournalVolNumPages{Geophysical Research Letters}{50}{15}{e2023GL104668}.
\newblock
\begin{APACrefURL} \url{https://agupubs.onlinelibrary.wiley.com/doi/abs/10.1029/2023GL104668} \end{APACrefURL}
\newblock
\APACrefnote{e2023GL104668 2023GL104668}
\newblock
\begin{APACrefDOI} \doi{https://doi.org/10.1029/2023GL104668} \end{APACrefDOI}
\PrintBackRefs{\CurrentBib}

\bibitem [\protect \citeauthoryear {%
Espinosa%
, Sheshadri%
, Cain%
, Gerber%
\BCBL {}\ \BBA {} DallaSanta%
}{%
Espinosa%
\ \protect \BOthers {.}}{%
{\protect \APACyear {2022}}%
}]{%
Espinosa2022}
\APACinsertmetastar {%
Espinosa2022}%
\begin{APACrefauthors}%
Espinosa, Z\BPBI I.%
, Sheshadri, A.%
, Cain, G\BPBI R.%
, Gerber, E\BPBI P.%
\BCBL {}\ \BBA {} DallaSanta, K\BPBI J.%
\end{APACrefauthors}%
\unskip\
\newblock
\APACrefYearMonthDay{2022}{}{}.
\newblock
{\BBOQ}\APACrefatitle {Machine learning gravity wave parameterization generalizes to capture the QBO and response to increased CO2} {Machine learning gravity wave parameterization generalizes to capture the qbo and response to increased co2}.{\BBCQ}
\newblock
\APACjournalVolNumPages{Geophysical Research Letters}{49}{8}{e2022GL098174}.
\PrintBackRefs{\CurrentBib}

\bibitem [\protect \citeauthoryear {%
Finkel%
, Gerber%
, Abbot%
\BCBL {}\ \BBA {} Weare%
}{%
Finkel%
\ \protect \BOthers {.}}{%
{\protect \APACyear {2023}}%
}]{%
Finkeletal2023}
\APACinsertmetastar {%
Finkeletal2023}%
\begin{APACrefauthors}%
Finkel, J.%
, Gerber, E\BPBI P.%
, Abbot, D\BPBI S.%
\BCBL {}\ \BBA {} Weare, J.%
\end{APACrefauthors}%
\unskip\
\newblock
\APACrefYearMonthDay{2023}{}{}.
\newblock
{\BBOQ}\APACrefatitle {Revealing the Statistics of Extreme Events Hidden in Short Weather Forecast Data} {Revealing the statistics of extreme events hidden in short weather forecast data}.{\BBCQ}
\newblock
\APACjournalVolNumPages{AGU Advances}{4}{2}{e2023AV000881}.
\newblock
\begin{APACrefURL} \url{https://agupubs.onlinelibrary.wiley.com/doi/abs/10.1029/2023AV000881} \end{APACrefURL}
\newblock
\APACrefnote{e2023AV000881 2023AV000881}
\newblock
\begin{APACrefDOI} \doi{https://doi.org/10.1029/2023AV000881} \end{APACrefDOI}
\PrintBackRefs{\CurrentBib}

\bibitem [\protect \citeauthoryear {%
Foster%
, Gagne%
\BCBL {}\ \BBA {} Whitt%
}{%
Foster%
\ \protect \BOthers {.}}{%
{\protect \APACyear {2021}}%
}]{%
Foster2021}
\APACinsertmetastar {%
Foster2021}%
\begin{APACrefauthors}%
Foster, D.%
, Gagne, D\BPBI J.%
\BCBL {}\ \BBA {} Whitt, D\BPBI B.%
\end{APACrefauthors}%
\unskip\
\newblock
\APACrefYearMonthDay{2021}{12}{}.
\newblock
{\BBOQ}\APACrefatitle {Probabilistic Machine Learning Estimation of Ocean Mixed Layer Depth From Dense Satellite and Sparse In Situ Observations} {Probabilistic machine learning estimation of ocean mixed layer depth from dense satellite and sparse in situ observations}.{\BBCQ}
\newblock
\APACjournalVolNumPages{Journal of Advances in Modeling Earth Systems}{13}{}{}.
\newblock
\begin{APACrefDOI} \doi{10.1029/2021MS002474} \end{APACrefDOI}
\PrintBackRefs{\CurrentBib}

\bibitem [\protect \citeauthoryear {%
Frezat%
, Sommer%
, Fablet%
, Balarac%
\BCBL {}\ \BBA {} Lguensat%
}{%
Frezat%
\ \protect \BOthers {.}}{%
{\protect \APACyear {2022}}%
}]{%
frezat2022posteriori}
\APACinsertmetastar {%
frezat2022posteriori}%
\begin{APACrefauthors}%
Frezat, H.%
, Sommer, J\BPBI L.%
, Fablet, R.%
, Balarac, G.%
\BCBL {}\ \BBA {} Lguensat, R.%
\end{APACrefauthors}%
\unskip\
\newblock
\APACrefYearMonthDay{2022}{}{}.
\newblock
{\BBOQ}\APACrefatitle {A posteriori learning for quasi-geostrophic turbulence parametrization} {A posteriori learning for quasi-geostrophic turbulence parametrization}.{\BBCQ}
\newblock
\APACjournalVolNumPages{arXiv preprint arXiv:2204.03911}{}{}{}.
\newblock
\begin{APACrefDOI} \doi{10.1029/2022MS003124} \end{APACrefDOI}
\PrintBackRefs{\CurrentBib}

\bibitem [\protect \citeauthoryear {%
Gagne%
, Christensen%
, Subramanian%
\BCBL {}\ \BBA {} Monahan%
}{%
Gagne%
\ \protect \BOthers {.}}{%
{\protect \APACyear {2020}}%
}]{%
gagne2020machine}
\APACinsertmetastar {%
gagne2020machine}%
\begin{APACrefauthors}%
Gagne, D\BPBI J.%
, Christensen, H\BPBI M.%
, Subramanian, A\BPBI C.%
\BCBL {}\ \BBA {} Monahan, A\BPBI H.%
\end{APACrefauthors}%
\unskip\
\newblock
\APACrefYearMonthDay{2020}{}{}.
\newblock
{\BBOQ}\APACrefatitle {Machine learning for stochastic parameterization: Generative adversarial networks in the Lorenz'96 model} {Machine learning for stochastic parameterization: Generative adversarial networks in the lorenz'96 model}.{\BBCQ}
\newblock
\APACjournalVolNumPages{Journal of Advances in Modeling Earth Systems}{12}{3}{e2019MS001896}.
\PrintBackRefs{\CurrentBib}

\bibitem [\protect \citeauthoryear {%
Gal%
\ \BBA {} Ghahramani%
}{%
Gal%
\ \BBA {} Ghahramani%
}{%
{\protect \APACyear {2016}}%
}]{%
gal2016dropout}
\APACinsertmetastar {%
gal2016dropout}%
\begin{APACrefauthors}%
Gal, Y.%
\BCBT {}\ \BBA {} Ghahramani, Z.%
\end{APACrefauthors}%
\unskip\
\newblock
\APACrefYearMonthDay{2016}{}{}.
\newblock
{\BBOQ}\APACrefatitle {Dropout as a bayesian approximation: Representing model uncertainty in deep learning} {Dropout as a bayesian approximation: Representing model uncertainty in deep learning}.{\BBCQ}
\newblock
\BIn{} \APACrefbtitle {international conference on machine learning} {international conference on machine learning}\ (\BPGS\ 1050--1059).
\PrintBackRefs{\CurrentBib}

\bibitem [\protect \citeauthoryear {%
Garcia%
, Smith%
, Kinnison%
, Álvaro de~la Cámara%
\BCBL {}\ \BBA {} Murphy%
}{%
Garcia%
\ \protect \BOthers {.}}{%
{\protect \APACyear {2017}}%
}]{%
GarciaOGW-WACCM2017}
\APACinsertmetastar {%
GarciaOGW-WACCM2017}%
\begin{APACrefauthors}%
Garcia, R\BPBI R.%
, Smith, A\BPBI K.%
, Kinnison, D\BPBI E.%
, Álvaro de~la Cámara%
\BCBL {}\ \BBA {} Murphy, D\BPBI J.%
\end{APACrefauthors}%
\unskip\
\newblock
\APACrefYearMonthDay{2017}{}{}.
\newblock
{\BBOQ}\APACrefatitle {Modification of the Gravity Wave Parameterization in the Whole Atmosphere Community Climate Model: Motivation and Results} {Modification of the gravity wave parameterization in the whole atmosphere community climate model: Motivation and results}.{\BBCQ}
\newblock
\APACjournalVolNumPages{Journal of the Atmospheric Sciences}{74}{1}{275 - 291}.
\newblock
\begin{APACrefURL} \url{https://journals.ametsoc.org/view/journals/atsc/74/1/jas-d-16-0104.1.xml} \end{APACrefURL}
\newblock
\begin{APACrefDOI} \doi{https://doi.org/10.1175/JAS-D-16-0104.1} \end{APACrefDOI}
\PrintBackRefs{\CurrentBib}

\bibitem [\protect \citeauthoryear {%
Geller%
\ \protect \BOthers {.}}{%
Geller%
\ \protect \BOthers {.}}{%
{\protect \APACyear {2013}}%
}]{%
geller2013comparison}
\APACinsertmetastar {%
geller2013comparison}%
\begin{APACrefauthors}%
Geller, M\BPBI A.%
, Alexander, M\BPBI J.%
, Love, P\BPBI T.%
, Bacmeister, J.%
, Ern, M.%
, Hertzog, A.%
\BDBL {}others%
\end{APACrefauthors}%
\unskip\
\newblock
\APACrefYearMonthDay{2013}{}{}.
\newblock
{\BBOQ}\APACrefatitle {A comparison between gravity wave momentum fluxes in observations and climate models} {A comparison between gravity wave momentum fluxes in observations and climate models}.{\BBCQ}
\newblock
\APACjournalVolNumPages{Journal of Climate}{26}{17}{6383--6405}.
\PrintBackRefs{\CurrentBib}

\bibitem [\protect \citeauthoryear {%
Gettelman%
\ \protect \BOthers {.}}{%
Gettelman%
\ \protect \BOthers {.}}{%
{\protect \APACyear {2021}}%
}]{%
Gettelman2021}
\APACinsertmetastar {%
Gettelman2021}%
\begin{APACrefauthors}%
Gettelman, A.%
, Gagne, D\BPBI J.%
, Chen, C\BPBI C.%
, Christensen, M\BPBI W.%
, Lebo, Z\BPBI J.%
, Morrison, H.%
\BCBL {}\ \BBA {} Gantos, G.%
\end{APACrefauthors}%
\unskip\
\newblock
\APACrefYearMonthDay{2021}{}{}.
\newblock
{\BBOQ}\APACrefatitle {Machine Learning the Warm Rain Process} {Machine learning the warm rain process}.{\BBCQ}
\newblock
\APACjournalVolNumPages{Journal of Advances in Modeling Earth Systems}{13}{}{}.
\newblock
\begin{APACrefDOI} \doi{10.1029/2020MS002268} \end{APACrefDOI}
\PrintBackRefs{\CurrentBib}

\bibitem [\protect \citeauthoryear {%
Gettelman%
\ \protect \BOthers {.}}{%
Gettelman%
\ \protect \BOthers {.}}{%
{\protect \APACyear {2019}}%
}]{%
Gettelman2019}
\APACinsertmetastar {%
Gettelman2019}%
\begin{APACrefauthors}%
Gettelman, A.%
, Mills, M\BPBI J.%
, Kinnison, D\BPBI E.%
, Garcia, R\BPBI R.%
, Smith, A\BPBI K.%
, Marsh, D\BPBI R.%
\BDBL {}Randel, W\BPBI J.%
\end{APACrefauthors}%
\unskip\
\newblock
\APACrefYearMonthDay{2019}{12}{}.
\newblock
{\BBOQ}\APACrefatitle {The Whole Atmosphere Community Climate Model Version 6 (WACCM6)} {The whole atmosphere community climate model version 6 (waccm6)}.{\BBCQ}
\newblock
\APACjournalVolNumPages{Journal of Geophysical Research: Atmospheres}{124}{}{12380-12403}.
\newblock
\begin{APACrefDOI} \doi{10.1029/2019JD030943} \end{APACrefDOI}
\PrintBackRefs{\CurrentBib}

\bibitem [\protect \citeauthoryear {%
Gordon%
\ \BBA {} Barnes%
}{%
Gordon%
\ \BBA {} Barnes%
}{%
{\protect \APACyear {2022}}%
}]{%
Gordon2022}
\APACinsertmetastar {%
Gordon2022}%
\begin{APACrefauthors}%
Gordon, E\BPBI M.%
\BCBT {}\ \BBA {} Barnes, E\BPBI A.%
\end{APACrefauthors}%
\unskip\
\newblock
\APACrefYearMonthDay{2022}{8}{}.
\newblock
{\BBOQ}\APACrefatitle {Incorporating Uncertainty Into a Regression Neural Network Enables Identification of Decadal State-Dependent Predictability in CESM2} {Incorporating uncertainty into a regression neural network enables identification of decadal state-dependent predictability in cesm2}.{\BBCQ}
\newblock
\APACjournalVolNumPages{Geophysical Research Letters}{49}{}{}.
\newblock
\begin{APACrefDOI} \doi{10.1029/2022GL098635} \end{APACrefDOI}
\PrintBackRefs{\CurrentBib}

\bibitem [\protect \citeauthoryear {%
Guan%
, Chattopadhyay%
, Subel%
\BCBL {}\ \BBA {} Hassanzadeh%
}{%
Guan%
\ \protect \BOthers {.}}{%
{\protect \APACyear {2022}}%
}]{%
guan2022stable}
\APACinsertmetastar {%
guan2022stable}%
\begin{APACrefauthors}%
Guan, Y.%
, Chattopadhyay, A.%
, Subel, A.%
\BCBL {}\ \BBA {} Hassanzadeh, P.%
\end{APACrefauthors}%
\unskip\
\newblock
\APACrefYearMonthDay{2022}{}{}.
\newblock
{\BBOQ}\APACrefatitle {Stable a posteriori {LES} of {2D} turbulence using convolutional neural networks: {B}ackscattering analysis and generalization to higher {R}e via transfer learning} {Stable a posteriori {LES} of {2D} turbulence using convolutional neural networks: {B}ackscattering analysis and generalization to higher {R}e via transfer learning}.{\BBCQ}
\newblock
\APACjournalVolNumPages{Journal of Computational Physics}{458}{}{111090}.
\PrintBackRefs{\CurrentBib}

\bibitem [\protect \citeauthoryear {%
Guan%
, Subel%
, Chattopadhyay%
\BCBL {}\ \BBA {} Hassanzadeh%
}{%
Guan%
\ \protect \BOthers {.}}{%
{\protect \APACyear {2023}}%
}]{%
guan2022learning}
\APACinsertmetastar {%
guan2022learning}%
\begin{APACrefauthors}%
Guan, Y.%
, Subel, A.%
, Chattopadhyay, A.%
\BCBL {}\ \BBA {} Hassanzadeh, P.%
\end{APACrefauthors}%
\unskip\
\newblock
\APACrefYearMonthDay{2023}{}{}.
\newblock
{\BBOQ}\APACrefatitle {Learning physics-constrained subgrid-scale closures in the small-data regime for stable and accurate LES} {Learning physics-constrained subgrid-scale closures in the small-data regime for stable and accurate les}.{\BBCQ}
\newblock
\APACjournalVolNumPages{Physica D: Nonlinear Phenomena}{443}{}{133568}.
\newblock
\begin{APACrefDOI} \doi{https://doi.org/10.1016/j.physd.2022.133568} \end{APACrefDOI}
\PrintBackRefs{\CurrentBib}

\bibitem [\protect \citeauthoryear {%
Guillaumin%
\ \BBA {} Zanna%
}{%
Guillaumin%
\ \BBA {} Zanna%
}{%
{\protect \APACyear {2021}}%
}]{%
guillaumin2021stochastic}
\APACinsertmetastar {%
guillaumin2021stochastic}%
\begin{APACrefauthors}%
Guillaumin, A\BPBI P.%
\BCBT {}\ \BBA {} Zanna, L.%
\end{APACrefauthors}%
\unskip\
\newblock
\APACrefYearMonthDay{2021}{}{}.
\newblock
{\BBOQ}\APACrefatitle {Stochastic-deep learning parameterization of ocean momentum forcing} {Stochastic-deep learning parameterization of ocean momentum forcing}.{\BBCQ}
\newblock
\APACjournalVolNumPages{Journal of Advances in Modeling Earth Systems}{13}{9}{e2021MS002534}.
\PrintBackRefs{\CurrentBib}

\bibitem [\protect \citeauthoryear {%
Hardiman%
\ \protect \BOthers {.}}{%
Hardiman%
\ \protect \BOthers {.}}{%
{\protect \APACyear {2023}}%
}]{%
hardiman2023machine}
\APACinsertmetastar {%
hardiman2023machine}%
\begin{APACrefauthors}%
Hardiman, S\BPBI C.%
, Scaife, A\BPBI A.%
, Niekerk, A\BPBI v.%
, Prudden, R.%
, Owen, A.%
, Adams, S\BPBI V.%
\BDBL {}Madge, S.%
\end{APACrefauthors}%
\unskip\
\newblock
\APACrefYearMonthDay{2023}{}{}.
\newblock
{\BBOQ}\APACrefatitle {Machine learning for non-orographic gravity waves in a climate model} {Machine learning for non-orographic gravity waves in a climate model}.{\BBCQ}
\newblock
\APACjournalVolNumPages{Artificial Intelligence for the Earth Systems}{}{}{}.
\PrintBackRefs{\CurrentBib}

\bibitem [\protect \citeauthoryear {%
Haynes%
, Lagerquist%
, McGraw%
, Musgrave%
\BCBL {}\ \BBA {} Ebert-Uphoff%
}{%
Haynes%
\ \protect \BOthers {.}}{%
{\protect \APACyear {2023}}%
}]{%
haynes2023creating}
\APACinsertmetastar {%
haynes2023creating}%
\begin{APACrefauthors}%
Haynes, K.%
, Lagerquist, R.%
, McGraw, M.%
, Musgrave, K.%
\BCBL {}\ \BBA {} Ebert-Uphoff, I.%
\end{APACrefauthors}%
\unskip\
\newblock
\APACrefYearMonthDay{2023}{}{}.
\newblock
{\BBOQ}\APACrefatitle {Creating and evaluating uncertainty estimates with neural networks for environmental-science applications} {Creating and evaluating uncertainty estimates with neural networks for environmental-science applications}.{\BBCQ}
\newblock
\APACjournalVolNumPages{Artificial Intelligence for the Earth Systems}{}{}{1--58}.
\PrintBackRefs{\CurrentBib}

\bibitem [\protect \citeauthoryear {%
Hertzog%
, Alexander%
\BCBL {}\ \BBA {} Plougonven%
}{%
Hertzog%
\ \protect \BOthers {.}}{%
{\protect \APACyear {2012}}%
}]{%
hertzog2012intermittency}
\APACinsertmetastar {%
hertzog2012intermittency}%
\begin{APACrefauthors}%
Hertzog, A.%
, Alexander, M\BPBI J.%
\BCBL {}\ \BBA {} Plougonven, R.%
\end{APACrefauthors}%
\unskip\
\newblock
\APACrefYearMonthDay{2012}{}{}.
\newblock
{\BBOQ}\APACrefatitle {On the intermittency of gravity wave momentum flux in the stratosphere} {On the intermittency of gravity wave momentum flux in the stratosphere}.{\BBCQ}
\newblock
\APACjournalVolNumPages{Journal of the Atmospheric Sciences}{69}{11}{3433--3448}.
\PrintBackRefs{\CurrentBib}

\bibitem [\protect \citeauthoryear {%
Hourdin%
\ \protect \BOthers {.}}{%
Hourdin%
\ \protect \BOthers {.}}{%
{\protect \APACyear {2017}}%
}]{%
hourdin2017art}
\APACinsertmetastar {%
hourdin2017art}%
\begin{APACrefauthors}%
Hourdin, F.%
, Mauritsen, T.%
, Gettelman, A.%
, Golaz, J\BHBI C.%
, Balaji, V.%
, Duan, Q.%
\BDBL {}others%
\end{APACrefauthors}%
\unskip\
\newblock
\APACrefYearMonthDay{2017}{}{}.
\newblock
{\BBOQ}\APACrefatitle {The art and science of climate model tuning} {The art and science of climate model tuning}.{\BBCQ}
\newblock
\APACjournalVolNumPages{Bulletin of the American Meteorological Society}{98}{3}{589--602}.
\PrintBackRefs{\CurrentBib}

\bibitem [\protect \citeauthoryear {%
Huang%
, Li%
, Loy%
\BCBL {}\ \BBA {} Tang%
}{%
Huang%
\ \protect \BOthers {.}}{%
{\protect \APACyear {2016}}%
}]{%
huang2016learning}
\APACinsertmetastar {%
huang2016learning}%
\begin{APACrefauthors}%
Huang, C.%
, Li, Y.%
, Loy, C\BPBI C.%
\BCBL {}\ \BBA {} Tang, X.%
\end{APACrefauthors}%
\unskip\
\newblock
\APACrefYearMonthDay{2016}{}{}.
\newblock
{\BBOQ}\APACrefatitle {Learning deep representation for imbalanced classification} {Learning deep representation for imbalanced classification}.{\BBCQ}
\newblock
\BIn{} \APACrefbtitle {Proceedings of the IEEE conference on computer vision and pattern recognition} {Proceedings of the ieee conference on computer vision and pattern recognition}\ (\BPGS\ 5375--5384).
\PrintBackRefs{\CurrentBib}

\bibitem [\protect \citeauthoryear {%
Iglesias-Suarez%
\ \protect \BOthers {.}}{%
Iglesias-Suarez%
\ \protect \BOthers {.}}{%
{\protect \APACyear {2023}}%
}]{%
iglesiassuarez2023causallyinformed}
\APACinsertmetastar {%
iglesiassuarez2023causallyinformed}%
\begin{APACrefauthors}%
Iglesias-Suarez, F.%
, Gentine, P.%
, Solino-Fernandez, B.%
, Beucler, T.%
, Pritchard, M.%
, Runge, J.%
\BCBL {}\ \BBA {} Eyring, V.%
\end{APACrefauthors}%
\unskip\
\newblock
\APACrefYearMonthDay{2023}{}{}.
\newblock
\APACrefbtitle {Causally-informed deep learning to improve climate models and projections.} {Causally-informed deep learning to improve climate models and projections.}
\PrintBackRefs{\CurrentBib}

\bibitem [\protect \citeauthoryear {%
Japkowicz%
\ \BBA {} Stephen%
}{%
Japkowicz%
\ \BBA {} Stephen%
}{%
{\protect \APACyear {2002}}%
}]{%
Japkowicz2002}
\APACinsertmetastar {%
Japkowicz2002}%
\begin{APACrefauthors}%
Japkowicz, N.%
\BCBT {}\ \BBA {} Stephen, S.%
\end{APACrefauthors}%
\unskip\
\newblock
\APACrefYearMonthDay{2002}{}{}.
\newblock
{\BBOQ}\APACrefatitle {The class imbalance problem: A systematic study} {The class imbalance problem: A systematic study}.{\BBCQ}
\newblock
\APACjournalVolNumPages{Intelligent Data Analysis}{6}{}{429-449}.
\newblock
\begin{APACrefURL} \url{https://doi.org/10.3233/IDA-2002-6504} \end{APACrefURL}
\newblock
\APACrefnote{5}
\newblock
\begin{APACrefDOI} \doi{10.3233/IDA-2002-6504} \end{APACrefDOI}
\PrintBackRefs{\CurrentBib}

\bibitem [\protect \citeauthoryear {%
Johnson%
\ \BBA {} Khoshgoftaar%
}{%
Johnson%
\ \BBA {} Khoshgoftaar%
}{%
{\protect \APACyear {2019}}%
}]{%
Johnson2019}
\APACinsertmetastar {%
Johnson2019}%
\begin{APACrefauthors}%
Johnson, J\BPBI M.%
\BCBT {}\ \BBA {} Khoshgoftaar, T\BPBI M.%
\end{APACrefauthors}%
\unskip\
\newblock
\APACrefYearMonthDay{2019}{Mar}{19}.
\newblock
{\BBOQ}\APACrefatitle {Survey on deep learning with class imbalance} {Survey on deep learning with class imbalance}.{\BBCQ}
\newblock
\APACjournalVolNumPages{Journal of Big Data}{6}{1}{27}.
\newblock
\begin{APACrefURL} \url{https://doi.org/10.1186/s40537-019-0192-5} \end{APACrefURL}
\newblock
\begin{APACrefDOI} \doi{10.1186/s40537-019-0192-5} \end{APACrefDOI}
\PrintBackRefs{\CurrentBib}

\bibitem [\protect \citeauthoryear {%
Kim%
, Bölöni%
, Borchert%
, Chun%
\BCBL {}\ \BBA {} Achatz%
}{%
Kim%
\ \protect \BOthers {.}}{%
{\protect \APACyear {2021}}%
}]{%
Kim2021JAS}
\APACinsertmetastar {%
Kim2021JAS}%
\begin{APACrefauthors}%
Kim, Y.%
, Bölöni, G.%
, Borchert, S.%
, Chun, H\BPBI Y.%
\BCBL {}\ \BBA {} Achatz, U.%
\end{APACrefauthors}%
\unskip\
\newblock
\APACrefYearMonthDay{2021}{}{}.
\newblock
{\BBOQ}\APACrefatitle {Toward transient subgrid-scale gravity wave representation in atmospheric models. Part II: Wave intermittency simulated with convective sources} {Toward transient subgrid-scale gravity wave representation in atmospheric models. part ii: Wave intermittency simulated with convective sources}.{\BBCQ}
\newblock
\APACjournalVolNumPages{Journal of the Atmospheric Sciences}{78}{}{}.
\newblock
\begin{APACrefDOI} \doi{10.1175/JAS-D-20-0066.1} \end{APACrefDOI}
\PrintBackRefs{\CurrentBib}

\bibitem [\protect \citeauthoryear {%
Kim%
, Eckermann%
\BCBL {}\ \BBA {} Chun%
}{%
Kim%
\ \protect \BOthers {.}}{%
{\protect \APACyear {2003}}%
}]{%
Kimetal2003}
\APACinsertmetastar {%
Kimetal2003}%
\begin{APACrefauthors}%
Kim, Y.%
, Eckermann, S\BPBI D.%
\BCBL {}\ \BBA {} Chun, H.%
\end{APACrefauthors}%
\unskip\
\newblock
\APACrefYearMonthDay{2003}{}{}.
\newblock
{\BBOQ}\APACrefatitle {An overview of the past, present and future of gravity‐wave drag parametrization for numerical climate and weather prediction models} {An overview of the past, present and future of gravity‐wave drag parametrization for numerical climate and weather prediction models}.{\BBCQ}
\newblock
\APACjournalVolNumPages{Atmosphere-Ocean}{41}{}{65-98}.
\newblock
\begin{APACrefURL} \url{https://doi.org/10.3137/ao.410105} \end{APACrefURL}
\newblock
\begin{APACrefDOI} \doi{10.3137/ao.410105} \end{APACrefDOI}
\PrintBackRefs{\CurrentBib}

\bibitem [\protect \citeauthoryear {%
Kingma%
\ \BBA {} Welling%
}{%
Kingma%
\ \BBA {} Welling%
}{%
{\protect \APACyear {2014}}%
}]{%
Kingma2014}
\APACinsertmetastar {%
Kingma2014}%
\begin{APACrefauthors}%
Kingma, D\BPBI P.%
\BCBT {}\ \BBA {} Welling, M.%
\end{APACrefauthors}%
\unskip\
\newblock
\APACrefYearMonthDay{2014}{}{}.
\newblock
{\BBOQ}\APACrefatitle {Auto-encoding variational bayes} {Auto-encoding variational bayes}.{\BBCQ}
\newblock
\APACjournalVolNumPages{2nd International Conference on Learning Representations, ICLR 2014 - Conference Track Proceedings}{}{}{}.
\PrintBackRefs{\CurrentBib}

\bibitem [\protect \citeauthoryear {%
K{\"o}hler%
, Green%
\BCBL {}\ \BBA {} Stephan%
}{%
K{\"o}hler%
\ \protect \BOthers {.}}{%
{\protect \APACyear {2023}}%
}]{%
kohler2023comparing}
\APACinsertmetastar {%
kohler2023comparing}%
\begin{APACrefauthors}%
K{\"o}hler, L.%
, Green, B.%
\BCBL {}\ \BBA {} Stephan, C\BPBI C.%
\end{APACrefauthors}%
\unskip\
\newblock
\APACrefYearMonthDay{2023}{}{}.
\newblock
{\BBOQ}\APACrefatitle {Comparing Loon Superpressure Balloon Observations of Gravity Waves in the Tropics With Global Storm-Resolving Models} {Comparing loon superpressure balloon observations of gravity waves in the tropics with global storm-resolving models}.{\BBCQ}
\newblock
\APACjournalVolNumPages{Journal of Geophysical Research: Atmospheres}{128}{15}{e2023JD038549}.
\PrintBackRefs{\CurrentBib}

\bibitem [\protect \citeauthoryear {%
Krasnopolsky%
, Fox-Rabinovitz%
\BCBL {}\ \BBA {} Chalikov%
}{%
Krasnopolsky%
\ \protect \BOthers {.}}{%
{\protect \APACyear {2005}}%
}]{%
Krasnopolsky2005}
\APACinsertmetastar {%
Krasnopolsky2005}%
\begin{APACrefauthors}%
Krasnopolsky, V\BPBI M.%
, Fox-Rabinovitz, M\BPBI S.%
\BCBL {}\ \BBA {} Chalikov, D\BPBI V.%
\end{APACrefauthors}%
\unskip\
\newblock
\APACrefYearMonthDay{2005}{}{}.
\newblock
{\BBOQ}\APACrefatitle {New Approach to Calculation of Atmospheric Model Physics: Accurate and Fast Neural Network Emulation of Longwave Radiation in a Climate Model} {New approach to calculation of atmospheric model physics: Accurate and fast neural network emulation of longwave radiation in a climate model}.{\BBCQ}
\newblock
\APACjournalVolNumPages{Monthly Weather Review}{133}{5}{1370 - 1383}.
\newblock
\begin{APACrefURL} \url{https://journals.ametsoc.org/view/journals/mwre/133/5/mwr2923.1.xml} \end{APACrefURL}
\newblock
\begin{APACrefDOI} \doi{https://doi.org/10.1175/MWR2923.1} \end{APACrefDOI}
\PrintBackRefs{\CurrentBib}

\bibitem [\protect \citeauthoryear {%
Krueger%
\ \protect \BOthers {.}}{%
Krueger%
\ \protect \BOthers {.}}{%
{\protect \APACyear {2021}}%
}]{%
krueger2021out}
\APACinsertmetastar {%
krueger2021out}%
\begin{APACrefauthors}%
Krueger, D.%
, Caballero, E.%
, Jacobsen, J\BHBI H.%
, Zhang, A.%
, Binas, J.%
, Zhang, D.%
\BDBL {}Courville, A.%
\end{APACrefauthors}%
\unskip\
\newblock
\APACrefYearMonthDay{2021}{}{}.
\newblock
{\BBOQ}\APACrefatitle {Out-of-distribution generalization via risk extrapolation (rex)} {Out-of-distribution generalization via risk extrapolation (rex)}.{\BBCQ}
\newblock
\BIn{} \APACrefbtitle {International Conference on Machine Learning} {International conference on machine learning}\ (\BPGS\ 5815--5826).
\PrintBackRefs{\CurrentBib}

\bibitem [\protect \citeauthoryear {%
Kruse%
\ \protect \BOthers {.}}{%
Kruse%
\ \protect \BOthers {.}}{%
{\protect \APACyear {2022}}%
}]{%
Kruse2022}
\APACinsertmetastar {%
Kruse2022}%
\begin{APACrefauthors}%
Kruse, C\BPBI G.%
, Alexander, M\BPBI J.%
, Hoffmann, L.%
, Niekerk, A\BPBI V.%
, Polichtchouk, I.%
, Bacmeister, J\BPBI T.%
\BDBL {}Stein, O.%
\end{APACrefauthors}%
\unskip\
\newblock
\APACrefYearMonthDay{2022}{}{}.
\newblock
{\BBOQ}\APACrefatitle {Observed and Modeled Mountain Waves from the Surface to the Mesosphere near the Drake Passage} {Observed and modeled mountain waves from the surface to the mesosphere near the drake passage}.{\BBCQ}
\newblock
\APACjournalVolNumPages{Journal of the Atmospheric Sciences}{79}{}{}.
\newblock
\begin{APACrefDOI} \doi{10.1175/JAS-D-21-0252.1} \end{APACrefDOI}
\PrintBackRefs{\CurrentBib}

\bibitem [\protect \citeauthoryear {%
Li%
\ \protect \BOthers {.}}{%
Li%
\ \protect \BOthers {.}}{%
{\protect \APACyear {2022}}%
}]{%
li2022uncertainty}
\APACinsertmetastar {%
li2022uncertainty}%
\begin{APACrefauthors}%
Li, X.%
, Dai, Y.%
, Ge, Y.%
, Liu, J.%
, Shan, Y.%
\BCBL {}\ \BBA {} Duan, L\BHBI Y.%
\end{APACrefauthors}%
\unskip\
\newblock
\APACrefYearMonthDay{2022}{}{}.
\newblock
{\BBOQ}\APACrefatitle {Uncertainty modeling for out-of-distribution generalization} {Uncertainty modeling for out-of-distribution generalization}.{\BBCQ}
\newblock
\APACjournalVolNumPages{arXiv preprint arXiv:2202.03958}{}{}{}.
\PrintBackRefs{\CurrentBib}

\bibitem [\protect \citeauthoryear {%
Ling%
\ \BBA {} Templeton%
}{%
Ling%
\ \BBA {} Templeton%
}{%
{\protect \APACyear {2015}}%
}]{%
Lingetal2015}
\APACinsertmetastar {%
Lingetal2015}%
\begin{APACrefauthors}%
Ling, J.%
\BCBT {}\ \BBA {} Templeton, J.%
\end{APACrefauthors}%
\unskip\
\newblock
\APACrefYearMonthDay{2015}{08}{}.
\newblock
{\BBOQ}\APACrefatitle {{Evaluation of machine learning algorithms for prediction of regions of high Reynolds averaged Navier Stokes uncertainty}} {{Evaluation of machine learning algorithms for prediction of regions of high Reynolds averaged Navier Stokes uncertainty}}.{\BBCQ}
\newblock
\APACjournalVolNumPages{Physics of Fluids}{27}{8}{}.
\newblock
\begin{APACrefURL} \url{https://doi.org/10.1063/1.4927765} \end{APACrefURL}
\newblock
\APACrefnote{085103}
\newblock
\begin{APACrefDOI} \doi{10.1063/1.4927765} \end{APACrefDOI}
\PrintBackRefs{\CurrentBib}

\bibitem [\protect \citeauthoryear {%
Liu%
\ \protect \BOthers {.}}{%
Liu%
\ \protect \BOthers {.}}{%
{\protect \APACyear {2016}}%
}]{%
LiuDNNextreme2016}
\APACinsertmetastar {%
LiuDNNextreme2016}%
\begin{APACrefauthors}%
Liu, Y.%
, Racah, E.%
, Correa, J.%
, Khosrowshahi, A.%
, Lavers, D.%
, Kunkel, K.%
\BDBL {}others%
\end{APACrefauthors}%
\unskip\
\newblock
\APACrefYearMonthDay{2016}{}{}.
\newblock
{\BBOQ}\APACrefatitle {Application of deep convolutional neural networks for detecting extreme weather in climate datasets} {Application of deep convolutional neural networks for detecting extreme weather in climate datasets}.{\BBCQ}
\newblock
\APACjournalVolNumPages{arXiv preprint arXiv:1605.01156}{}{}{}.
\PrintBackRefs{\CurrentBib}

\bibitem [\protect \citeauthoryear {%
Lopez-Gomez%
, McGovern%
, Agrawal%
\BCBL {}\ \BBA {} Hickey%
}{%
Lopez-Gomez%
\ \protect \BOthers {.}}{%
{\protect \APACyear {2022}}%
}]{%
Gomez2022HeatWave}
\APACinsertmetastar {%
Gomez2022HeatWave}%
\begin{APACrefauthors}%
Lopez-Gomez, I.%
, McGovern, A.%
, Agrawal, S.%
\BCBL {}\ \BBA {} Hickey, J.%
\end{APACrefauthors}%
\unskip\
\newblock
\APACrefYearMonthDay{2022}{}{}.
\newblock
{\BBOQ}\APACrefatitle {Global Extreme Heat Forecasting Using Neural Weather Models} {Global extreme heat forecasting using neural weather models}.{\BBCQ}
\newblock
\APACjournalVolNumPages{Artificial Intelligence for the Earth Systems}{}{}{1 - 41}.
\newblock
\begin{APACrefURL} \url{https://journals.ametsoc.org/view/journals/aies/aop/AIES-D-22-0035.1/AIES-D-22-0035.1.xml} \end{APACrefURL}
\newblock
\begin{APACrefDOI} \doi{10.1175/AIES-D-22-0035.1} \end{APACrefDOI}
\PrintBackRefs{\CurrentBib}

\bibitem [\protect \citeauthoryear {%
Lu%
, Jin%
, Pang%
, Zhang%
\BCBL {}\ \BBA {} Karniadakis%
}{%
Lu%
\ \protect \BOthers {.}}{%
{\protect \APACyear {2021}}%
}]{%
lu2021learning}
\APACinsertmetastar {%
lu2021learning}%
\begin{APACrefauthors}%
Lu, L.%
, Jin, P.%
, Pang, G.%
, Zhang, Z.%
\BCBL {}\ \BBA {} Karniadakis, G\BPBI E.%
\end{APACrefauthors}%
\unskip\
\newblock
\APACrefYearMonthDay{2021}{}{}.
\newblock
{\BBOQ}\APACrefatitle {Learning nonlinear operators via DeepONet based on the universal approximation theorem of operators} {Learning nonlinear operators via deeponet based on the universal approximation theorem of operators}.{\BBCQ}
\newblock
\APACjournalVolNumPages{Nature machine intelligence}{3}{3}{218--229}.
\PrintBackRefs{\CurrentBib}

\bibitem [\protect \citeauthoryear {%
Maalouf%
\ \BBA {} Siddiqi%
}{%
Maalouf%
\ \BBA {} Siddiqi%
}{%
{\protect \APACyear {2014}}%
}]{%
maalouf2014weighted}
\APACinsertmetastar {%
maalouf2014weighted}%
\begin{APACrefauthors}%
Maalouf, M.%
\BCBT {}\ \BBA {} Siddiqi, M.%
\end{APACrefauthors}%
\unskip\
\newblock
\APACrefYearMonthDay{2014}{}{}.
\newblock
{\BBOQ}\APACrefatitle {Weighted logistic regression for large-scale imbalanced and rare events data} {Weighted logistic regression for large-scale imbalanced and rare events data}.{\BBCQ}
\newblock
\APACjournalVolNumPages{Knowledge-Based Systems}{59}{}{142--148}.
\PrintBackRefs{\CurrentBib}

\bibitem [\protect \citeauthoryear {%
Maalouf%
\ \BBA {} Trafalis%
}{%
Maalouf%
\ \BBA {} Trafalis%
}{%
{\protect \APACyear {2011}}%
}]{%
maalouf2011robust}
\APACinsertmetastar {%
maalouf2011robust}%
\begin{APACrefauthors}%
Maalouf, M.%
\BCBT {}\ \BBA {} Trafalis, T\BPBI B.%
\end{APACrefauthors}%
\unskip\
\newblock
\APACrefYearMonthDay{2011}{}{}.
\newblock
{\BBOQ}\APACrefatitle {Robust weighted kernel logistic regression in imbalanced and rare events data} {Robust weighted kernel logistic regression in imbalanced and rare events data}.{\BBCQ}
\newblock
\APACjournalVolNumPages{Computational Statistics \& Data Analysis}{55}{1}{168--183}.
\PrintBackRefs{\CurrentBib}

\bibitem [\protect \citeauthoryear {%
Maddox%
, Izmailov%
, Garipov%
, Vetrov%
\BCBL {}\ \BBA {} Wilson%
}{%
Maddox%
\ \protect \BOthers {.}}{%
{\protect \APACyear {2019}}%
}]{%
maddox2019simple}
\APACinsertmetastar {%
maddox2019simple}%
\begin{APACrefauthors}%
Maddox, W\BPBI J.%
, Izmailov, P.%
, Garipov, T.%
, Vetrov, D\BPBI P.%
\BCBL {}\ \BBA {} Wilson, A\BPBI G.%
\end{APACrefauthors}%
\unskip\
\newblock
\APACrefYearMonthDay{2019}{}{}.
\newblock
{\BBOQ}\APACrefatitle {A Simple Baseline for Bayesian Uncertainty in Deep Learning} {A simple baseline for bayesian uncertainty in deep learning}.{\BBCQ}
\newblock
\APACjournalVolNumPages{Advances in Neural Information Processing Systems}{32}{}{}.
\PrintBackRefs{\CurrentBib}

\bibitem [\protect \citeauthoryear {%
Mamalakis%
, Barnes%
\BCBL {}\ \BBA {} Ebert-Uphoff%
}{%
Mamalakis%
\ \protect \BOthers {.}}{%
{\protect \APACyear {2022}}%
}]{%
mamalakis2022investigating}
\APACinsertmetastar {%
mamalakis2022investigating}%
\begin{APACrefauthors}%
Mamalakis, A.%
, Barnes, E\BPBI A.%
\BCBL {}\ \BBA {} Ebert-Uphoff, I.%
\end{APACrefauthors}%
\unskip\
\newblock
\APACrefYearMonthDay{2022}{}{}.
\newblock
{\BBOQ}\APACrefatitle {Investigating the fidelity of explainable artificial intelligence methods for applications of convolutional neural networks in geoscience} {Investigating the fidelity of explainable artificial intelligence methods for applications of convolutional neural networks in geoscience}.{\BBCQ}
\newblock
\APACjournalVolNumPages{Artificial Intelligence for the Earth Systems}{1}{4}{e220012}.
\PrintBackRefs{\CurrentBib}

\bibitem [\protect \citeauthoryear {%
Matsuoka%
\ \protect \BOthers {.}}{%
Matsuoka%
\ \protect \BOthers {.}}{%
{\protect \APACyear {2020}}%
}]{%
Matsuoka2020}
\APACinsertmetastar {%
Matsuoka2020}%
\begin{APACrefauthors}%
Matsuoka, D.%
, Watanabe, S.%
, Sato, K.%
, Kawazoe, S.%
, Yu, W.%
\BCBL {}\ \BBA {} Easterbrook, S.%
\end{APACrefauthors}%
\unskip\
\newblock
\APACrefYearMonthDay{2020}{}{}.
\newblock
{\BBOQ}\APACrefatitle {Application of deep learning to estimate atmospheric gravity wave parameters in reanalysis data sets} {Application of deep learning to estimate atmospheric gravity wave parameters in reanalysis data sets}.{\BBCQ}
\newblock
\APACjournalVolNumPages{Geophysical Research Letters}{47}{19}{e2020GL089436}.
\PrintBackRefs{\CurrentBib}

\bibitem [\protect \citeauthoryear {%
Maulik%
, San%
, Rasheed%
\BCBL {}\ \BBA {} Vedula%
}{%
Maulik%
\ \protect \BOthers {.}}{%
{\protect \APACyear {2019}}%
}]{%
maulik2019subgrid}
\APACinsertmetastar {%
maulik2019subgrid}%
\begin{APACrefauthors}%
Maulik, R.%
, San, O.%
, Rasheed, A.%
\BCBL {}\ \BBA {} Vedula, P.%
\end{APACrefauthors}%
\unskip\
\newblock
\APACrefYearMonthDay{2019}{}{}.
\newblock
{\BBOQ}\APACrefatitle {Subgrid modelling for two-dimensional turbulence using neural networks} {Subgrid modelling for two-dimensional turbulence using neural networks}.{\BBCQ}
\newblock
\APACjournalVolNumPages{Journal of Fluid Mechanics}{858}{}{122--144}.
\PrintBackRefs{\CurrentBib}

\bibitem [\protect \citeauthoryear {%
McGovern%
\ \protect \BOthers {.}}{%
McGovern%
\ \protect \BOthers {.}}{%
{\protect \APACyear {2019}}%
}]{%
McGovern2019}
\APACinsertmetastar {%
McGovern2019}%
\begin{APACrefauthors}%
McGovern, A.%
, Lagerquist, R.%
, Gagne, D\BPBI J.%
, Jergensen, G\BPBI E.%
, Elmore, K\BPBI L.%
, Homeyer, C\BPBI R.%
\BCBL {}\ \BBA {} Smith, T.%
\end{APACrefauthors}%
\unskip\
\newblock
\APACrefYearMonthDay{2019}{}{}.
\newblock
{\BBOQ}\APACrefatitle {Making the black box more transparent: Understanding the physical implications of machine learning} {Making the black box more transparent: Understanding the physical implications of machine learning}.{\BBCQ}
\newblock
\APACjournalVolNumPages{Bulletin of the American Meteorological Society}{100}{}{}.
\newblock
\begin{APACrefDOI} \doi{10.1175/BAMS-D-18-0195.1} \end{APACrefDOI}
\PrintBackRefs{\CurrentBib}

\bibitem [\protect \citeauthoryear {%
Miller%
\ \protect \BOthers {.}}{%
Miller%
\ \protect \BOthers {.}}{%
{\protect \APACyear {2021}}%
}]{%
miller2021accuracy}
\APACinsertmetastar {%
miller2021accuracy}%
\begin{APACrefauthors}%
Miller, J\BPBI P.%
, Taori, R.%
, Raghunathan, A.%
, Sagawa, S.%
, Koh, P\BPBI W.%
, Shankar, V.%
\BDBL {}Schmidt, L.%
\end{APACrefauthors}%
\unskip\
\newblock
\APACrefYearMonthDay{2021}{}{}.
\newblock
{\BBOQ}\APACrefatitle {Accuracy on the line: on the strong correlation between out-of-distribution and in-distribution generalization} {Accuracy on the line: on the strong correlation between out-of-distribution and in-distribution generalization}.{\BBCQ}
\newblock
\BIn{} \APACrefbtitle {International Conference on Machine Learning} {International conference on machine learning}\ (\BPGS\ 7721--7735).
\PrintBackRefs{\CurrentBib}

\bibitem [\protect \citeauthoryear {%
Miloshevich%
, Cozian%
, Abry%
, Borgnat%
\BCBL {}\ \BBA {} Bouchet%
}{%
Miloshevich%
\ \protect \BOthers {.}}{%
{\protect \APACyear {2023}}%
}]{%
PhysRevFluids.8.040501}
\APACinsertmetastar {%
PhysRevFluids.8.040501}%
\begin{APACrefauthors}%
Miloshevich, G.%
, Cozian, B.%
, Abry, P.%
, Borgnat, P.%
\BCBL {}\ \BBA {} Bouchet, F.%
\end{APACrefauthors}%
\unskip\
\newblock
\APACrefYearMonthDay{2023}{Apr}{}.
\newblock
{\BBOQ}\APACrefatitle {Probabilistic forecasts of extreme heatwaves using convolutional neural networks in a regime of lack of data} {Probabilistic forecasts of extreme heatwaves using convolutional neural networks in a regime of lack of data}.{\BBCQ}
\newblock
\APACjournalVolNumPages{Phys. Rev. Fluids}{8}{}{040501}.
\newblock
\begin{APACrefURL} \url{https://link.aps.org/doi/10.1103/PhysRevFluids.8.040501} \end{APACrefURL}
\newblock
\begin{APACrefDOI} \doi{10.1103/PhysRevFluids.8.040501} \end{APACrefDOI}
\PrintBackRefs{\CurrentBib}

\bibitem [\protect \citeauthoryear {%
Nagarajan%
, Andreassen%
\BCBL {}\ \BBA {} Neyshabur%
}{%
Nagarajan%
\ \protect \BOthers {.}}{%
{\protect \APACyear {2020}}%
}]{%
nagarajan2020understanding}
\APACinsertmetastar {%
nagarajan2020understanding}%
\begin{APACrefauthors}%
Nagarajan, V.%
, Andreassen, A.%
\BCBL {}\ \BBA {} Neyshabur, B.%
\end{APACrefauthors}%
\unskip\
\newblock
\APACrefYearMonthDay{2020}{}{}.
\newblock
{\BBOQ}\APACrefatitle {Understanding the failure modes of out-of-distribution generalization} {Understanding the failure modes of out-of-distribution generalization}.{\BBCQ}
\newblock
\APACjournalVolNumPages{arXiv preprint arXiv:2010.15775}{}{}{}.
\PrintBackRefs{\CurrentBib}

\bibitem [\protect \citeauthoryear {%
O'Gorman%
\ \BBA {} Dwyer%
}{%
O'Gorman%
\ \BBA {} Dwyer%
}{%
{\protect \APACyear {2018}}%
}]{%
GormanDwyer2018}
\APACinsertmetastar {%
GormanDwyer2018}%
\begin{APACrefauthors}%
O'Gorman, P\BPBI A.%
\BCBT {}\ \BBA {} Dwyer, J\BPBI G.%
\end{APACrefauthors}%
\unskip\
\newblock
\APACrefYearMonthDay{2018}{}{}.
\newblock
{\BBOQ}\APACrefatitle {Using Machine Learning to Parameterize Moist Convection: Potential for Modeling of Climate, Climate Change, and Extreme Events} {Using machine learning to parameterize moist convection: Potential for modeling of climate, climate change, and extreme events}.{\BBCQ}
\newblock
\APACjournalVolNumPages{Journal of Advances in Modeling Earth Systems}{10}{10}{2548-2563}.
\newblock
\begin{APACrefURL} \url{https://agupubs.onlinelibrary.wiley.com/doi/abs/10.1029/2018MS001351} \end{APACrefURL}
\newblock
\begin{APACrefDOI} \doi{https://doi.org/10.1029/2018MS001351} \end{APACrefDOI}
\PrintBackRefs{\CurrentBib}

\bibitem [\protect \citeauthoryear {%
Oh%
, Rehg%
, Balch%
\BCBL {}\ \BBA {} Dellaert%
}{%
Oh%
\ \protect \BOthers {.}}{%
{\protect \APACyear {2005}}%
}]{%
oh2005data}
\APACinsertmetastar {%
oh2005data}%
\begin{APACrefauthors}%
Oh, S\BPBI M.%
, Rehg, J\BPBI M.%
, Balch, T.%
\BCBL {}\ \BBA {} Dellaert, F.%
\end{APACrefauthors}%
\unskip\
\newblock
\APACrefYearMonthDay{2005}{}{}.
\newblock
{\BBOQ}\APACrefatitle {Data-driven MCMC for learning and inference in switching linear dynamic systems} {Data-driven mcmc for learning and inference in switching linear dynamic systems}.{\BBCQ}
\newblock
\BIn{} \APACrefbtitle {Proceedings of the national conference on artificial intelligence} {Proceedings of the national conference on artificial intelligence}\ (\BVOL~20, \BPG~944).
\PrintBackRefs{\CurrentBib}

\bibitem [\protect \citeauthoryear {%
Ovadia%
\ \protect \BOthers {.}}{%
Ovadia%
\ \protect \BOthers {.}}{%
{\protect \APACyear {2019}}%
}]{%
ovadia2019can}
\APACinsertmetastar {%
ovadia2019can}%
\begin{APACrefauthors}%
Ovadia, Y.%
, Fertig, E.%
, Ren, J.%
, Nado, Z.%
, Sculley, D.%
, Nowozin, S.%
\BDBL {}Snoek, J.%
\end{APACrefauthors}%
\unskip\
\newblock
\APACrefYearMonthDay{2019}{}{}.
\newblock
{\BBOQ}\APACrefatitle {Can you trust your model's uncertainty? evaluating predictive uncertainty under dataset shift} {Can you trust your model's uncertainty? evaluating predictive uncertainty under dataset shift}.{\BBCQ}
\newblock
\APACjournalVolNumPages{Advances in neural information processing systems}{32}{}{}.
\PrintBackRefs{\CurrentBib}

\bibitem [\protect \citeauthoryear {%
Pahlavan%
, Hassanzadeh%
\BCBL {}\ \BBA {} Alexander%
}{%
Pahlavan%
\ \protect \BOthers {.}}{%
{\protect \APACyear {2023}}%
}]{%
pahlavan2023explainable}
\APACinsertmetastar {%
pahlavan2023explainable}%
\begin{APACrefauthors}%
Pahlavan, H\BPBI A.%
, Hassanzadeh, P.%
\BCBL {}\ \BBA {} Alexander, M\BPBI J.%
\end{APACrefauthors}%
\unskip\
\newblock
\APACrefYearMonthDay{2023}{}{}.
\newblock
{\BBOQ}\APACrefatitle {Explainable Offline-Online Training of Neural Networks for Parameterizations: A 1D Gravity Wave-QBO Testbed in the Small-data Regime} {Explainable offline-online training of neural networks for parameterizations: A 1d gravity wave-qbo testbed in the small-data regime}.{\BBCQ}
\newblock
\APACjournalVolNumPages{arXiv preprint arXiv:2309.09024}{}{}{}.
\PrintBackRefs{\CurrentBib}

\bibitem [\protect \citeauthoryear {%
Palmer%
}{%
Palmer%
}{%
{\protect \APACyear {2019}}%
}]{%
palmer2019stochastic}
\APACinsertmetastar {%
palmer2019stochastic}%
\begin{APACrefauthors}%
Palmer, T.%
\end{APACrefauthors}%
\unskip\
\newblock
\APACrefYearMonthDay{2019}{}{}.
\newblock
{\BBOQ}\APACrefatitle {Stochastic weather and climate models} {Stochastic weather and climate models}.{\BBCQ}
\newblock
\APACjournalVolNumPages{Nature Reviews Physics}{1}{7}{463--471}.
\PrintBackRefs{\CurrentBib}

\bibitem [\protect \citeauthoryear {%
Polichtchouk%
, Van~Niekerk%
\BCBL {}\ \BBA {} Wedi%
}{%
Polichtchouk%
\ \protect \BOthers {.}}{%
{\protect \APACyear {2023}}%
}]{%
polichtchouk2023resolved}
\APACinsertmetastar {%
polichtchouk2023resolved}%
\begin{APACrefauthors}%
Polichtchouk, I.%
, Van~Niekerk, A.%
\BCBL {}\ \BBA {} Wedi, N.%
\end{APACrefauthors}%
\unskip\
\newblock
\APACrefYearMonthDay{2023}{}{}.
\newblock
{\BBOQ}\APACrefatitle {Resolved Gravity Waves in the Extratropical Stratosphere: Effect of Horizontal Resolution Increase from O (10) to O (1) km} {Resolved gravity waves in the extratropical stratosphere: Effect of horizontal resolution increase from o (10) to o (1) km}.{\BBCQ}
\newblock
\APACjournalVolNumPages{Journal of the Atmospheric Sciences}{80}{2}{473--486}.
\PrintBackRefs{\CurrentBib}

\bibitem [\protect \citeauthoryear {%
Prein%
\ \protect \BOthers {.}}{%
Prein%
\ \protect \BOthers {.}}{%
{\protect \APACyear {2015}}%
}]{%
prein2015review}
\APACinsertmetastar {%
prein2015review}%
\begin{APACrefauthors}%
Prein, A\BPBI F.%
, Langhans, W.%
, Fosser, G.%
, Ferrone, A.%
, Ban, N.%
, Goergen, K.%
\BDBL {}others%
\end{APACrefauthors}%
\unskip\
\newblock
\APACrefYearMonthDay{2015}{}{}.
\newblock
{\BBOQ}\APACrefatitle {A review on regional convection-permitting climate modeling: Demonstrations, prospects, and challenges} {A review on regional convection-permitting climate modeling: Demonstrations, prospects, and challenges}.{\BBCQ}
\newblock
\APACjournalVolNumPages{Reviews of geophysics}{53}{2}{323--361}.
\PrintBackRefs{\CurrentBib}

\bibitem [\protect \citeauthoryear {%
Psaros%
, Meng%
, Zou%
, Guo%
\BCBL {}\ \BBA {} Karniadakis%
}{%
Psaros%
\ \protect \BOthers {.}}{%
{\protect \APACyear {2023}}%
}]{%
psaros2023uncertainty}
\APACinsertmetastar {%
psaros2023uncertainty}%
\begin{APACrefauthors}%
Psaros, A\BPBI F.%
, Meng, X.%
, Zou, Z.%
, Guo, L.%
\BCBL {}\ \BBA {} Karniadakis, G\BPBI E.%
\end{APACrefauthors}%
\unskip\
\newblock
\APACrefYearMonthDay{2023}{}{}.
\newblock
{\BBOQ}\APACrefatitle {Uncertainty quantification in scientific machine learning: Methods, metrics, and comparisons} {Uncertainty quantification in scientific machine learning: Methods, metrics, and comparisons}.{\BBCQ}
\newblock
\APACjournalVolNumPages{Journal of Computational Physics}{477}{}{111902}.
\PrintBackRefs{\CurrentBib}

\bibitem [\protect \citeauthoryear {%
Qi%
\ \BBA {} Majda%
}{%
Qi%
\ \BBA {} Majda%
}{%
{\protect \APACyear {2020}}%
}]{%
QiMajda2020}
\APACinsertmetastar {%
QiMajda2020}%
\begin{APACrefauthors}%
Qi, D.%
\BCBT {}\ \BBA {} Majda, A\BPBI J.%
\end{APACrefauthors}%
\unskip\
\newblock
\APACrefYearMonthDay{2020}{}{}.
\newblock
{\BBOQ}\APACrefatitle {Using machine learning to predict extreme events in complex systems} {Using machine learning to predict extreme events in complex systems}.{\BBCQ}
\newblock
\APACjournalVolNumPages{Proceedings of the National Academy of Sciences}{117}{1}{52-59}.
\newblock
\begin{APACrefURL} \url{https://www.pnas.org/doi/abs/10.1073/pnas.1917285117} \end{APACrefURL}
\newblock
\begin{APACrefDOI} \doi{10.1073/pnas.1917285117} \end{APACrefDOI}
\PrintBackRefs{\CurrentBib}

\bibitem [\protect \citeauthoryear {%
Ramachandran%
, Zoph%
\BCBL {}\ \BBA {} Le%
}{%
Ramachandran%
\ \protect \BOthers {.}}{%
{\protect \APACyear {2017}}%
}]{%
ramachandran2017searching}
\APACinsertmetastar {%
ramachandran2017searching}%
\begin{APACrefauthors}%
Ramachandran, P.%
, Zoph, B.%
\BCBL {}\ \BBA {} Le, Q\BPBI V.%
\end{APACrefauthors}%
\unskip\
\newblock
\APACrefYearMonthDay{2017}{}{}.
\newblock
{\BBOQ}\APACrefatitle {Searching for activation functions} {Searching for activation functions}.{\BBCQ}
\newblock
\APACjournalVolNumPages{arXiv preprint arXiv:1710.05941}{}{}{}.
\PrintBackRefs{\CurrentBib}

\bibitem [\protect \citeauthoryear {%
Rasp%
}{%
Rasp%
}{%
{\protect \APACyear {2020}}%
}]{%
Rasp2020online}
\APACinsertmetastar {%
Rasp2020online}%
\begin{APACrefauthors}%
Rasp, S.%
\end{APACrefauthors}%
\unskip\
\newblock
\APACrefYearMonthDay{2020}{}{}.
\newblock
{\BBOQ}\APACrefatitle {Coupled online learning as a way to tackle instabilities and biases in neural network parameterizations: general algorithms and Lorenz~96 case study (v1.0)} {Coupled online learning as a way to tackle instabilities and biases in neural network parameterizations: general algorithms and lorenz~96 case study (v1.0)}.{\BBCQ}
\newblock
\APACjournalVolNumPages{Geoscientific Model Development}{13}{5}{2185--2196}.
\newblock
\begin{APACrefURL} \url{https://gmd.copernicus.org/articles/13/2185/2020/} \end{APACrefURL}
\newblock
\begin{APACrefDOI} \doi{10.5194/gmd-13-2185-2020} \end{APACrefDOI}
\PrintBackRefs{\CurrentBib}

\bibitem [\protect \citeauthoryear {%
Rasp%
, Pritchard%
\BCBL {}\ \BBA {} Gentine%
}{%
Rasp%
\ \protect \BOthers {.}}{%
{\protect \APACyear {2018}}%
}]{%
rasp2018deep}
\APACinsertmetastar {%
rasp2018deep}%
\begin{APACrefauthors}%
Rasp, S.%
, Pritchard, M\BPBI S.%
\BCBL {}\ \BBA {} Gentine, P.%
\end{APACrefauthors}%
\unskip\
\newblock
\APACrefYearMonthDay{2018}{}{}.
\newblock
{\BBOQ}\APACrefatitle {Deep learning to represent subgrid processes in climate models} {Deep learning to represent subgrid processes in climate models}.{\BBCQ}
\newblock
\APACjournalVolNumPages{Proceedings of the National Academy of Sciences}{115}{39}{9684--9689}.
\PrintBackRefs{\CurrentBib}

\bibitem [\protect \citeauthoryear {%
Reichstein%
\ \protect \BOthers {.}}{%
Reichstein%
\ \protect \BOthers {.}}{%
{\protect \APACyear {2019}}%
}]{%
Reichstein2019}
\APACinsertmetastar {%
Reichstein2019}%
\begin{APACrefauthors}%
Reichstein, M.%
, Camps-Valls, G.%
, Stevens, B.%
, Jung, M.%
, Denzler, J.%
, Carvalhais, N.%
\BCBL {}\ \BBA {} {Prabhat}.%
\end{APACrefauthors}%
\unskip\
\newblock
\APACrefYearMonthDay{2019}{Feb}{01}.
\newblock
{\BBOQ}\APACrefatitle {Deep learning and process understanding for data-driven Earth system science} {Deep learning and process understanding for data-driven earth system science}.{\BBCQ}
\newblock
\APACjournalVolNumPages{Nature}{566}{7743}{195-204}.
\newblock
\begin{APACrefURL} \url{https://doi.org/10.1038/s41586-019-0912-1} \end{APACrefURL}
\newblock
\begin{APACrefDOI} \doi{10.1038/s41586-019-0912-1} \end{APACrefDOI}
\PrintBackRefs{\CurrentBib}

\bibitem [\protect \citeauthoryear {%
Richter%
\ \protect \BOthers {.}}{%
Richter%
\ \protect \BOthers {.}}{%
{\protect \APACyear {2022}}%
}]{%
Richter2022}
\APACinsertmetastar {%
Richter2022}%
\begin{APACrefauthors}%
Richter, J\BPBI H.%
, Butchart, N.%
, Kawatani, Y.%
, Bushell, A\BPBI C.%
, Holt, L.%
, Serva, F.%
\BDBL {}Yukimoto, S.%
\end{APACrefauthors}%
\unskip\
\newblock
\APACrefYearMonthDay{2022}{4}{}.
\newblock
{\BBOQ}\APACrefatitle {Response of the Quasi-Biennial Oscillation to a warming climate in global climate models} {Response of the quasi-biennial oscillation to a warming climate in global climate models}.{\BBCQ}
\newblock
\APACjournalVolNumPages{Quarterly Journal of the Royal Meteorological Society}{148}{}{1490-1518}.
\newblock
\begin{APACrefDOI} \doi{10.1002/qj.3749} \end{APACrefDOI}
\PrintBackRefs{\CurrentBib}

\bibitem [\protect \citeauthoryear {%
Richter%
, Sassi%
\BCBL {}\ \BBA {} Garcia%
}{%
Richter%
\ \protect \BOthers {.}}{%
{\protect \APACyear {2010}}%
}]{%
Richter2010}
\APACinsertmetastar {%
Richter2010}%
\begin{APACrefauthors}%
Richter, J\BPBI H.%
, Sassi, F.%
\BCBL {}\ \BBA {} Garcia, R\BPBI R.%
\end{APACrefauthors}%
\unskip\
\newblock
\APACrefYearMonthDay{2010}{}{}.
\newblock
{\BBOQ}\APACrefatitle {Toward a physically based gravity wave source parameterization in a general circulation model} {Toward a physically based gravity wave source parameterization in a general circulation model}.{\BBCQ}
\newblock
\APACjournalVolNumPages{Journal of the Atmospheric Sciences}{67}{}{}.
\newblock
\begin{APACrefDOI} \doi{10.1175/2009JAS3112.1} \end{APACrefDOI}
\PrintBackRefs{\CurrentBib}

\bibitem [\protect \citeauthoryear {%
Ross%
, Li%
, Perezhogin%
, Fernandez-Granda%
\BCBL {}\ \BBA {} Zanna%
}{%
Ross%
\ \protect \BOthers {.}}{%
{\protect \APACyear {2022}}%
}]{%
ross2022benchmarking}
\APACinsertmetastar {%
ross2022benchmarking}%
\begin{APACrefauthors}%
Ross, A\BPBI S.%
, Li, Z.%
, Perezhogin, P.%
, Fernandez-Granda, C.%
\BCBL {}\ \BBA {} Zanna, L.%
\end{APACrefauthors}%
\unskip\
\newblock
\APACrefYearMonthDay{2022}{}{}.
\newblock
{\BBOQ}\APACrefatitle {Benchmarking of machine learning ocean subgrid parameterizations in an idealized model} {Benchmarking of machine learning ocean subgrid parameterizations in an idealized model}.{\BBCQ}
\newblock
\APACjournalVolNumPages{Earth and Space Science Open Archive}{}{}{43}.
\newblock
\begin{APACrefURL} \url{https://doi.org/10.1002/essoar.10511742.1} \end{APACrefURL}
\newblock
\begin{APACrefDOI} \doi{10.1002/essoar.10511742.1} \end{APACrefDOI}
\PrintBackRefs{\CurrentBib}

\bibitem [\protect \citeauthoryear {%
Sato%
\ \protect \BOthers {.}}{%
Sato%
\ \protect \BOthers {.}}{%
{\protect \APACyear {2009}}%
}]{%
Sato2009}
\APACinsertmetastar {%
Sato2009}%
\begin{APACrefauthors}%
Sato, K.%
, Watanabe, S.%
, Kawatani, Y.%
, Tomikawa, Y.%
, Miyazaki, K.%
\BCBL {}\ \BBA {} Takahashi, M.%
\end{APACrefauthors}%
\unskip\
\newblock
\APACrefYearMonthDay{2009}{}{}.
\newblock
{\BBOQ}\APACrefatitle {On the origins of mesospheric gravity waves} {On the origins of mesospheric gravity waves}.{\BBCQ}
\newblock
\APACjournalVolNumPages{Geophysical Research Letters}{36}{}{}.
\newblock
\begin{APACrefDOI} \doi{10.1029/2009GL039908} \end{APACrefDOI}
\PrintBackRefs{\CurrentBib}

\bibitem [\protect \citeauthoryear {%
Schneider%
, Jeevanjee%
\BCBL {}\ \BBA {} Socolow%
}{%
Schneider%
\ \protect \BOthers {.}}{%
{\protect \APACyear {2021}}%
}]{%
schneider2021accelerating}
\APACinsertmetastar {%
schneider2021accelerating}%
\begin{APACrefauthors}%
Schneider, T.%
, Jeevanjee, N.%
\BCBL {}\ \BBA {} Socolow, R.%
\end{APACrefauthors}%
\unskip\
\newblock
\APACrefYearMonthDay{2021}{}{}.
\newblock
{\BBOQ}\APACrefatitle {Accelerating progress in climate science} {Accelerating progress in climate science}.{\BBCQ}
\newblock
\APACjournalVolNumPages{Physics Today}{74}{6}{44--51}.
\PrintBackRefs{\CurrentBib}

\bibitem [\protect \citeauthoryear {%
Schneider%
, Lan%
, Stuart%
\BCBL {}\ \BBA {} Teixeira%
}{%
Schneider%
\ \protect \BOthers {.}}{%
{\protect \APACyear {2017}}%
}]{%
Schneider2017}
\APACinsertmetastar {%
Schneider2017}%
\begin{APACrefauthors}%
Schneider, T.%
, Lan, S.%
, Stuart, A.%
\BCBL {}\ \BBA {} Teixeira, J.%
\end{APACrefauthors}%
\unskip\
\newblock
\APACrefYearMonthDay{2017}{}{}.
\newblock
{\BBOQ}\APACrefatitle {Earth System Modeling 2.0: A Blueprint for Models That Learn From Observations and Targeted High-Resolution Simulations} {Earth system modeling 2.0: A blueprint for models that learn from observations and targeted high-resolution simulations}.{\BBCQ}
\newblock
\APACjournalVolNumPages{Geophysical Research Letters}{44}{}{}.
\newblock
\begin{APACrefDOI} \doi{10.1002/2017GL076101} \end{APACrefDOI}
\PrintBackRefs{\CurrentBib}

\bibitem [\protect \citeauthoryear {%
Scinocca%
\ \BBA {} McFarlane%
}{%
Scinocca%
\ \BBA {} McFarlane%
}{%
{\protect \APACyear {2000}}%
}]{%
Scinocca2000}
\APACinsertmetastar {%
Scinocca2000}%
\begin{APACrefauthors}%
Scinocca, J\BPBI F.%
\BCBT {}\ \BBA {} McFarlane, N\BPBI A.%
\end{APACrefauthors}%
\unskip\
\newblock
\APACrefYearMonthDay{2000}{}{}.
\newblock
{\BBOQ}\APACrefatitle {The parametrization of drag induced by stratified flow over anisotropic orography} {The parametrization of drag induced by stratified flow over anisotropic orography}.{\BBCQ}
\newblock
\APACjournalVolNumPages{Quarterly Journal of the Royal Meteorological Society}{126}{}{}.
\newblock
\begin{APACrefDOI} \doi{10.1002/qj.49712656802} \end{APACrefDOI}
\PrintBackRefs{\CurrentBib}

\bibitem [\protect \citeauthoryear {%
Seifert%
\ \BBA {} Rasp%
}{%
Seifert%
\ \BBA {} Rasp%
}{%
{\protect \APACyear {2020}}%
}]{%
SeifertRasp2020}
\APACinsertmetastar {%
SeifertRasp2020}%
\begin{APACrefauthors}%
Seifert, A.%
\BCBT {}\ \BBA {} Rasp, S.%
\end{APACrefauthors}%
\unskip\
\newblock
\APACrefYearMonthDay{2020}{}{}.
\newblock
{\BBOQ}\APACrefatitle {Potential and Limitations of Machine Learning for Modeling Warm-Rain Cloud Microphysical Processes} {Potential and limitations of machine learning for modeling warm-rain cloud microphysical processes}.{\BBCQ}
\newblock
\APACjournalVolNumPages{Journal of Advances in Modeling Earth Systems}{12}{12}{e2020MS002301}.
\newblock
\begin{APACrefURL} \url{https://agupubs.onlinelibrary.wiley.com/doi/abs/10.1029/2020MS002301} \end{APACrefURL}
\newblock
\APACrefnote{e2020MS002301 10.1029/2020MS002301}
\newblock
\begin{APACrefDOI} \doi{https://doi.org/10.1029/2020MS002301} \end{APACrefDOI}
\PrintBackRefs{\CurrentBib}

\bibitem [\protect \citeauthoryear {%
Shamekh%
, Lamb%
, Huang%
\BCBL {}\ \BBA {} Gentine%
}{%
Shamekh%
\ \protect \BOthers {.}}{%
{\protect \APACyear {2023}}%
}]{%
ShamekhGentine2023PNAS}
\APACinsertmetastar {%
ShamekhGentine2023PNAS}%
\begin{APACrefauthors}%
Shamekh, S.%
, Lamb, K\BPBI D.%
, Huang, Y.%
\BCBL {}\ \BBA {} Gentine, P.%
\end{APACrefauthors}%
\unskip\
\newblock
\APACrefYearMonthDay{2023}{}{}.
\newblock
{\BBOQ}\APACrefatitle {Implicit learning of convective organization explains precipitation stochasticity} {Implicit learning of convective organization explains precipitation stochasticity}.{\BBCQ}
\newblock
\APACjournalVolNumPages{Proceedings of the National Academy of Sciences}{120}{20}{e2216158120}.
\newblock
\begin{APACrefURL} \url{https://www.pnas.org/doi/abs/10.1073/pnas.2216158120} \end{APACrefURL}
\newblock
\begin{APACrefDOI} \doi{10.1073/pnas.2216158120} \end{APACrefDOI}
\PrintBackRefs{\CurrentBib}

\bibitem [\protect \citeauthoryear {%
Shen%
\ \protect \BOthers {.}}{%
Shen%
\ \protect \BOthers {.}}{%
{\protect \APACyear {2021}}%
}]{%
shen2021towards}
\APACinsertmetastar {%
shen2021towards}%
\begin{APACrefauthors}%
Shen, Z.%
, Liu, J.%
, He, Y.%
, Zhang, X.%
, Xu, R.%
, Yu, H.%
\BCBL {}\ \BBA {} Cui, P.%
\end{APACrefauthors}%
\unskip\
\newblock
\APACrefYearMonthDay{2021}{}{}.
\newblock
{\BBOQ}\APACrefatitle {Towards out-of-distribution generalization: A survey} {Towards out-of-distribution generalization: A survey}.{\BBCQ}
\newblock
\APACjournalVolNumPages{arXiv preprint arXiv:2108.13624}{}{}{}.
\PrintBackRefs{\CurrentBib}

\bibitem [\protect \citeauthoryear {%
Song%
\ \BBA {} Roh%
}{%
Song%
\ \BBA {} Roh%
}{%
{\protect \APACyear {2021}}%
}]{%
songetal2021radiation}
\APACinsertmetastar {%
songetal2021radiation}%
\begin{APACrefauthors}%
Song, H\BHBI J.%
\BCBT {}\ \BBA {} Roh, S.%
\end{APACrefauthors}%
\unskip\
\newblock
\APACrefYearMonthDay{2021}{}{}.
\newblock
{\BBOQ}\APACrefatitle {Improved Weather Forecasting Using Neural Network Emulation for Radiation Parameterization} {Improved weather forecasting using neural network emulation for radiation parameterization}.{\BBCQ}
\newblock
\APACjournalVolNumPages{Journal of Advances in Modeling Earth Systems}{13}{10}{e2021MS002609}.
\newblock
\begin{APACrefURL} \url{https://agupubs.onlinelibrary.wiley.com/doi/abs/10.1029/2021MS002609} \end{APACrefURL}
\newblock
\APACrefnote{e2021MS002609 2021MS002609}
\newblock
\begin{APACrefDOI} \doi{https://doi.org/10.1029/2021MS002609} \end{APACrefDOI}
\PrintBackRefs{\CurrentBib}

\bibitem [\protect \citeauthoryear {%
Srivastava%
, Hinton%
, Krizhevsky%
, Sutskever%
\BCBL {}\ \BBA {} Salakhutdinov%
}{%
Srivastava%
\ \protect \BOthers {.}}{%
{\protect \APACyear {2014}}%
}]{%
srivastava2014dropout}
\APACinsertmetastar {%
srivastava2014dropout}%
\begin{APACrefauthors}%
Srivastava, N.%
, Hinton, G.%
, Krizhevsky, A.%
, Sutskever, I.%
\BCBL {}\ \BBA {} Salakhutdinov, R.%
\end{APACrefauthors}%
\unskip\
\newblock
\APACrefYearMonthDay{2014}{}{}.
\newblock
{\BBOQ}\APACrefatitle {Dropout: a simple way to prevent neural networks from overfitting} {Dropout: a simple way to prevent neural networks from overfitting}.{\BBCQ}
\newblock
\APACjournalVolNumPages{The journal of machine learning research}{15}{1}{1929--1958}.
\PrintBackRefs{\CurrentBib}

\bibitem [\protect \citeauthoryear {%
Stensrud%
}{%
Stensrud%
}{%
{\protect \APACyear {2007}}%
}]{%
Stensrud2007}
\APACinsertmetastar {%
Stensrud2007}%
\begin{APACrefauthors}%
Stensrud, D\BPBI J.%
\end{APACrefauthors}%
\unskip\
\newblock
\APACrefYear{2007}.
\newblock
\APACrefbtitle {Parameterization Schemes: Keys to Understanding Numerical Weather Prediction Models} {Parameterization schemes: Keys to understanding numerical weather prediction models}.
\newblock
\APACaddressPublisher{}{Cambridge University Press}.
\newblock
\begin{APACrefURL} \url{https://www.cambridge.org/core/product/identifier/9780511812590/type/book} \end{APACrefURL}
\newblock
\begin{APACrefDOI} \doi{10.1017/CBO9780511812590} \end{APACrefDOI}
\PrintBackRefs{\CurrentBib}

\bibitem [\protect \citeauthoryear {%
Subel%
, Chattopadhyay%
, Guan%
\BCBL {}\ \BBA {} Hassanzadeh%
}{%
Subel%
\ \protect \BOthers {.}}{%
{\protect \APACyear {2021}}%
}]{%
subel2021data}
\APACinsertmetastar {%
subel2021data}%
\begin{APACrefauthors}%
Subel, A.%
, Chattopadhyay, A.%
, Guan, Y.%
\BCBL {}\ \BBA {} Hassanzadeh, P.%
\end{APACrefauthors}%
\unskip\
\newblock
\APACrefYearMonthDay{2021}{}{}.
\newblock
{\BBOQ}\APACrefatitle {Data-driven subgrid-scale modeling of forced {B}urgers turbulence using deep learning with generalization to higher {R}eynolds numbers via transfer learning} {Data-driven subgrid-scale modeling of forced {B}urgers turbulence using deep learning with generalization to higher {R}eynolds numbers via transfer learning}.{\BBCQ}
\newblock
\APACjournalVolNumPages{Physics of Fluids}{33}{3}{031702}.
\PrintBackRefs{\CurrentBib}

\bibitem [\protect \citeauthoryear {%
Subel%
, Guan%
, Chattopadhyay%
\BCBL {}\ \BBA {} Hassanzadeh%
}{%
Subel%
\ \protect \BOthers {.}}{%
{\protect \APACyear {2023}}%
}]{%
subel2023explaining}
\APACinsertmetastar {%
subel2023explaining}%
\begin{APACrefauthors}%
Subel, A.%
, Guan, Y.%
, Chattopadhyay, A.%
\BCBL {}\ \BBA {} Hassanzadeh, P.%
\end{APACrefauthors}%
\unskip\
\newblock
\APACrefYearMonthDay{2023}{}{}.
\newblock
{\BBOQ}\APACrefatitle {Explaining the physics of transfer learning in data-driven turbulence modeling} {Explaining the physics of transfer learning in data-driven turbulence modeling}.{\BBCQ}
\newblock
\APACjournalVolNumPages{PNAS nexus}{2}{3}{pgad015}.
\PrintBackRefs{\CurrentBib}

\bibitem [\protect \citeauthoryear {%
Y.~Sun%
, Wong%
\BCBL {}\ \BBA {} Kamel%
}{%
Y.~Sun%
\ \protect \BOthers {.}}{%
{\protect \APACyear {2009}}%
}]{%
Sunetal2009DB}
\APACinsertmetastar {%
Sunetal2009DB}%
\begin{APACrefauthors}%
Sun, Y.%
, Wong, A\BPBI K.%
\BCBL {}\ \BBA {} Kamel, M\BPBI S.%
\end{APACrefauthors}%
\unskip\
\newblock
\APACrefYearMonthDay{2009}{}{}.
\newblock
{\BBOQ}\APACrefatitle {Classification of imbalanced data: A review} {Classification of imbalanced data: A review}.{\BBCQ}
\newblock
\APACjournalVolNumPages{International journal of pattern recognition and artificial intelligence}{23}{04}{687--719}.
\PrintBackRefs{\CurrentBib}

\bibitem [\protect \citeauthoryear {%
Y\BPBI Q.~Sun%
, Hassanzadeh%
, Alexander%
\BCBL {}\ \BBA {} Kruse%
}{%
Y\BPBI Q.~Sun%
\ \protect \BOthers {.}}{%
{\protect \APACyear {2023}}%
}]{%
Sunetal2023WRF}
\APACinsertmetastar {%
Sunetal2023WRF}%
\begin{APACrefauthors}%
Sun, Y\BPBI Q.%
, Hassanzadeh, P.%
, Alexander, M\BPBI J.%
\BCBL {}\ \BBA {} Kruse, C\BPBI G.%
\end{APACrefauthors}%
\unskip\
\newblock
\APACrefYearMonthDay{2023}{}{}.
\newblock
{\BBOQ}\APACrefatitle {Quantifying 3D Gravity Wave Drag in a Library of Tropical Convection-Permitting Simulations for Data-Driven Parameterizations} {Quantifying 3d gravity wave drag in a library of tropical convection-permitting simulations for data-driven parameterizations}.{\BBCQ}
\newblock
\APACjournalVolNumPages{Journal of Advances in Modeling Earth Systems}{15}{5}{e2022MS003585}.
\PrintBackRefs{\CurrentBib}

\bibitem [\protect \citeauthoryear {%
Tan%
\ \protect \BOthers {.}}{%
Tan%
\ \protect \BOthers {.}}{%
{\protect \APACyear {2018}}%
}]{%
Tanetal2018}
\APACinsertmetastar {%
Tanetal2018}%
\begin{APACrefauthors}%
Tan, C.%
, Sun, F.%
, Kong, T.%
, Zhang, W.%
, Yang, C.%
\BCBL {}\ \BBA {} Liu, C.%
\end{APACrefauthors}%
\unskip\
\newblock
\APACrefYearMonthDay{2018}{}{}.
\newblock
{\BBOQ}\APACrefatitle {A survey on deep transfer learning} {A survey on deep transfer learning}.{\BBCQ}
\newblock
\BIn{} \APACrefbtitle {Artificial Neural Networks and Machine Learning--ICANN 2018: 27th International Conference on Artificial Neural Networks, Rhodes, Greece, October 4-7, 2018, Proceedings, Part III 27} {Artificial neural networks and machine learning--icann 2018: 27th international conference on artificial neural networks, rhodes, greece, october 4-7, 2018, proceedings, part iii 27}\ (\BPGS\ 270--279).
\PrintBackRefs{\CurrentBib}

\bibitem [\protect \citeauthoryear {%
Wedi%
\ \protect \BOthers {.}}{%
Wedi%
\ \protect \BOthers {.}}{%
{\protect \APACyear {2020}}%
}]{%
wedi2020baseline}
\APACinsertmetastar {%
wedi2020baseline}%
\begin{APACrefauthors}%
Wedi, N\BPBI P.%
, Polichtchouk, I.%
, Dueben, P.%
, Anantharaj, V\BPBI G.%
, Bauer, P.%
, Boussetta, S.%
\BDBL {}others%
\end{APACrefauthors}%
\unskip\
\newblock
\APACrefYearMonthDay{2020}{}{}.
\newblock
{\BBOQ}\APACrefatitle {A baseline for global weather and climate simulations at 1 km resolution} {A baseline for global weather and climate simulations at 1 km resolution}.{\BBCQ}
\newblock
\APACjournalVolNumPages{Journal of Advances in Modeling Earth Systems}{12}{11}{e2020MS002192}.
\PrintBackRefs{\CurrentBib}

\bibitem [\protect \citeauthoryear {%
D.~Wu%
\ \protect \BOthers {.}}{%
D.~Wu%
\ \protect \BOthers {.}}{%
{\protect \APACyear {2021}}%
}]{%
wu2021quantifying}
\APACinsertmetastar {%
wu2021quantifying}%
\begin{APACrefauthors}%
Wu, D.%
, Gao, L.%
, Xiong, X.%
, Chinazzi, M.%
, Vespignani, A.%
, Ma, Y\BHBI A.%
\BCBL {}\ \BBA {} Yu, R.%
\end{APACrefauthors}%
\unskip\
\newblock
\APACrefYearMonthDay{2021}{}{}.
\newblock
\APACrefbtitle {Quantifying Uncertainty in Deep Spatiotemporal Forecasting.} {Quantifying uncertainty in deep spatiotemporal forecasting.}
\PrintBackRefs{\CurrentBib}

\bibitem [\protect \citeauthoryear {%
G.~Wu%
\ \BBA {} Chang%
}{%
G.~Wu%
\ \BBA {} Chang%
}{%
{\protect \APACyear {2003}}%
}]{%
WuChang2003}
\APACinsertmetastar {%
WuChang2003}%
\begin{APACrefauthors}%
Wu, G.%
\BCBT {}\ \BBA {} Chang, E\BPBI Y.%
\end{APACrefauthors}%
\unskip\
\newblock
\APACrefYearMonthDay{2003}{}{}.
\newblock
{\BBOQ}\APACrefatitle {Adaptive Feature-Space Conformal Transformation for Imbalanced-Data Learning} {Adaptive feature-space conformal transformation for imbalanced-data learning}.{\BBCQ}
\newblock
\BIn{} \APACrefbtitle {Proceedings of the Twentieth International Conference on International Conference on Machine Learning} {Proceedings of the twentieth international conference on international conference on machine learning}\ (\BPG~816–823).
\newblock
\APACaddressPublisher{}{AAAI Press}.
\PrintBackRefs{\CurrentBib}

\bibitem [\protect \citeauthoryear {%
Ye%
\ \protect \BOthers {.}}{%
Ye%
\ \protect \BOthers {.}}{%
{\protect \APACyear {2021}}%
}]{%
ye2021towards}
\APACinsertmetastar {%
ye2021towards}%
\begin{APACrefauthors}%
Ye, H.%
, Xie, C.%
, Cai, T.%
, Li, R.%
, Li, Z.%
\BCBL {}\ \BBA {} Wang, L.%
\end{APACrefauthors}%
\unskip\
\newblock
\APACrefYearMonthDay{2021}{}{}.
\newblock
{\BBOQ}\APACrefatitle {Towards a theoretical framework of out-of-distribution generalization} {Towards a theoretical framework of out-of-distribution generalization}.{\BBCQ}
\newblock
\APACjournalVolNumPages{Advances in Neural Information Processing Systems}{34}{}{23519--23531}.
\PrintBackRefs{\CurrentBib}

\bibitem [\protect \citeauthoryear {%
Yosinski%
, Clune%
, Bengio%
\BCBL {}\ \BBA {} Lipson%
}{%
Yosinski%
\ \protect \BOthers {.}}{%
{\protect \APACyear {2014}}%
}]{%
yosinski2014transferable}
\APACinsertmetastar {%
yosinski2014transferable}%
\begin{APACrefauthors}%
Yosinski, J.%
, Clune, J.%
, Bengio, Y.%
\BCBL {}\ \BBA {} Lipson, H.%
\end{APACrefauthors}%
\unskip\
\newblock
\APACrefYearMonthDay{2014}{}{}.
\newblock
{\BBOQ}\APACrefatitle {How transferable are features in deep neural networks?} {How transferable are features in deep neural networks?}{\BBCQ}
\newblock
\APACjournalVolNumPages{Advances in neural information processing systems}{27}{}{}.
\PrintBackRefs{\CurrentBib}

\bibitem [\protect \citeauthoryear {%
Yuval%
\ \BBA {} O’Gorman%
}{%
Yuval%
\ \BBA {} O’Gorman%
}{%
{\protect \APACyear {2020}}%
}]{%
yuval2020stable}
\APACinsertmetastar {%
yuval2020stable}%
\begin{APACrefauthors}%
Yuval, J.%
\BCBT {}\ \BBA {} O’Gorman, P\BPBI A.%
\end{APACrefauthors}%
\unskip\
\newblock
\APACrefYearMonthDay{2020}{}{}.
\newblock
{\BBOQ}\APACrefatitle {Stable machine-learning parameterization of subgrid processes for climate modeling at a range of resolutions} {Stable machine-learning parameterization of subgrid processes for climate modeling at a range of resolutions}.{\BBCQ}
\newblock
\APACjournalVolNumPages{Nature communications}{11}{1}{1--10}.
\PrintBackRefs{\CurrentBib}

\bibitem [\protect \citeauthoryear {%
C.~Zhang%
\ \protect \BOthers {.}}{%
C.~Zhang%
\ \protect \BOthers {.}}{%
{\protect \APACyear {2023}}%
}]{%
zhang2023implementation}
\APACinsertmetastar {%
zhang2023implementation}%
\begin{APACrefauthors}%
Zhang, C.%
, Perezhogin, P.%
, Gultekin, C.%
, Adcroft, A.%
, Fernandez-Granda, C.%
\BCBL {}\ \BBA {} Zanna, L.%
\end{APACrefauthors}%
\unskip\
\newblock
\APACrefYearMonthDay{2023}{}{}.
\newblock
\APACrefbtitle {Implementation and Evaluation of a Machine Learned Mesoscale Eddy Parameterization into a Numerical Ocean Circulation Model.} {Implementation and evaluation of a machine learned mesoscale eddy parameterization into a numerical ocean circulation model.}
\PrintBackRefs{\CurrentBib}

\bibitem [\protect \citeauthoryear {%
D.~Zhang%
, Ahuja%
, Xu%
, Wang%
\BCBL {}\ \BBA {} Courville%
}{%
D.~Zhang%
\ \protect \BOthers {.}}{%
{\protect \APACyear {2021}}%
}]{%
zhang2021can}
\APACinsertmetastar {%
zhang2021can}%
\begin{APACrefauthors}%
Zhang, D.%
, Ahuja, K.%
, Xu, Y.%
, Wang, Y.%
\BCBL {}\ \BBA {} Courville, A.%
\end{APACrefauthors}%
\unskip\
\newblock
\APACrefYearMonthDay{2021}{}{}.
\newblock
{\BBOQ}\APACrefatitle {Can subnetwork structure be the key to out-of-distribution generalization?} {Can subnetwork structure be the key to out-of-distribution generalization?}{\BBCQ}
\newblock
\BIn{} \APACrefbtitle {International Conference on Machine Learning} {International conference on machine learning}\ (\BPGS\ 12356--12367).
\PrintBackRefs{\CurrentBib}

\bibitem [\protect \citeauthoryear {%
Zhu%
, Zabaras%
, Koutsourelakis%
\BCBL {}\ \BBA {} Perdikaris%
}{%
Zhu%
\ \protect \BOthers {.}}{%
{\protect \APACyear {2019}}%
}]{%
zhu2019physics}
\APACinsertmetastar {%
zhu2019physics}%
\begin{APACrefauthors}%
Zhu, Y.%
, Zabaras, N.%
, Koutsourelakis, P\BHBI S.%
\BCBL {}\ \BBA {} Perdikaris, P.%
\end{APACrefauthors}%
\unskip\
\newblock
\APACrefYearMonthDay{2019}{}{}.
\newblock
{\BBOQ}\APACrefatitle {Physics-constrained deep learning for high-dimensional surrogate modeling and uncertainty quantification without labeled data} {Physics-constrained deep learning for high-dimensional surrogate modeling and uncertainty quantification without labeled data}.{\BBCQ}
\newblock
\APACjournalVolNumPages{Journal of Computational Physics}{394}{}{56--81}.
\PrintBackRefs{\CurrentBib}

\end{thebibliography}

%Reference citation instructions and examples:
%
% Please use ONLY \cite and \citeA for reference citations.
% \cite for parenthetical references
% ...as shown in recent studies (Simpson et al., 2019)
% \citeA for in-text citations
% ...Simpson et al. (2019) have shown...
%
%
%...as shown by \citeA{jskilby}.
%...as shown by \citeA{lewin76}, \citeA{carson86}, \citeA{bartoldy02}, and \citeA{rinaldi03}.
%...has been shown \cite{jskilbye}.
%...has been shown \cite{lewin76,carson86,bartoldy02,rinaldi03}.
%... \cite <i.e.>[]{lewin76,carson86,bartoldy02,rinaldi03}.
%...has been shown by \cite <e.g.,>[and others]{lewin76}.
%
% apacite uses < > for prenotes and [ ] for postnotes
% DO NOT use other cite commands (e.g., \citet, \citep, \citeyear, \citealp, etc.).
% \nocite is okay to use to add references from your Supporting Information
%

\end{document}